\documentclass[prd,aps,floats,superscriptaddress,nofootinbib,floatfix,notitlepage]{revtex4-1}
\usepackage{exscale}                  
\usepackage[intlimits]{amsmath}       
\usepackage{amsfonts}
\usepackage{amssymb,amscd}
\usepackage{wrapfig}
\usepackage{graphicx}
\usepackage{array}
\usepackage{bbm}
\usepackage{multirow}
\usepackage[T1]{fontenc}
\usepackage{times}
\usepackage[caption=false]{subfig}
\usepackage[x11names]{xcolor}
\usepackage{epstopdf}
\usepackage{slashed}
\usepackage{braket}
\usepackage{mathtools}
\usepackage{pstricks}
\usepackage{epstopdf}
\usepackage{ulem}
\usepackage{comment}

\newcommand{\sh}[1]{#1\hskip-7pt \diagup}
\newcommand{\Sh}[1]{#1\hskip-10pt \diagup}

\newcommand{\beware}{\marginpar{\bf beware}}
\newcommand{\nf}{N_{f}\,}
\newcommand{\nc}{N_{c}\,}
\newcommand{\cc}{\langle \bar{\psi} \psi \rangle}
\newcommand{\dc}{\langle \widetilde{\bar{\psi} \psi} \rangle}
\newcommand{\qed}{$\mbox{QED}_3$}

\newcommand{\be}{\begin{equation}}
\newcommand{\ee}{\end{equation}}
\newcommand{\bea}{\begin{eqnarray}}
\newcommand{\eea}{\end{eqnarray}}
\newcommand{\mc}[1]{\mathcal{#1}} 
\newcommand{\ms}[1]{\mathscr{#1}}
\newcommand{\mb}[1]{\mbox{#1}} 
\newcommand{\mf}[1]{\mathsf{#1}} 
\newcommand{\nn}{\nonumber}

\newcommand{\commentCF}[1]{{\color{blue}[#1]}}
\newcommand{\commentJB}[1]{{\color{green!85!black!85!red}[#1]}}
\newcommand{\commentNM}[1]{{\color{red}[#1]}}

\newcommand{\lint}[2]{\int\limits^{#2}\frac{ d^4#1}{(2\pi)^4}} 
\newcommand{\upd}[1]{ d#1\text{ }}
\newcommand{\diff}[2]{\frac{ d#1}{ d#2}}
\newcommand{\pdiff}[0]{\partial}
\newcommand{\Eq}[1]{Eq.~(\ref{#1})}
\newcommand{\Eqs}[2]{Eqs.~(\ref{#1})-(\ref{#2})}
\newcommand{\Fig}[1]{Fig.~(\ref{#1})}
\newcommand{\Figs}[2]{Fig.~(\ref{#1})-(\ref{#2})}
\newcommand{\tphi}[0]{\tilde{\varphi}}
\newcommand{\dint}[1]{ d^4 #1\text{ }}
\newcommand{\Tr}[1]{\text{tr} \left[ #1 \right]  }
\newcommand{\HRule}{\rule{\linewidth}{0.5mm}}

\begin{document}

\title{Dynamical quark mass generation in a strong external magnetic field}

\author{Niklas~Mueller}
\affiliation{Institut f\"ur Theoretische Physik, 
 Universit\"at Giessen, 35392 Giessen, Germany}
\affiliation{Institut f\"{u}r Theoretische Physik, Universit\"{a}t
  Heidelberg, 69120 Heidelberg, Germany}
\author{Jacqueline~A.~Bonnet}
\affiliation{Institut f\"ur Theoretische Physik, 
 Universit\"at Giessen, 35392 Giessen, Germany}
\author{Christian~S.~Fischer}
\affiliation{Institut f\"ur Theoretische Physik, 
 Universit\"at Giessen, 35392 Giessen, Germany}

\begin{abstract}
We investigate the effect of a strong magnetic field on dynamical chiral symmetry breaking 
in quenched and unquenched QCD. To this end we apply the Ritus formalism to the coupled set 
of (truncated) Dyson-Schwinger equations for the quark and gluon propagator under the 
presence of an external constant Abelian magnetic field. We work with an approximation
that is trustworthy for large fields $eH > \Lambda_{QCD}^2$ but is not restricted to the
lowest Landau level. We confirm the linear rise of the quark condensate with large external 
field previously found in other studies and observe the transition to the asymptotic power 
law at extremely large fields. We furthermore quantify the validity of the lowest
Landau level approximation and find substantial quantitative differences to the full
calculation even at very large fields. We discuss unquenching effects in the strong field  
propagators, condensate and the magnetic polarization of the vacuum. We find a
significant weakening of magnetic catalysis caused by the back reaction of quarks 
on the Yang-Mills sector. Our results support explanations of the inverse magnetic 
catalysis found in recent lattice studies due to unquenching effects.
 
\end{abstract}

\maketitle

\section{Introduction \label{sec:introduction}}

The study of the influence of Abelian background magnetic fields onto the fundamental properties
of QCD, confinement and dynamical chiral symmetry breaking is a topic of ever growing interest. 
Strong time-dependent magnetic fields may play an important role in the early universe and in 
the initial stages of heavy ion collisions as well as in the interior of dense neutron stars. 
Magnetic fields influence the thermodynamics of QCD, thereby adding an additional dimension to 
the phase diagram and presumably changing the phases of matter found in the latter. From a purely 
theoretical point of view, by tuning an external magnetic field and studying the reaction of QCD,
one obtains important insights into the structure of strongly interacting matter, see e.g. 
\cite{Shovkovy:2012zn,Gatto:2012sp,D'Elia:2012tr,Fukushima:2012vr} and references therein. 

The influence of magnetic fields onto strongly interacting systems has been investigated
in many approaches in the past years. Model calculations involved the quark-meson and 
Nambu-Jona-Lasinio models, see e.g. \cite{Frasca:2011zn,Gatto:2012sp,Fraga:2013ova}. The functional 
renormalisation group has been invoked to quantify the effects of fluctuations beyond
the mean field level \cite{Skokov:2011ib,Fukushima:2012xw,Kojo:2012js,Andersen:2012bq} and 
lattice gauge theory delivered interesting results at zero and finite temperature 
\cite{Buividovich:2008wf,Braguta:2010ej,D'Elia:2011zu,Bali:2011qj,Bali:2012zg,Bali:2012jv,%
Ilgenfritz:2012fw,Ilgenfritz:2013ara,Bali:2013esa,Bonati:2013vba}.  

With respect to chiral symmetry breaking, an important property of fermionic systems has been 
pointed out in Ref.~\cite{Gusynin:1995nb}. {\it Magnetic catalysis} describes the effect that
a non-vanishing external magnetic field induces a dynamically generated fermion mass, even if the
generic interaction strength of the fermionic theory is so small that the system is chirally 
symmetric otherwise. It is currently debated whether this effect persists or is replaced 
by {\it inverse magnetic catalysis} at temperatures around and above the chiral crossover 
of QCD \cite{D'Elia:2011zu,Bali:2012zg,Fukushima:2012kc,Ilgenfritz:2013ara,Bonati:2013vba}. 
Under debate is also the issue of potential condensation of vector mesons in strong 
magnetic field, see \cite{Chernodub:2011mc,Braguta:2011hq,Hidaka:2012mz} and Refs. therein.
Moreover, the reaction of the electric charges of the quarks inside hadrons to a magnetic
field can be used to probe the corresponding color forces. This is the underlying physical 
idea behind the recently proposed dual Wilson loop \cite{Bruckmann:2011zx}.  

In this work, we employ a functional approach to continuum QCD using Dyson-Schwinger equations
(DSEs) to study the influence of magnetic fields onto the quark propagation and the chiral condensate 
\cite{Lee:1997zj,Miransky:2002rp,Leung:2005xz,Ayala:2006sv,Rojas:2008sg,Murguia:2009fs,Watson:2013ghq}. 
While our long-term goal is to describe the phase diagram of QCD under influence of 
the magnetic field, in the following, we restrict ourselves to zero temperature and chemical potential. 
We start out with the quenched theory, using the full Ritus eigenfunction formalism suitable for
strong external fields. In this respect, our study is complementary to Ref.~\cite{Watson:2013ghq}, 
where results for the limit of small fields have been discussed. We describe the corresponding 
formalism in some detail in section \ref{sec:QuantummagneticFields} and present results for the 
quenched theory in section \ref{res:quenched}. We then provide the formalism for the treatment 
of the unquenched theory in section \ref{sec:unquenched} and discuss results for the gluon
and quark propagators, the condensate and the magnetic polarization of the vacuum in 
section \ref{res:unquenched}. We summarize our conclusions in section \ref{summary}.

%
%
\section{Continuum QCD in an external magnetic field}\label{sec:QuantummagneticFields}
\subsection[Fermion Eigenfunctions]{Fermion Eigenfunctions in an Abelian Background Field}\label{fermion-eigen}
There are different ways of treating a quantum field theory in an external $U(1)$ field. 
One is to introduce sources to which the charged fields of the theory can couple. This makes 
particular sense if the Abelian field is weak enough to be treated perturbatively and if it
vanishes asymptotically. There are, however, interesting systems with strong magnetic fields
like neutron stars or heavy ion collisions, where the effects of the magnetic fields must be
included to all orders. Furthermore, magnetically non-interacting asymptotic states can only 
be constructed in special cases, as for example for a field that covers only a finite volume.
For very strong fields, one also has to take into account the breaking of Poincar\'{e} 
invariance, rendering the well known expansion in Fourier modes in a perturbative attempt useless.  

The case of a strong external magnetic field is quasi classical and one can treat the interaction 
of the charges with the background Abelian field statistically by solving the equations of motion.
The resulting eigenfunctions can be used for expansions within the field theory one is interested 
in. The advantages of such a procedure are obvious: by transforming into the eigensystem of the 
particle in the background field, one obtains equations of motion which include the background 
field, but are formally equivalent to the one of a free particle, see \Eq{dirac1} below. This 
leads to a set of new Feynman rules that include the interaction with the external field to 
every order. In order to make the paper self-contained, we summarize this procedure in the following.

One begins with the Dirac equation of a fermion in an arbitrary external $U(1)$-valued gauge field
\begin{equation}
(\mb i\gamma\cdot\Pi+m)\Psi(x)=0\label{dirac1}
\end{equation}    
where $\Pi_\mu=\partial_\mu+\mb ieA_\mu(x)$ is the covariant derivative with electric charge $e$. Without 
loss of generality, this charge can be set to one. For simplicity, let us consider $A_\mu(x)=(0,0,Hx,0)$
corresponding to a constant magnetic field along the z-direction. It was shown by Ritus 
\cite{Ritus:1972ky,Ritus:1989sg} that the fermion two-point Green's function can only depend on 
four independent Lorentz scalar structures
\begin{equation}
\gamma\Pi,\quad\sigma F,\quad(F\Pi)^2,\quad \gamma^5FF^*,\label{operator}
\end{equation}
with indices omitted that are being summed over. Hereby, $F$ is the field strength tensor 
of the magnetic field, $F^*$ is its dual and $\sigma^{\mu\nu}=-\mb i/2[\gamma^\mu,\gamma^\nu]$. 
All the operators in \Eq{operator} commute with $(\gamma\Pi)^2=\Pi^2-\frac{1}{2}e\sigma F$, 
thus one is left with solving the eigenvalue equation
\begin{equation}
(\gamma\Pi)^2E_p=p^2E_p\label{eq:eigenvalue}
\end{equation}
with the generic eigenvalue $p^2$ to be determined. Other operators commuting with $\gamma\Pi$
are $\mb i\partial_0$, $\mb i\partial_3$ and $\mb i\partial_2$, corresponding to the eigenvalues 
$p_\parallel=(p_0,p_3)$ and $p_2$. Thus, the eigenfunctions in 0, 2- and 3-direction are still plane 
waves, whereas the 1-direction resembles a harmonic oscillator.

There is still one further operator denoted by 
\begin{equation}
\mathcal{H}=-(\gamma\Pi)^2+\Pi_0^2=\Pi_1^2+\Pi_2^2-eH\Sigma^3,
\end{equation}
that has the same eigenfunctions as the ones in \Eq{operator}. Here, $\Sigma^3$ is the third Pauli 
spin matrix given by $\Sigma^3=\sigma^{12}$. Furthermore, $\mathcal{H}E_p=kE_p$ and the eigenfunctions 
$E_p$ are of the form
\begin{equation}
E_p=E_{p,\sigma}\Delta(\sigma),
\end{equation}
where $\Delta(\sigma)=\frac{1}{2}(1+\sigma\Sigma^3)$ is the spin projector along the z-axis with the
eigenvalues $\sigma=\pm1$. With the above knowledge, the eigenfunctions can be written as 
\begin{equation}
E_{p,\sigma}=N_\sigma e^{\mb i(p_0x_0-p_2x_2-p_3x_3)}F_{k,p_2,\sigma}\,.
\end{equation}
Here, $F_{k,p_2,\sigma}$ is an unknown scalar function and $N_\sigma$ a generic normalization. 
This ansatz can be plugged into \Eq{eq:eigenvalue} and solved. An instructive derivation can 
be found in Ref.~\cite{Murguia:2009fs}. One obtains
\begin{eqnarray}
E_{p,\sigma}(x)&=&N(n)e^{\mb i(p_0x_0-p_2x_2-p_3x_3)}D_n(\rho)\,, \hspace*{10mm} 
\rho = \sqrt{2|eH|}\left(x_1 - \frac{p_2}{eH}\right)\,, \hspace*{10mm} 
N(n)=\frac{\left(4\pi\left| eH\right| \right)^\frac{1}{4}}{\sqrt{n!}},\nonumber
\end{eqnarray}
where $D_n(\rho)$ are the parabolic cylinder functions, which can be expressed in terms of Hermite polynomials
\begin{equation}
D_n(x)=2^{-n/2}e^{-x^2/4}H_n(x/\sqrt{2})
\end{equation}
of order $n=l+\frac{\sigma}{2}sgn(eH)-\frac{1}{2},$
where the positive integer $l$ labels the Landau level.  
Furthermore, between the eigenvalues one finds the relation
\begin{equation}
p^2=p_0^2-p_3^2-k\,,\hspace*{10mm}
k=|eH|(2n+1)+\sigma |eH|,\label{eigenvalues}
\end{equation}
and realizes that $n$ are the eigenvalues of a harmonic oscillator with $n\in \mathbb{N}_0$.
In fact, $l$ is the total angular momentum quantum for each 
Landau level, realized by two spin directions. 
Except for the lowest eigenvalue (the lowest Landau level), every fermionic energy value is 
degenerate with respect to two spin orientations differing by $\pm 1$. Furthermore, the 
transition between two adjacent energy levels (note that these are fermionic eigenstates) is 
identical to a bosonic spin one transition of a harmonic oscillator. One can regroup the 
eigenvalues $n$ and $\sigma$ and replace them by the quantum number $l\in \mathbb{N}_0$ in order to label states of different energy. This regrouping is 
shown exemplarily for the first few Landau levels in the following table,
\renewcommand{\arraystretch}{1.5}
\begin{center}
 \begin{tabular}{|c|c|c|}
\hline
 $l$ \,\, &\,\,  $p_{\perp}=\sqrt{2|eH|l}$\,\,    & $\sqrt{k} = \sqrt{|eH|(2n+1) +\sigma |eH|}$  \\\hline\hline
 0        &    0                       	          & $\left\{ n= 0 \,\, \sigma = -1\right.$\\\hline
 1        & $\sqrt{2|eH|} $            		  & $\left\{\begin{array}{cc}
						      n = 0 \,   &\sigma = +1 \\
						      n = 1  \,  &\sigma = -1
						      \end{array} \right.$\\\hline
 2        & $\sqrt{4|eH|} $            		  & $\left\{\begin{array}{cc}
						      n = 1 \, &  \sigma = +1 \\
						      n = 2  \,  &\sigma = -1
						      \end{array} \right.$    \\\hline
\vdots	  &     \vdots                            & \vdots \\\hline					      
						     
\end{tabular}.
\end{center}
\vspace*{3mm}
The eigenvalue $\sqrt{k}$ can be replaced by  $p_\perp=\sqrt{2|eH|l}$, which has the dimension of momentum.    
The complete set of eigenvalues corresponding to the Ritus eigenfunctions is $(p_0,p_3,p_2,l)$ 
or equivalently the "pseudo-momenta" $(p_0,0,p_\perp,p_3)$. 
Effectively, the magnetic field reduces the problem to 2+1 dimensions, breaking the Euclidean
$O(4)$ symmetry to an $O_\parallel(2)\otimes O_\perp(2)$. The $O_\perp(2)$ symmetry represents 
the gauge freedom, for one could have chosen a different vector potential giving the same 
magnetic field (but a different definition of $p_\perp$). For $l=0$ we have 
$p_\perp=0$, so that the problem is in fact 1+1 dimensional on the lowest Landau level (LLL). 

It has been shown that the resulting Ritus basis is orthonormal and complete \cite{Leung:2005xz,Lee:1997zj},
\begin{eqnarray}
\int d^4x\,\bar{E}_p(x)E_{p'}(x)&=&(2\pi)^4\delta^{(4)}(p-p')\, \Pi(l)\\
\mathclap{\displaystyle\int}\mathclap{\textstyle\sum}\text{ }\frac{ d^4p}{(2\pi)^4}E_p(x)\bar{E}_p(y)&=&(2\pi)^4\delta^{(4)}(x-y), \hspace*{1cm}\mbox{with}\hspace*{1cm}
\mathclap{\displaystyle\int}\mathclap{\textstyle\sum}\text{ }\frac{ d^4p}{(2\pi)^4}=\sum\limits_{l=0}^\infty\int\frac{ d^2p_{\|}}{(2\pi)^4}\int\limits_{-\infty}^{\infty}  dp_2
\label{complete}
\end{eqnarray}
and 
\begin{equation}
\Pi(l)=\left\{\begin{array}{cl} \Delta(\text{sgn}(eH)) & l=0\\ 1 & l>0 \end{array}\right. \,.
\end{equation}

Let us further discuss the properties of an expansion in such eigenfunctions. First, by construction, 
the equations of motion for a fermion in the Ritus basis are formally identical with that of a 
free particle. Hence, they can be used to add a quantum theory, such as QCD, on top. 
The Dirac propagator and the fermion self energy are diagonal in this basis, thus also the self 
energy $\Sigma(x,x')$ satisfies an eigenvalue equation with eigenvalue $\Sigma(p)$ 
\begin{equation}
\int\dint{x'}\Sigma(x,x')E_p(x')=E_p(x)\Sigma(p).
\end{equation}  
There is, however, also a technical difficulty that comes with this method. Since neutral 
particles such as photons and gluons still have plane waves as eigenfunctions, complications 
arise whenever they couple to charged particles. Coupling particles that live in different 
eigenspaces renders the form of the vertex in (pseudo-)momentum space complicated, as shall 
be seen below. As a result, momentum conservation at those vertices is not what one is used 
to from covariant field theory. Whereas physical momentum is not conserved, because of 
the loss of translational invariance caused by the external field, the pseudo-momentum Ritus 
eigenvalues $(p_0,p_3,p_2,l)$ are conserved along every fermion line.

In the following section, we will derive the Dyson-Schwinger equation for the quark propagator in 
Ritus functions. We will not be concerned too much with the distinction between momentum and 
pseudo-momentum, for particles will always be expanded in their eigenbasis and it should be clear 
from the context which eigenvalue is referred to.

\subsection{Quark Dyson-Schwinger Equation in a Background Magnetic field}\label{sec:simpleTesnor}
In order to write down the quark Dyson-Schwinger equation (DSE), we need to expand the fermion fields 
in terms of Ritus eigenfunctions instead of the usual plane wave Fourier representation as discussed 
above. The gluon fields do not couple to the magnetic background field and are still to be 
expanded in plane waves. This is similar with or without quark back-coupling effects to the 
Yang-Mills sector of the theory, since quarks appear in closed loops only. Thus, the fully dressed
gluon remains diagonal in Fourier space. However, in the unquenched theory, the gluons feel the magnetic 
field due to the modification of the vacuum, filled by charged quark anti-quark pairs. The resulting
splitting of the gluon propagator in longitudinal and transverse pieces (with respect to the magnetic 
field) will be discussed later in section \ref{sec:unquenched}. In this section we will treat the 
quenched case, where the gluon remains isotropic. 

Because of these two eigensystems involved, the DSE in a background magnetic field needs a 
systematic investigation. We will therefore follow \cite{Lee:1997zj,Leung:2005xz} and start from the 
DSE in 
position space in order to derive the equation in (pseudo-)momentum space from first principles. 
This leads to a set of modified Feynman rules describing a quantum theory in a background 
magnetic field, with the background treated statistically and to every order implicitly already 
in the propagators and vertices of the theory. 
The magnetic field is considered as constant and, for convenience, directed along the z-axis, 
with $A_\mu=(0,0,Hx,0)$ as before. In principle, also non-constant arbitrary fields can be 
treated within this method, provided one is able to solve for the eigenfunctions. One example,  
where the Ritus eigenfunctions can be found analytically, is an exponentially decaying magnetic 
field, discussed in Ref.~\cite{Murguia:2009fs}.

The Dyson--Schwinger equation in position space and with local interaction is given by
\begin{equation}
S^{-1}(x,y)=S^{-1}_0(x,y)+\Sigma(x,y)\,,
\end{equation}
where the quark self energy reads
\begin{equation}
\Sigma(x,y)=\mb i \,g^2 \,C_F\, \gamma^\mu \,S(x,y) \,\Gamma^\nu(y) \,D_{\mu\nu}(x,y),
\end{equation}
with $C_F\delta_{ij}=(T^aT^a)_{ij}$, and $T$ the $SU(3)$ generators in the fundamental 
representation. Color indices are omitted in the following. One can now expand this equation 
in terms of Ritus eigenfunctions. By multiplying with $\bar{E}_p(x)$ from the left and 
$E_{p'}(y)$ from the right (where $p$ and $p'$ denote the incoming and outgoing pseudo-momenta) 
the integration over x and y yields
\begin{equation}
\int d^4x\,d^4y\text{ }\bar{E}_p(x)S^{-1}(x,y)E_{p'}(y)=
\int d^4x\,d^4y\text{ }\bar{E}_p(x)S_0^{-1}(x,y)E_{p'}(y)
+\int d^4x\,d^4y\text{ }\bar{E}_p(x)\Sigma(x,y)E_{p'}(y)\label{eq:DSE-pos_int}.
\end{equation}
Using the completeness relation \Eq{complete} one obtains
\begin{equation}
(2\pi)^4\delta^{(4)}(p-p')\Pi(l)\left[A_{\|}(p)\mb i\gamma p_{\|}+A_{\perp}(p)i\gamma p_{\perp}+B(p)\right]
=(2\pi)^4\delta^{(4)}(p-p')\Pi(l)\left[\gamma p+m\right]
+\Sigma(p,p')\label{DSEbla},
\end{equation}
where $A_{\|}(p), A_{\perp}(p)$ and $B(p)$ are vector and scalar dressing functions of the quark 
propagator in (pseudo-)momentum space, whereas $\Sigma(p,p')$ denotes the self energy. The 
momentum vectors parallel and perpendicular to the magnetic field direction are denoted by 
$p_{\|}=(p_0,0,0,p_3)^T$ and $p_{\perp}=(0,0,p_2,0)^T $. The self energy term is implicitly 
proportional to $\delta^{(4)}(p-p')\Pi(l)$, a property that will later 
also show up explicitly. The DSE for the quark self energy in the Ritus eigenbasis is then given by
\begin{equation}
\Sigma(p,p')=g^2C_F\int d^4x\text{ } d^4y\text{ }\bar{E}_p(x)\text{ }\gamma^\mu 
S(x,y)\Gamma^\nu(y)D_{\mu\nu}(x,y)E_{p'}(y)\,.\label{eq:selfenergyRitus}
\end{equation}
To evaluate this expression, it is necessary to use the representation of the fermion propagator 
in Ritus eigenfunctions. The eigenvalues of the fermion in an external magnetic field in the 
configuration given above are $(p_0,p_3,p_2,l)$ where $l$ labels the Landau level.
The quantum number $p_2$ is still a "good" (referring to the Fourier eigenfunction) 
quantum number. However, as seen from the previous section, the energy of the fermion is degenerate 
with respect to this eigenvalue. The momentum $p_2$ merely fixes the origin of the $x_1$ component of 
our quantum harmonic oscillator system. The momenta of the fermions are $p_{\|}$ and $p_{\perp}$ 
or $(p_0,\sqrt{2|eH|l},0,p_3)$. The fermion propagator in Ritus representation is given by
\begin{equation}
S(x,y)=\,\,\mathclap{\displaystyle\int}\mathclap{\textstyle\sum}\text{ }\frac{d^4q}{(2\pi)^4}
\text{ }E_q(x)\frac{1}{\mb i\gamma \cdot q_{\|}\,A_{\|}(q)+\mb i\gamma \cdot q_{\perp}\,A_{\perp}(q)+B(q)}\bar{E}_q(y),
\label{eq:quarkRitus}
\end{equation}  
where the sum/integral is over the eigenvalues $(p_0,p_3,p_2,l)$ as given in \Eq{complete}. 
The integration over $p_2$ accounts for the degeneracy of states of one Landau level.
Before proceeding, one should notice that the form of \Eq{eq:quarkRitus} is used here in 
analogy to the vacuum case, accounting for the anisotropy by introducing separate dressing 
functions for the transverse and longitudinal components. In principle, due to the appearance 
of further Lorenz structures ($\propto F_{\mu\nu}$), the fermion propagator could possess 
a richer tensor structure. However, as argued in Ref.~\cite{Leung:2005xz}, any other spin 
dependent tensor 
structures violate a remaining $Z(2)$ symmetry of the system by rendering the position of a putative
pole structure in the quark propagator dependent on the direction of the external field. In our 
numerical investigation of the system, we find support for this point of view. When we take into 
account the additional structures, we obtain non-trivial solutions for  
$A_{\|}(p), A_{\perp}(p),B(p)$ only together with zero dressing functions in these additional 
structures. For brevity, we therefore omit these structures here and in the following from the start.

The isotropic Fourier representation of the Landau gauge gluon, as it is used in the quenched 
approximation, is given by
\begin{equation}
D_{\mu\nu}(x,y)=\int\frac{ d^4k}{(2\pi)^4}\, e^{\mb ik(x-y)} D(k^2) P_{\mu\nu},\label{eq:gluonFourier}
\end{equation}
where the integration is a momentum integration in the conventional sense since the gluon 
is still diagonal in Fourier space. The gluon propagator function $D(k^2)$ is related to the 
dressing function $Z(k^2)$ via $D(k^2)=Z(k^2)/k^2$ and 
$P_{\mu\nu} = \delta_{\mu \nu} - k_\mu k_\nu /k^2$ is the transverse projector.
By plugging \Eq{eq:quarkRitus} and \Eq{eq:gluonFourier} into \Eq{eq:selfenergyRitus}, one obtains
\begin{eqnarray}
\nonumber \Sigma(p,p')=g^2C_F\text{ } \,\,\mathclap{\displaystyle\int}\mathclap{\textstyle\sum}
\frac{ d^4q}{(2\pi)^4}\int\frac{ d^4k}{(2\pi)^4}\int d^4x \,d^4y&\Bigg\{\bar{E}_p(x)\gamma^\mu 
E_q(x)\frac{1}{A_{\|}(q)\mb i\gamma \cdot q_{\|}+A_{\perp}(q)\mb i\gamma \cdot q_{\perp}+B(q)}\bar{E}_q(y)
\Gamma^\nu E_{p'}(y) e^{\mb ik(x-y)} D(k^2) P_{\mu \nu}
\Bigg\}\,.\label{eq:dse-self-raw}
\end{eqnarray}
The quenched gluon propagator $D(k^2)$ is very well known, both from lattice calculations and 
solutions of the corresponding DSEs (without background magnetic field) 
\cite{Leinweber:1998uu,Cucchieri:2007md,Bogolubsky:2009dc,Aguilar:2008xm,Fischer:2008uz,Huber:2012kd}.
In order to facilitate our later treatment of the unquenched gluon DSE, we use the lattice results 
of \cite{Fischer:2010fx} as input for our study. The dressed quark-gluon vertex is a much more 
difficult object, which is not known in detail even for the case of vanishing background 
fields. For the purpose of this study and in order to make the equations 
tractable, we resort to a simple ansatz of the form $\Gamma^\nu \, \rightarrow\, \gamma^\nu \Gamma(k^2)$,
where $\Gamma(k^2)$ is taken to be independent of the magnetic field. The explicit form of this 
ansatz is discussed below and given in appendix \ref{gluonvertex}. Naturally, these approximations
should be improved in future work. However, the following sections will mainly be focussed on the 
necessary techniques and qualitative behavior of the Dyson--Schwinger equations in a magnetic 
background. Other truncations might complicate some calculations done here, but the general 
features may be expected to be similar.
 
The integral over x, involving a product of Ritus- and Fourier eigenfunctions, is given by 
\begin{equation}
\int d^4x\text{ }\bar{E}_p(x)\gamma^\mu E_q(x)e^{\mb ikx}\label{eq:vertexint}.
\end{equation}
A similar integral for y remains to be done. If we had only particles in Ritus or in Fourier
eigenfunctions at the quark-gluon vertex, this integral would be trivial 
due to the fact that those two systems are complete orthonormal vector spaces. 
We would simply obtain delta functions ensuring eigenvalue/momentum conservation. 
In our case however,  we have interacting particles that are 
diagonal in different bases. Nevertheless, the integral in \Eq{eq:vertexint} can be done 
analytically. This involves the Fourier transform of a product of parabolic cylinder functions 
and yields
\begin{eqnarray}
\int d^4x\, \bar{E}_p(x)\gamma^\mu E_q(x)e^{\mb ikx}&=&(2\pi)^4\delta^{(3)}(q+k-p) \,
e^{-k^2_\perp/4|eH|} e^{\mb ik_1(q_2+p_2)/2eH}\\
&&\times\sum\limits_{\sigma_1,\sigma_2=\pm } 
\frac{e^{\mb i \text{sgn}(eH)\left(n(\sigma_1,l)-n(\sigma_2,l_q)\right)\phi}}
{\sqrt{n(\sigma_1,l)!n(\sigma_2,l_q)!}} 
J_{n(\sigma_1,l)n(\sigma_2,l_q)} (k_\perp)\,\, \Delta(\sigma_1)\gamma^\mu\Delta(\sigma_2)\nonumber
\end{eqnarray}
with the abbreviations
\begin{eqnarray}
k_\perp^2 =k_1^2+k_2^2\,, \hspace*{10mm}
n(\sigma,l)=l+\frac{\sigma}{2}\text{sgn}(eH)-\frac{1}{2}\,, \hspace*{10mm}
\phi &=&\arctan(k_2/k_1)\label{eq:n}.
\end{eqnarray}
Furthermore,
\begin{equation}
J_{n_1n_2}\equiv \sum\limits_{m=0}^{\text{min}(n_1,n_2)}\frac{n_1!n_2!}{m!(n_1-m)!(n_2-m)!}
\left(\mb i\text{sgn}(eH)k_\perp\frac{\sqrt{2|eH|}}{2eH} \right)^{n_1+n_2-2m} \,.
\end{equation}
Composing all the bits and pieces, the quark self energy now reads
\begin{eqnarray}
\Sigma(p,p')&=&(2\pi)^4\delta^{(3)}(p-p')g^2C_F \sum \limits_{l_q}\int\frac{ d^2q_\parallel}{(2\pi)^4} 
\int\limits_{-\infty}^\infty dq_2 \int\limits_{-\infty}^\infty dk_1 e^{-k_\perp^2/2\left|eH\right|}
\sum\limits_{\sigma_1,\sigma_2,\sigma_3,\sigma_4}
\frac{e^{\mb i\text{sgn}(eH)(n_1-n_2+n_3-n_4)\phi}}{\sqrt{n_1!n_2!n_3!n_4!}}\label{eq:self:energy-full} \\
&&\times J_{n_1n_2}(k_\perp)J_{n_3n_4}(k_\perp) \Delta(\sigma_1)\gamma^\mu \Delta(\sigma_2)
\frac{1}{A_{\|}(q)\mb i\gamma \cdot q_{\|}+A_{\perp}(q)\mb i\gamma \cdot q_{\perp}+B(q)}
\Delta(\sigma_3)\gamma^\nu\Delta(\sigma_4)
P^{\mu\nu}(k)\Gamma(k^2)D(k^2)\,.\nonumber 
\end{eqnarray}
Here, the sum over $l_q$ counts the Landau levels and the spin projection sums over 
$\sigma_{1\dots4}$ realize their degeneracies. 
The expression (\ref{eq:self:energy-full}) is exact with respect to the treatment of 
the magnetic field. Unfortunately, 
it is extremely difficult to solve numerically. The reason for that lies in the form of the 
functions $J_{nm}$, as can be seen when using an alternative derivation. Starting from 
\Eq{eq:dse-self-raw}, it can be shown that \Eq{eq:self:energy-full} is identical to
\begin{eqnarray}\label{eq:self-Laguerre}
\Sigma(p,p') &=& (2\pi)^4\delta^{(3)}(p-p')g^2C_F \sum\limits_{l_q}\int\frac{ d^2q_\parallel}{(2\pi)^4} 
\int\limits_{-\infty}^\infty dq_2 \int\limits_{-\infty}^\infty dk_1
e^{-k_\perp^2/2\left|eH\right|} \\&&\times
\sum\limits_{\sigma_1,\sigma_2,\sigma_3,\sigma_4} e^{\mb i\text{sgn} (eH)(n_1-n_2+n_3-n_4)\phi}n_1!n_3! 
\left(\mb i\frac{k_\perp}{\sqrt{2\left|eH\right|}} \right)^{n_2-n_1+n_4-n_3}
 L_{n_1}^{n_2-n_1}\left( \frac{k_\perp^2}{2\left|eH\right|}\right)L_{n_3}^{n_4-n_3}
 \left( \frac{k_\perp^2}{2\left|eH\right|}\right) \nonumber\\
 &&\times \,\Delta(\sigma_1)\gamma^\mu \Delta(\sigma_2)\frac{1}{A_{\|}(q)\mb i\gamma \cdot q_{\|}+
 A_{\perp}(q)\mb i\gamma \cdot q_{\perp}+B(q)}\Delta(\sigma_3)\gamma^\nu\Delta(\sigma_4)
P^{\mu\nu}(k)\Gamma(k^2)D(k^2), \nonumber 
\end{eqnarray}
where $L_n^m(x)$ are the generalized Laguerre polynomials. Therefore, solving \Eq{eq:self-Laguerre}
numerically involves an integration routine that is precise for an integrand that behaves like a 
polynomial of order $n$. According to \Eq{eq:n}, $n$ is proportional to the number of Landau levels $l$, 
lying arbitrary close to each other for small $eH$, and hence a numerical treatment of the above 
expression is extremely hard. This is unfortunate for QCD, where in general the gluon dressing 
function or the quark gluon vertex are not known analytically and numerical approaches are 
the only available tool.

An approximation of the above expressions for small magnetic fields is discussed 
in \cite{Watson:2013ghq}. Here, we follow the opposite strategy and consider the case where the 
magnetic field is sufficiently large \cite{Leung:2005xz}. To this end, note that the integrand 
in the quark self energy is given as a 
function of $k_\perp/2\left|eH\right|$, where large values of $k_\perp$ are essentially suppressed.
If the magnetic field is large, only terms up to the smallest order in $k_\perp/2\left|eH\right|$ 
need to be kept. We adopt this approximation in the following, keeping in mind that our results will not
be reliable in the small field limit. 

In this approximation, the vertex is simplified drastically and given by \cite{Leung:2005xz}
\begin{equation}\label{approx}
J_{nm}(k_\perp)\,\rightarrow\, \frac{[\text{max}(n,m)]!}{\left| n-m \right| !} (\mb ik_\perp  / \sqrt{2 \left| eH\right| })^{ \left| n-m \right| }\,\rightarrow \,n !\,\delta_{nm}.
\end{equation}
One then has
\begin{equation}
\int d^4x\,\bar{E}_p(x)\gamma^\mu E_q(x)e^{\mb ikx} = (2\pi)^4\delta^{(3)}(q+k-p)\, 
e^{-k^2_\perp/4|eH|}e^{\mb ik_1(q_2+p_2)/2eH} \sum\limits_{\sigma_1,\sigma_2 }
\delta_{n(\sigma_1,l)n(\sigma_2,l_q)}\,\Delta(\sigma_1)\gamma^\mu\Delta(\sigma_2)\label{eq:vertex-Simple}
\end{equation}
and thus
\begin{equation}
\begin{split}
\Sigma(p,p')=(2\pi)^4\delta^{(3)}(p-p')\mb ig^2C_F\sum\limits_{l_q=0}^\infty\int\frac{ d^2q_{\|}}{(2\pi)^4}
\int\limits_{-\infty}^{\infty} dq_2\int\limits_{-\infty}^{\infty} dk_1\text{ }e^{-k_\perp^2/2|eH|}
\sum\limits_{\sigma _1 \sigma _2 \sigma _3 \sigma_4}
\delta_{n(\sigma_1,l)n(\sigma_2,l_q)}\delta_{n(\sigma_3,l_q)n(\sigma_4,l')}\\\times
\Delta(\sigma_1)\gamma^\mu\Delta(\sigma_2)\frac{1}{A_{\|}(q)\gamma \cdot q_{\|}+A_{\perp}(q)\gamma \cdot q_{\perp}+B(q)}\Delta(\sigma_3)\gamma^\nu\Delta(\sigma_4)D(k^2)\Gamma(k^2)P^{\mu\nu}(k)\label{eq:fullselfRiyues}
\end{split}
\end{equation}
As compared to the case of zero background field, which would yield a factor 
$(2\pi)^4\delta^{(4)}(q+k-p)\gamma^\mu$ in front of the integral, here only the momenta 
$\delta^{(3)}(q+k-p)\equiv\delta(q_0+k_0-p_0)\delta(q_3+k_3-p_3)\delta(q_2+k_2-p_2)$ are conserved
as already discussed above. In the integrand, the additional factor $\delta_{n(\sigma_1,l)n(\sigma_2,l_q)}$
allows for transitions between adjacent Landau levels. This does not come as a surprise, since 
the eigensystem of the quark in an Abelian background field is supersymmetric in the sense that 
transitions between two neighbouring Landau levels constitute spin one transitions (i.e. from 
$\pm1/2$ to $\mp1/2$). Gluons, perpendicular wrt. the magnetic field, will therefore induce such 
transitions, whereas longitudinal gluons do not change the Landau level of incoming and outgoing 
quarks at the vertex. The $U(1)$ field breaks the initial $O(4)$ symmetry to an $O(2)$ symmetry in 
the t-z-plane. This explains the modification of the vertex
$$
\gamma^\mu\rightarrow\Delta(\sigma_1)\gamma^\mu\Delta(\sigma_2),
$$
seen in \Eq{eq:fullselfRiyues} as compared to the zero field case. 

These considerations can be easily generalized to unquenched QCD. The only difference is the 
appearance of an anisotropy in the gluon dressing functions, accounting for the modified behavior 
of the gluon polarization. We will discuss this below in section \ref{sec:unquenched}.

The relations 
\begin{eqnarray}
\Delta(\sigma)\gamma^\mu_{\|}=\gamma^\mu_{\|}\Delta(\sigma)\,,
\hspace*{1cm}
\Delta(\sigma)\gamma^\mu_{\perp}=\gamma^\mu_{\perp}\Delta(-\sigma)\,,
\hspace*{1cm}
\Delta(\sigma_a)\Delta(\sigma_b)=\Delta(\sigma_a)\delta_{ab}
\end{eqnarray}
are useful to decompose the vertex into two contributions
\begin{equation}
\Delta(\sigma_1)\gamma^\mu\Delta(\sigma_2)=\Delta(\sigma_1)\left(\gamma_{\|}^\mu
+\gamma_{\perp}^\mu\right)\Delta(\sigma_2)=\delta_{\sigma_1,\sigma_2}\Delta(\sigma_1)
\gamma_{\|}^\mu+\delta_{\sigma_1,-\sigma_2}\Delta(\sigma_1)
\gamma_{\perp}^\mu.\label{eq:vertex-decomp}
\end{equation}
Furthermore, tracing over the spin projector gives
\begin{equation}
\sum\limits_\sigma\text{Tr} \left[ \Delta(\sigma) \right] \rightarrow\chi(l)=\left\{\begin{array}{cl} 4, & l>0\\ 2, & l=0 \end{array}\right. \label{eqchi}
\end{equation}
since for $l=0$ the fermion can only have $\sigma=\text{sgn}(eH)$.
After performing the traces in the quark DSE and using the abbreviation
$\int_q \equiv \int\frac{d^2q_{\|}}{(2\pi)^4}\,\int\limits_{-\infty}^{\infty} dq_2\, dk_1\,$ 
we obtain
\begin{eqnarray}
B(p)|_{l_p=l}&=&Z_2 m+Z_{1f} g^2 C_F\int_q
\left\{\left(\left.\frac{B(q)}{B^2(q)+A_{\|}^2(q)q_{\|}^2+A_\perp^2(q)q_\perp^2} \right)\right|_{l_q=l}
e^{-k_\perp^2/2|eH|}\left(2-\frac{k_{\|}^2}{k^2} \right)
D(k^2)\Gamma(k^2)\right\} \\
&&+\,
\frac{g^2C_F}{p_\parallel^2}\frac{2}{\chi(l)}\sum_{l_q= l\pm 1}\int_q
\Bigg\{\left.\left(\frac{B(q)}{B(q)^2(q)+A_{\|}^2(q)q_{\|}^2+A_\perp^2(q)q_\perp^2}\right)\right|_{l_q}
e^{-k_\perp^2/2|eH|}\left( 2-\frac{k_\perp^2}{k^2}\right)
D(k^2)\Gamma(k^2)\Bigg\}\nonumber\label{eq:B_simple}
\end{eqnarray}
where $k_2=q_2-p_2$. Although $p_2$ appears explicitly here, it can be seen from the 
form of the integrand that the final result does not depend on it, as expected. Without 
loss of generality, we will set $p_2=0$ therefore. For the vector dressing functions we find,
\begin{eqnarray}
A_{\parallel}(p)|_{l_p=l}&=& Z_2-Z_{1f}\frac{g^2C_F}{p_\parallel^2}\int_q
\Bigg\{\left.\left(\frac{A_\parallel(q)}{B(q)^2(q)+A_{\|}^2(q)q_{\|}^2+A_\perp^2(q)q_\perp^2}\right)\right|_{l_q=l}
e^{-k_\perp^2/2|eH|} K_1(p,q) D(k^2)\Gamma(k^2)\Bigg\}\\
&&+\,
\frac{g^2C_F}{p_\parallel^2}\frac{2}{\chi(l)}\sum_{l_q=l \pm 1}\int_q
\Bigg\{\left.\left(\frac{A_\parallel(q)}{B(q)^2(q)+A_{\|}^2(q)q_{\|}^2+A_\perp^2(q)q_\perp^2}\right)\right|_{l_q}
\,e^{-k_\perp^2/2|eH|}
K_2(p,q) D(k^2)\Gamma(k^2)\Bigg\} \nonumber
\end{eqnarray}
with kernels
\begin{eqnarray}
K_1(p,q) &=&  p_\parallel q_\parallel\cos(\varphi)\frac{k_\parallel^2}{k^2} -
2\frac{(q_\parallel p_\parallel\cos(\varphi)-p_\parallel^2)
(q_\parallel^2-q_\parallel p_\parallel\cos(\varphi))}{k^2} \nonumber\\
K_2(p,q) &=& \left( 2-\frac{k_\perp^2}{k^2}\right)p_\parallel q_\parallel\cos(\varphi)
\end{eqnarray}
and $\cos(\varphi)= \frac{\vec{p_{\|}}\cdot \vec{q_{\|}}}{|p_{\|}||q_{\|}|}$. Furthermore,
\begin{eqnarray}
A_{\perp}(p)|_{l_p=l}&=& Z_2+Z_{1f}\frac{g^2C_F}{p_\parallel^2}\int_q
\Bigg\{\left.\left(\frac{A_\perp(q)}{B(q)^2(q)+A_{\|}^2(q)q_{\|}^2+A_\perp^2(q)q_\perp^2}\right)\right|_{l_q=l}
e^{-k_\perp^2/2|eH|}\left(2-\frac{k_{\|}^2}{k^2} \right)p_\perp q_\perp D(k^2)\Gamma(k^2)\Bigg\}
\nonumber\\
&-& \frac{g^2C_F}{p_\parallel^2}\frac{2}{\chi(l)}\sum_{l_q=l \pm 1}\int_q \Bigg\{\left.\left(
\frac{A_\perp(q)}{B(q)^2(q)+A_{\|}^2(q)q_{\|}^2+A_\perp^2(q)q_\perp^2}\right)\right|_{l_q}
e^{-k_\perp^2/2|eH|}\frac{k_1^2-k_2^2}{k^2}
p_\perp q_\perp D(k^2)\Gamma(k^2)\Bigg\}\label{eq:A_T_simple},
\end{eqnarray}
where $\chi(l)$ is given by \Eq{eqchi}. The renormalisation factors of the quark propagator 
and the quark-gluon vertex are denoted by $Z_2$ and $Z_{1f}$. 
Note that the contributions to the self-energy consist of two terms. The first one 
describes the radiation and emission of a "longitudinal" gluon which is polarized in 
the z-t-plane, as indicated by $=g_\parallel^{\mu\nu}-k_\parallel^\mu k_\parallel^\nu/k^2$. 
Such a gluon does not induce transitions between Landau levels. However, the second 
term corresponds to Landau level transitions, it is accompanied by a gluon 
$\propto g_\perp^{\mu\nu}-k_\perp^\mu k_\perp^\nu/k^2$. In the latter case the gluon 
emission can either increase or decrease the Landau level of the internal quark, except 
for the case of the lowest Landau level, where only a transition up to the second Landau 
level can happen (for there are no negative Landau levels). This decomposition is a 
direct result of \Eq{eq:vertex-decomp}. Mixed terms, such as 
$\Delta(\sigma_1)\gamma_\parallel\Delta(\sigma_2)\dots\Delta(\sigma_3)\gamma_\perp \Delta(\sigma_4)$, 
do not appear as they would violate conservation of the Ritus eigenvalues.

\Eqs{eq:B_simple}{eq:A_T_simple} can be solved numerically once the dressed gluon propagator
and the quark-gluon vertex have been determined. As explained above, for the gluon propagator,
we employ a fit to the lattice results given in Ref.~\cite{Fischer:2010fx}; for the vertex
we use an ansatz that satisfies both, the correct ultraviolet running from resummed
perturbation theory (with vanishing external field) and an approximate Slavnov-Taylor
identity in the infrared, see Ref.~\cite{Fischer:2012vc} for details. The ansatz naturally
takes into account an infrared enhancement of the quark-gluon interaction as discussed in
\cite{Kojo:2012js}; the explicit form is given in appendix \ref{gluonvertex}. 
\Eqs{eq:B_simple}{eq:A_T_simple} are then solved 
numerically on logarithmic integration grids using standard numerical methods. 
The sum over the Landau levels is 
carried out explicitly up to a sufficiently large number of discrete Landau levels and 
the remaining part of the sum is treated as an integral. Due to the dependence 
$p_\perp\propto\sqrt{2\left| eH \right|l}$, the density of Landau levels per energy 
interval grows and the error due to this approximation can be neglected, once the 
level spacing is sufficiently small. In practice, we use 80-100 Landau levels in the 
explicit summation part. 

\section{Results for quenched QCD}\label{res:quenched}

\begin{figure}[b!]
\subfloat[$B(p)$ {[}GeV{]} \,\,($eH=$ 0.5 GeV$^2$)]
{\includegraphics[width=5.8cm]{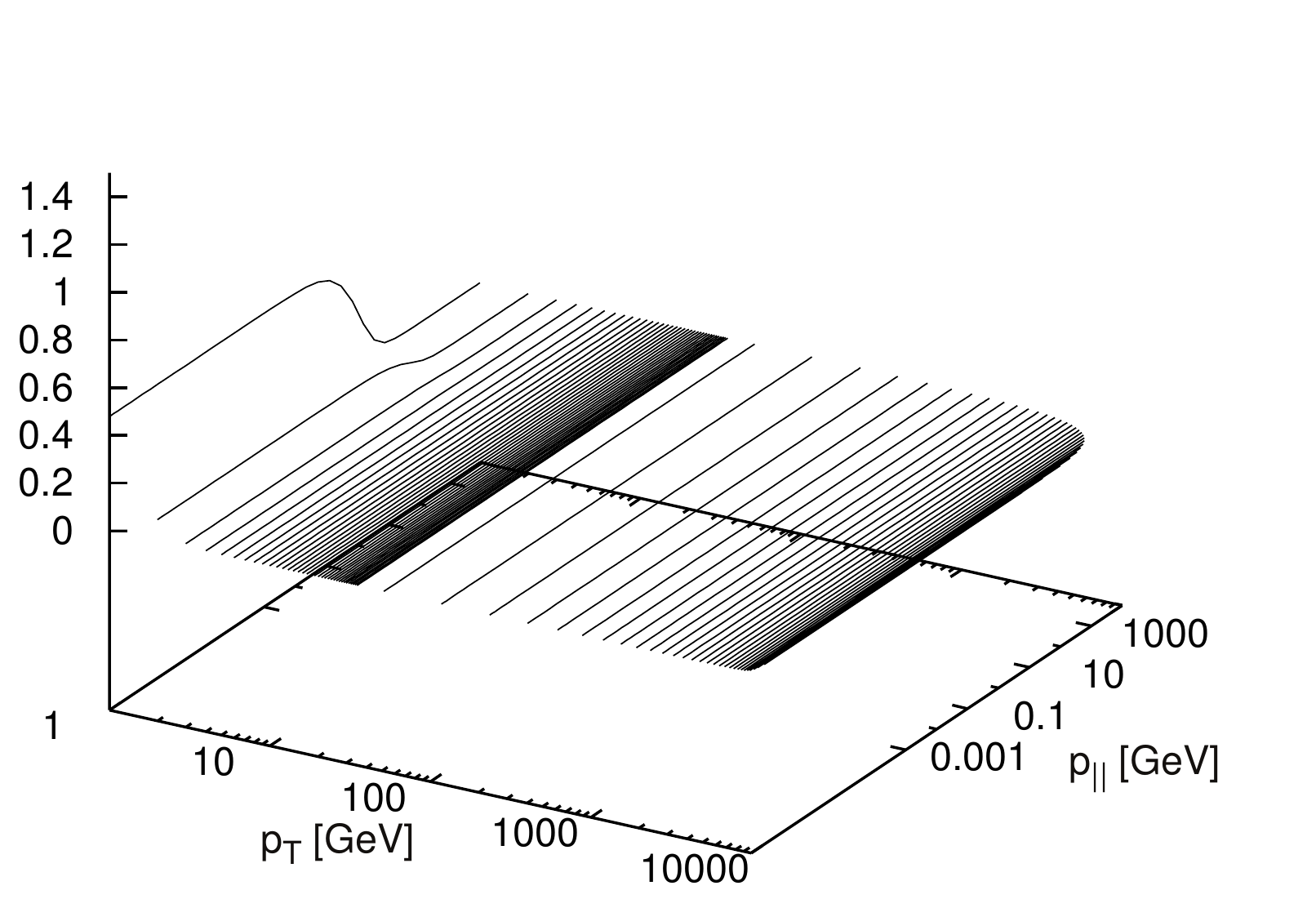}}
\subfloat[$A_\parallel(p)$ \,\,($eH=$ 0.5 GeV$^2$)]
{\includegraphics[width=5.8cm]{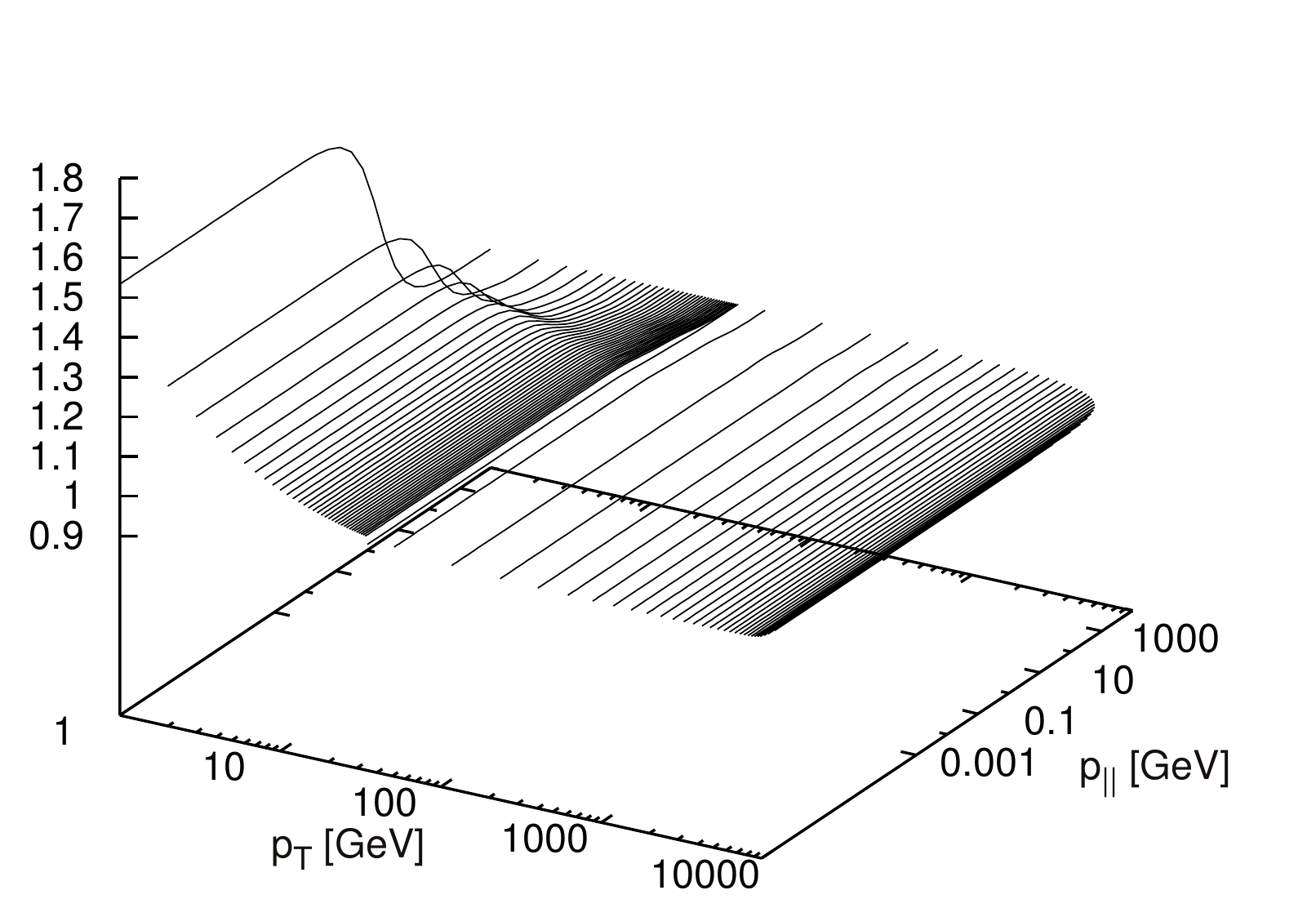}}
\subfloat[$A_\perp(p)$ \,\,($eH=$ 0.5 GeV$^2$)]
{\includegraphics[width=5.8cm]{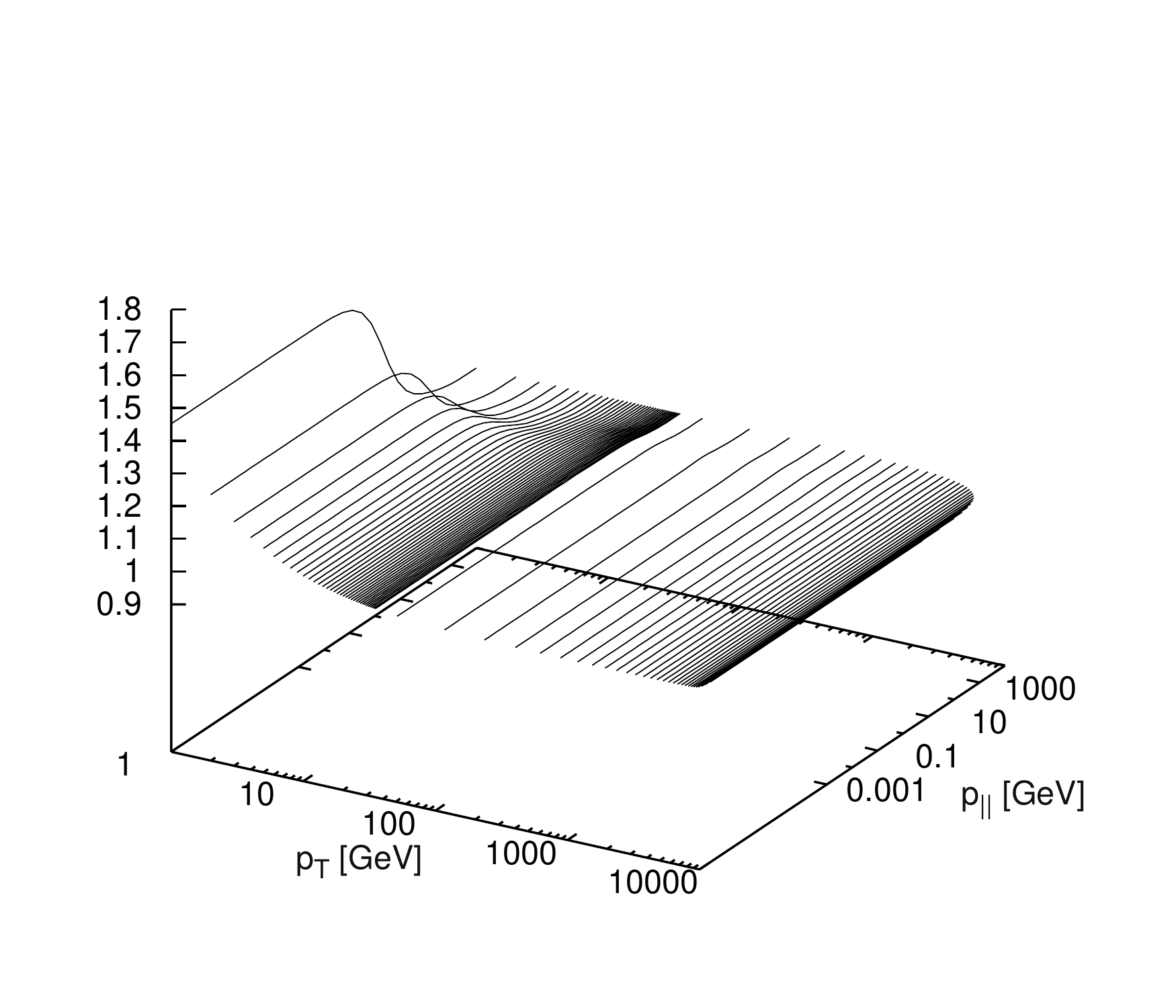}}\\
\subfloat[$B(p)$ {[}GeV{]} \,\,($eH=$ 1 GeV$^2$)]
{\includegraphics[width=5.8cm]{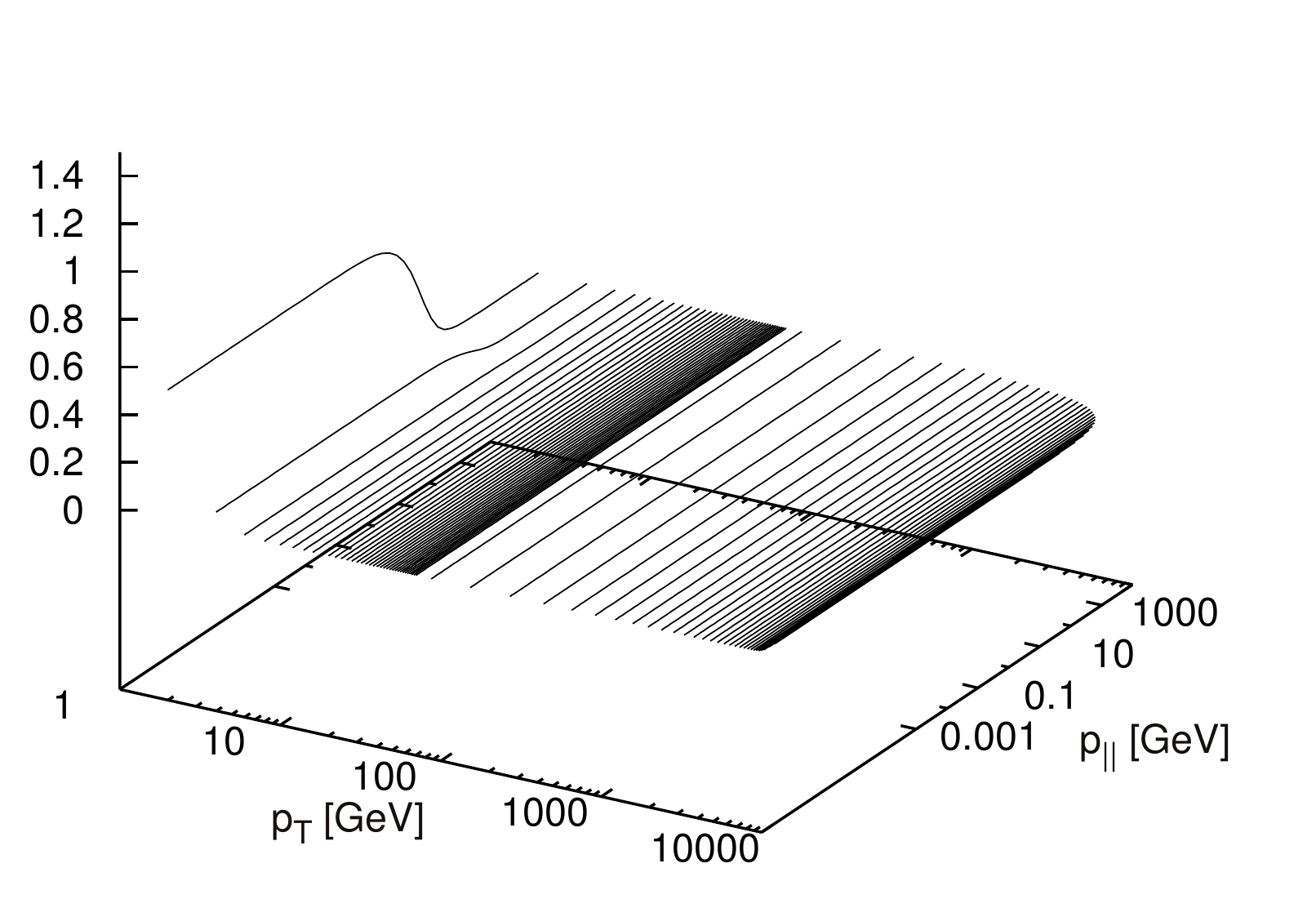}}
\subfloat[$A_\parallel(p)$ \,\,($eH=$ 1 GeV$^2$)]
{\includegraphics[width=5.8cm]{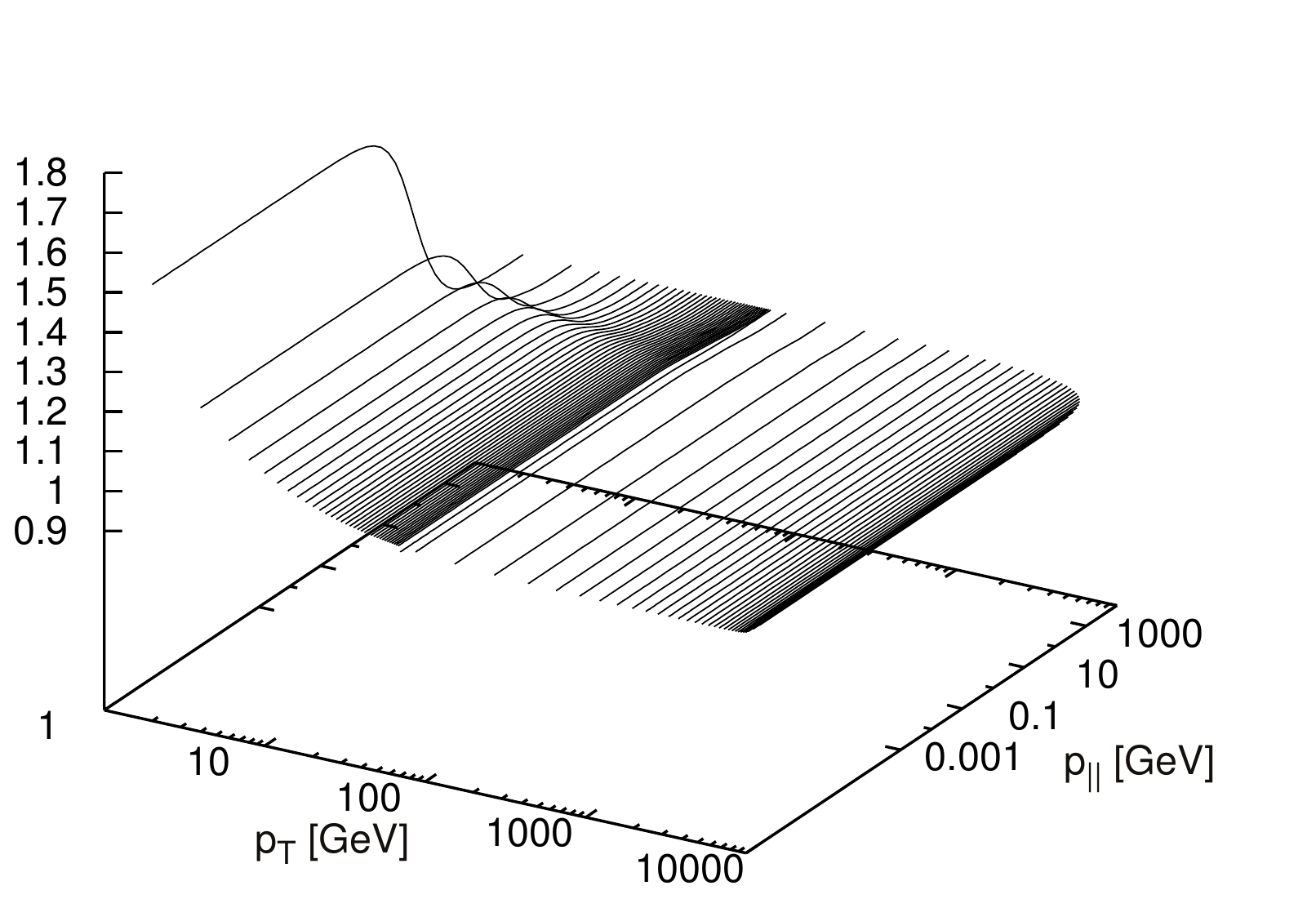}}
\subfloat[$A_\perp(p)$ \,\,($eH=$ 1 GeV$^2$)]
{\includegraphics[width=5.8cm]{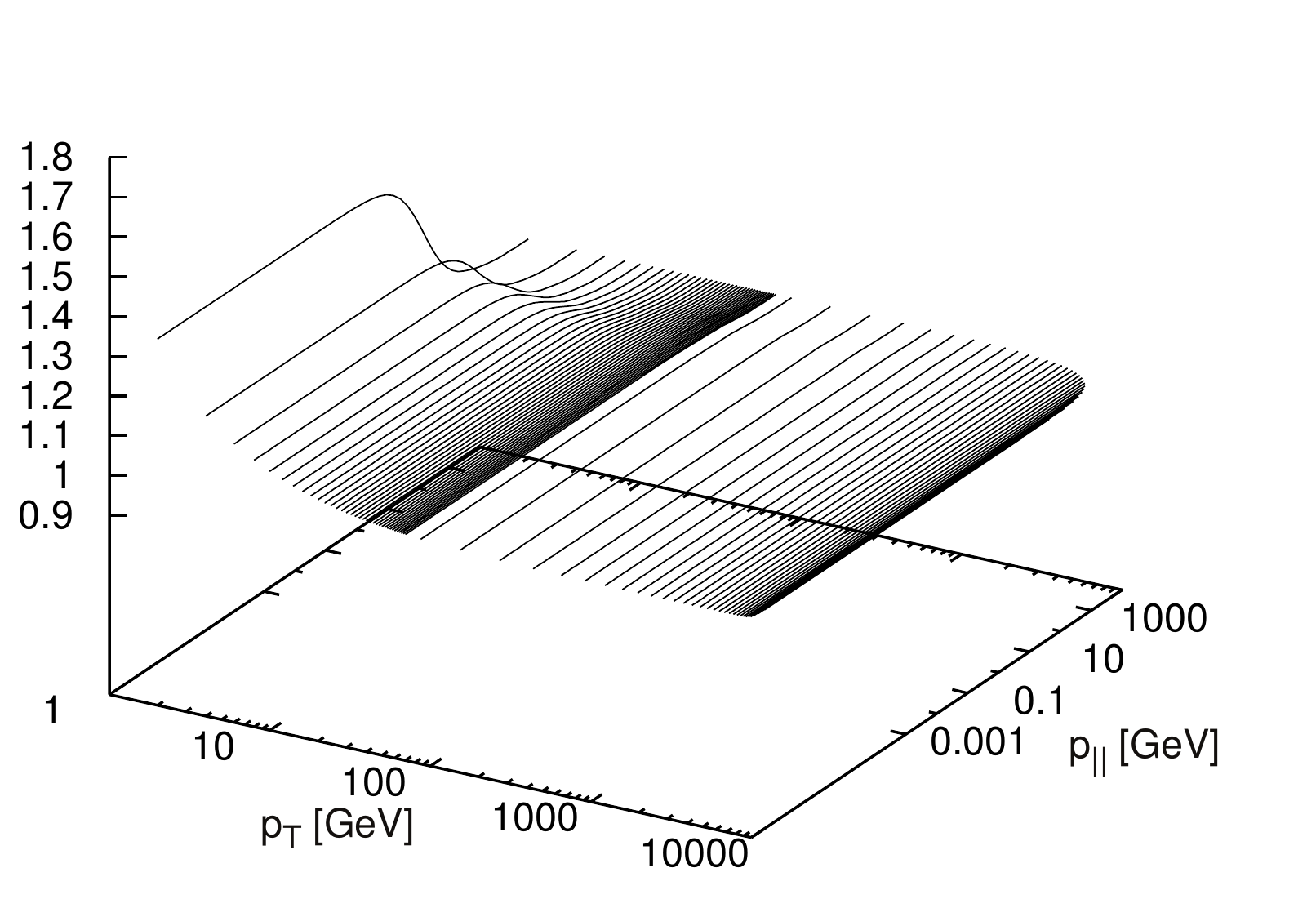}}\\
\subfloat[$B(p)$ {[}GeV{]} \,\,($eH=$ 4 GeV$^2$)]
{\includegraphics[width=5.8cm]{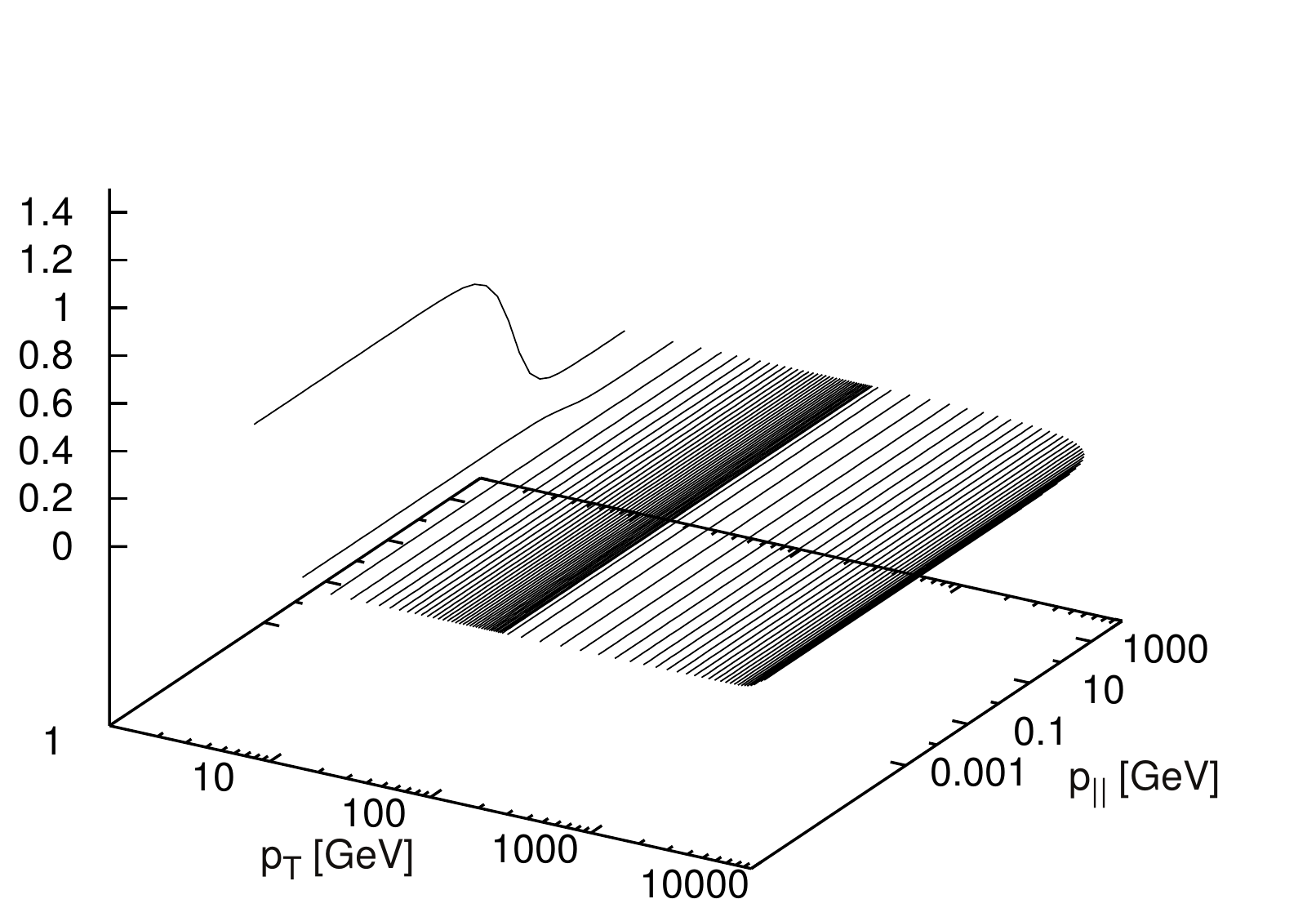}}
\subfloat[$A_\parallel(p)$ \,\,($eH=$ 4 GeV$^2$)]
{\includegraphics[width=5.8cm]{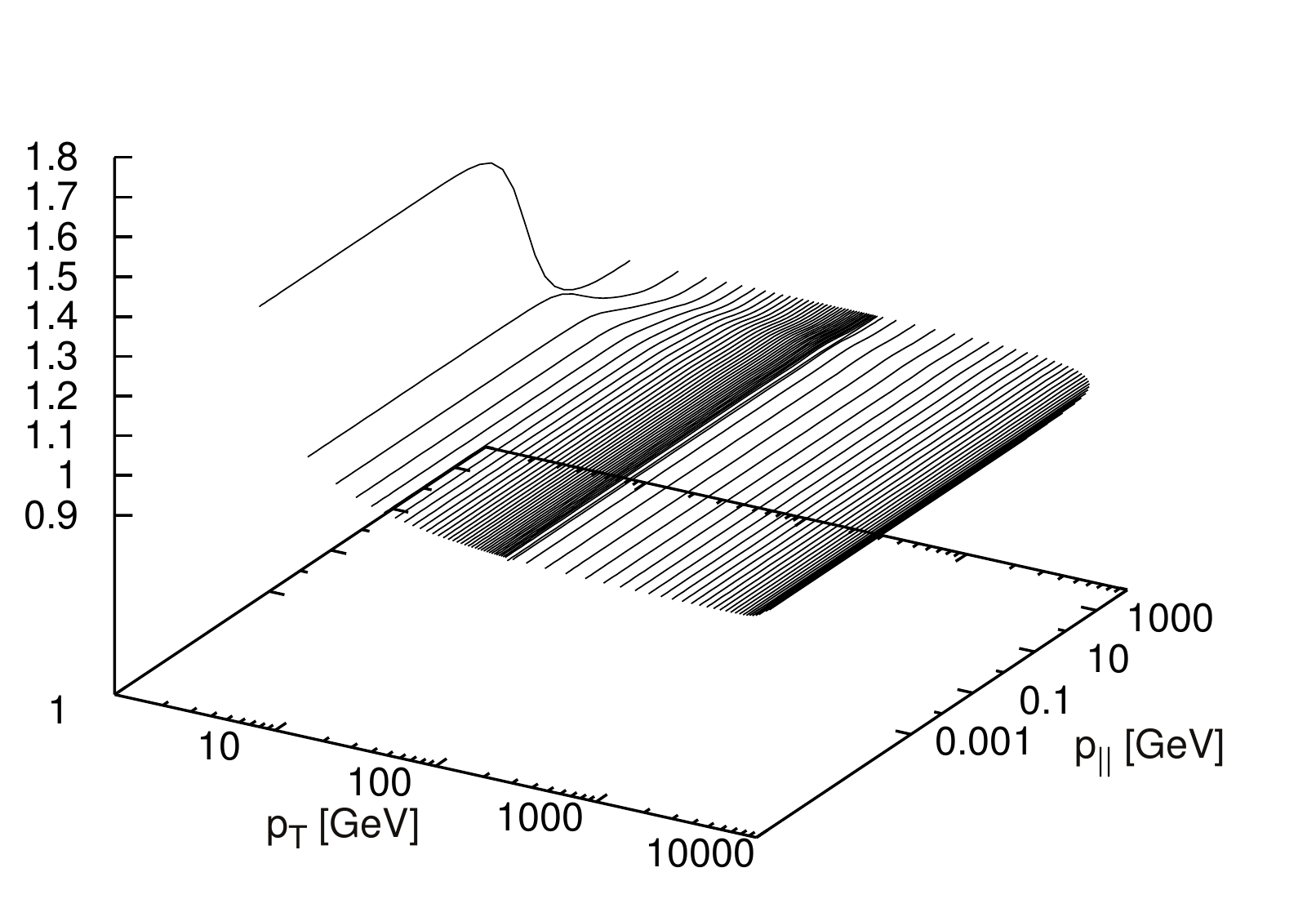}}
\subfloat[$A_\perp(p)$ \,\,($eH=$ 4 GeV$^2$)]
{\includegraphics[width=5.8cm]{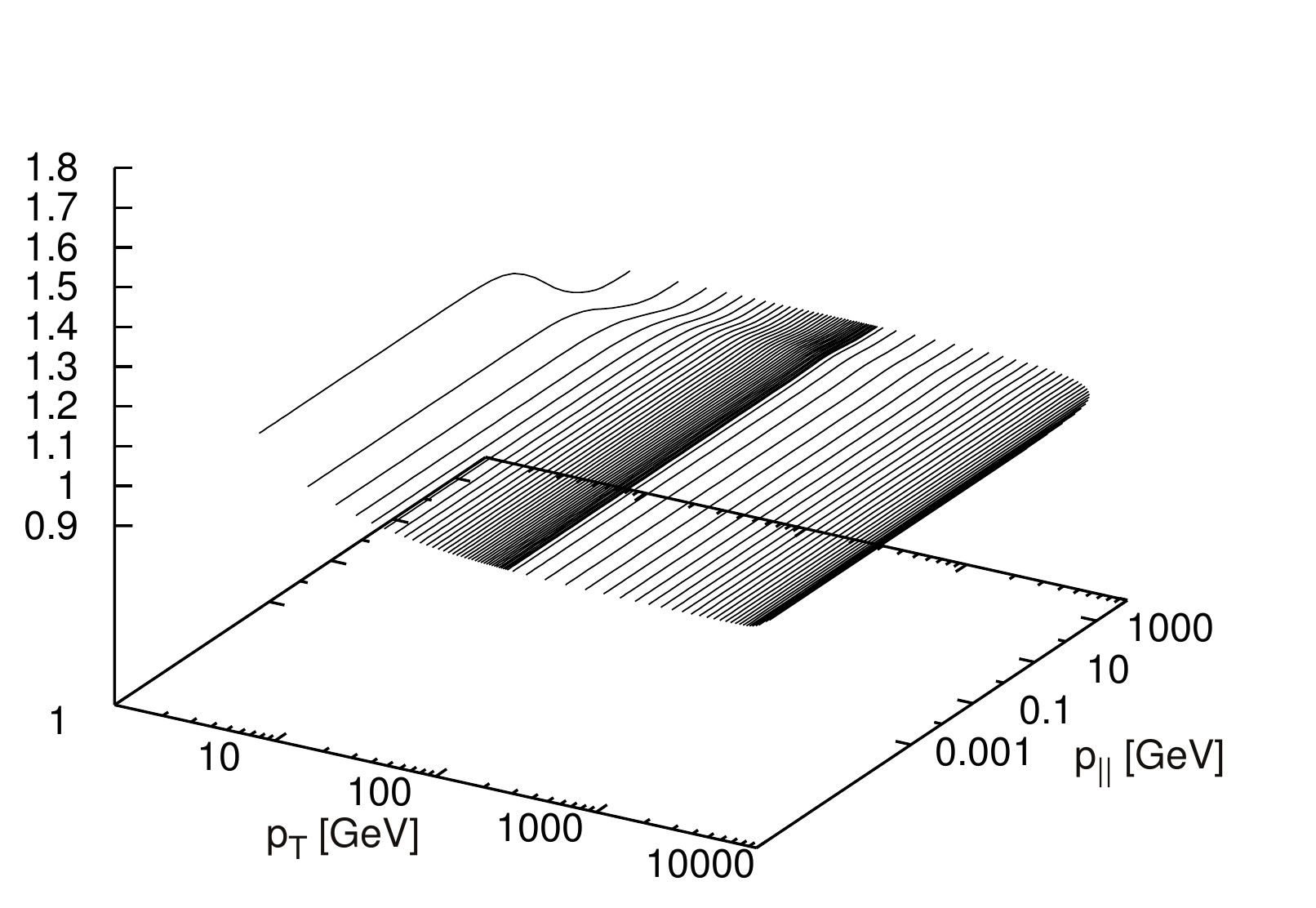}}
\caption{Quark dressing functions for $eH=$ 0.5 $\text{GeV}^2$ (first line), $eH=$ 1 $\text{GeV}^2$ 
(second line) and $eH=$ 4 $\text{GeV}^2$ (third line). Shown are the individual Landau levels
as a function of $p_\parallel^2$ and $p_\perp^2$ starting with the second lowest level $l=1$. 
}
\label{fig:2D-dressing-Simple-1GeV}
\end{figure}

\begin{figure}[t!]
\includegraphics[scale=.32]{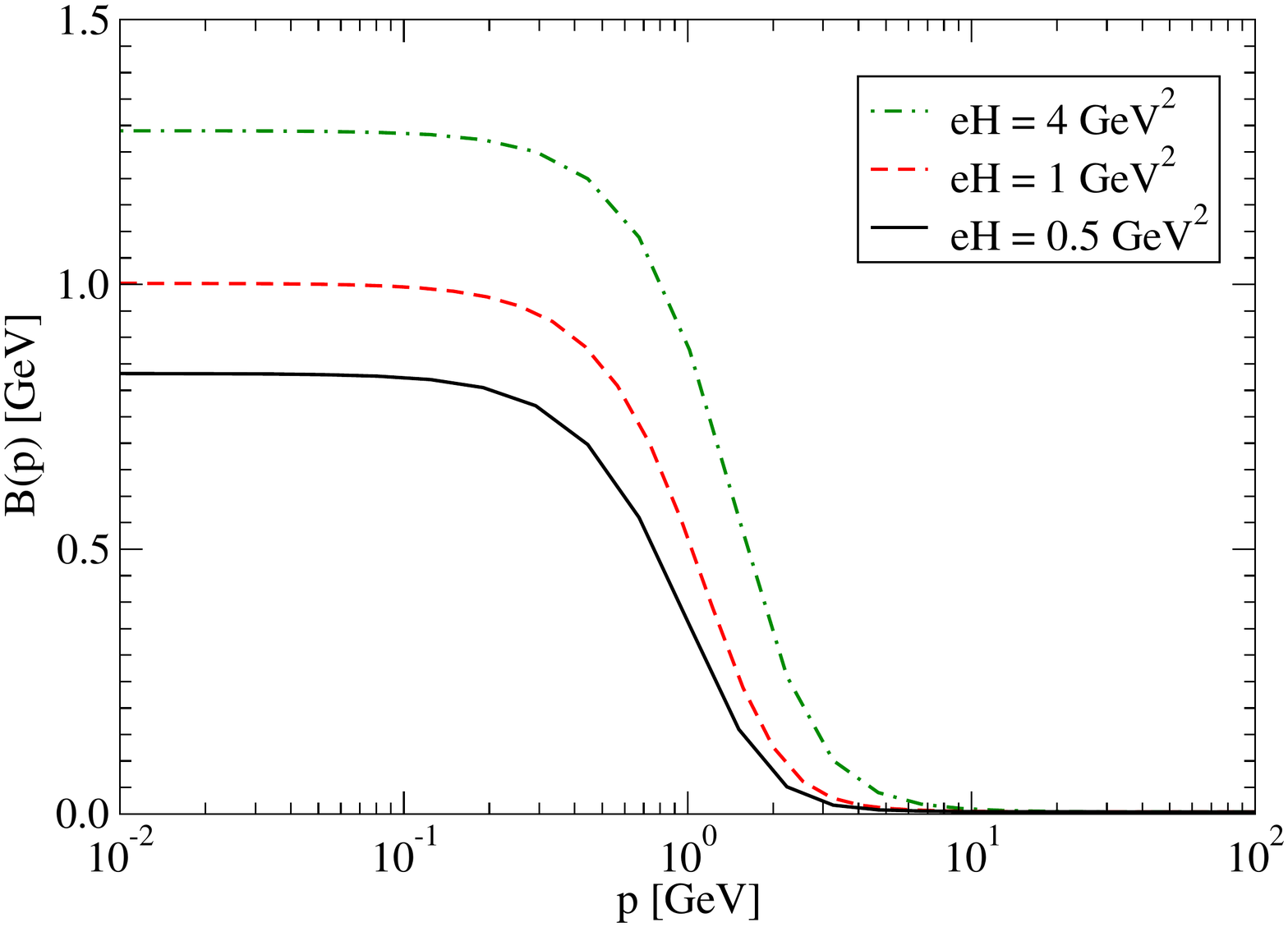}\hfill
\includegraphics[scale=.32]{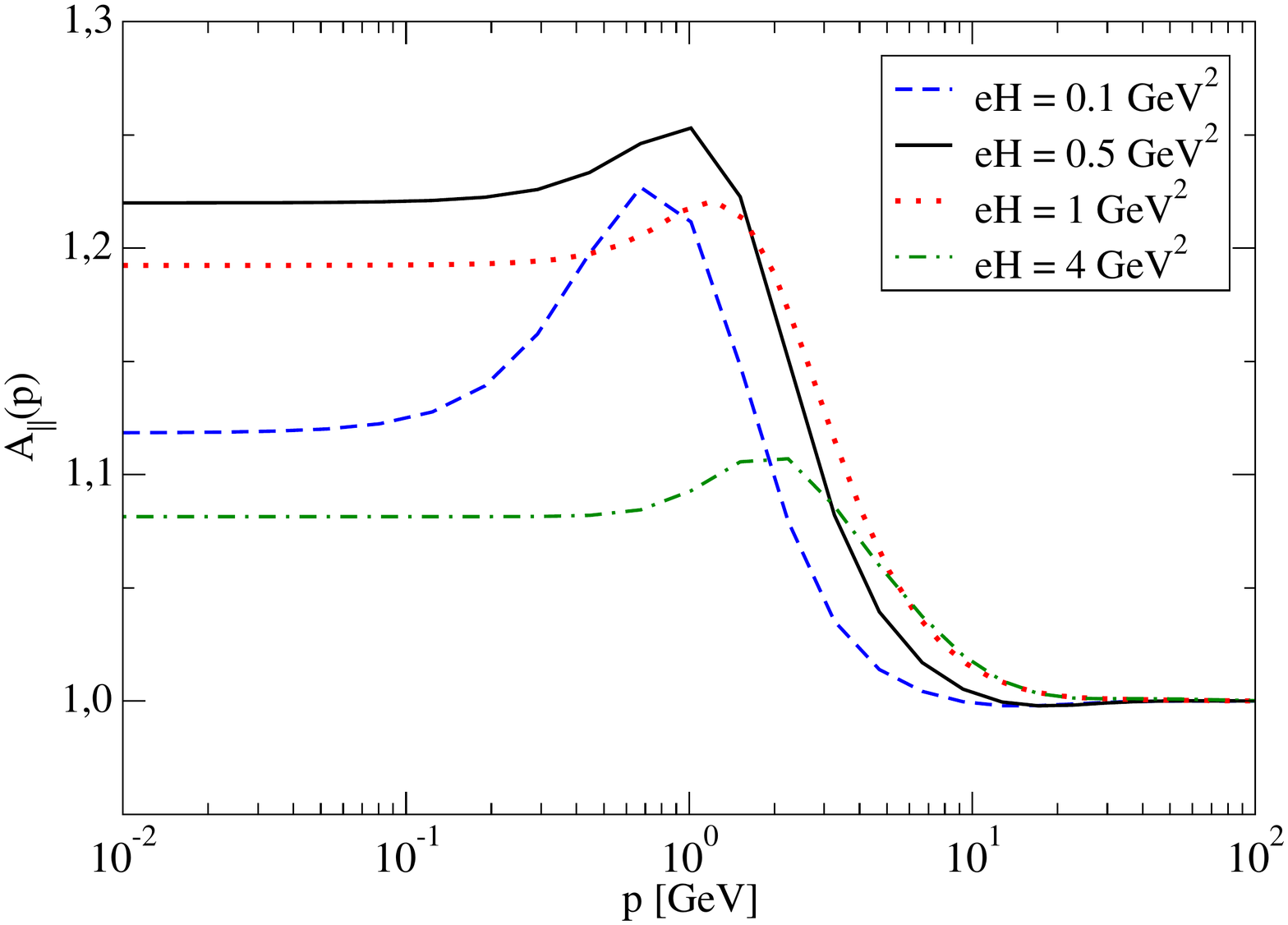}
\caption{Quenched dressing functions $B$ and $A_\parallel$ of the quark propagator for 
different magnetic fields as a function of $p\equiv p_\parallel$ 
at a bare quark mass of m = 3.7 MeV at $\mu =$ 100 GeV. Shown is the 
result for the lowest Landau level from a full calculation taking all 
Landau levels into account. The  
dressing function $A_\perp$ is not defined on the lowest Landau level.}
\label{fig:simpleTensor-B}
\end{figure}

Our results for the quark dressing functions $A_{\perp}(p_\perp,p_\parallel),
A_{\parallel}(p_\perp,p_\parallel)$ and $B(p_\perp,p_\parallel)$ are shown in 
\Fig{fig:2D-dressing-Simple-1GeV} for magnetic fields between 0.5 $\text{GeV}^2$ and 
4 $\text{GeV}^2$ and a renormalized current quark mass of $m=3.7$ MeV at $\mu$ = 100 GeV. 
Solid lines represent the single Landau levels, starting with $l=1$. The lowest Landau 
levels of $B$ and $A_{\parallel}$ are shown in \Fig{fig:simpleTensor-B}; the  
dressing function $A_\perp$ is not defined on the lowest Landau level. Whereas the two 
larger values of the magnetic field are clearly in a region where we trust our approximation, 
the lowest value is in the region where the approximation may start to break down, see below. 
In general, we clearly see the influence of the magnetic field on the dressing functions. 
For the large fields, the lowest Landau level dominates. Note, however, that the dressing 
functions of the second lowest Landau level are still large and also higher Landau levels 
are visibly different from their bare values $B=m$ and $A_{\perp}=A_\parallel=1$, even at 
the extremely large field of $eH = 4 \,\mbox{GeV}^2$ shown in the plot and beyond (we 
performed calculations up to $eH = 50 \,\mbox{GeV}^2$). 
This is true for the scalar dressing 
functions, but even more for the two vector components $A_{\perp}$ and $A_\parallel$. 
Thus, although an approximation using the lowest Landau level may capture essential 
qualitative features, significant quantitative corrections remain even for large fields. 
We will discuss this again below. 

The effects of the magnetic field on dynamical mass generation can be best seen at the 
lowest Landau level of the scalar dressing function in the left diagram of 
\Fig{fig:simpleTensor-B}. On the one hand, we find the typical increase of the dressing 
function with growing magnetic field indicative for magnetic catalysis. On the other 
hand, the scale at which the mid-momentum OPE-behavior $B(p) \sim 1/p^2$ sets in is 
shifted considerably into the direction of the larger momenta. This is indicative of 
the additional external scale $eH$ introduced into the system by the strong magnetic 
field. An interesting non-linear dependence on $eH$ can be seen in the infrared momentum 
region of $A_\parallel$. For the lowest Landau level, shown in the right diagram of 
\Fig{fig:simpleTensor-B}, we find that the dressing function $A_\parallel(p_\parallel=0)$
first rises with growing field, then reaches a maximum and drops again for very large fields.
The growth for small fields is similar to the one found in Ref.~\cite{Watson:2013ghq}, where
a different approximation of the quark-DSE has been used. We find that within our truncation 
the value of $A_\parallel(p_\parallel=0)$ is maximal around $|eH| \approx 0.5 \,\text{GeV}^2$ 
and then decreases monotonically with growing field, even crossing the $A_\parallel(0)=1$ line 
around $|eH| \approx 12 \,\text{GeV}^2$. Thus, for very large fields, the dressing
function even becomes smaller than one in the infrared. We will see later
in section \ref{res:unquenched} that this is an artifact of the quenched
approximation. For the higher Landau levels the non-linear behavior of 
$A_\parallel$ for fields around $|eH| \approx 0.5 \,\text{GeV}^2$ is also present, 
but these dressing functions remain larger than one even at extremely large fields.  

We now study the change of the quark condensate with the magnetic field. Using again the 
expansion in terms of Ritus eigenfunctions, the corresponding expression is given by
\begin{equation}
-\braket{\bar{q}q}=Z_2 \lim_{x\rightarrow 0}tr S(x,0) = Z_2 N_c 
\frac{eH}{2\pi^2}\sum\limits_{l_q=0}^\infty \frac{\chi(l_q)}{2}
\int\limits_0^\infty dq_\parallel q_\parallel
\left.\left(\frac{B(q)}{B^2(q)+q_\parallel^2 A_\parallel^2(q)+q_\perp^2 A_\perp^2(q)}\right)\right|_{l_q}.
\label{eq:chiralCond-Simple}
\end{equation}
Below, we discuss results both in the chiral limit and at a finite quark mass roughly
corresponding to an up-quark. At finite bare mass, the quark condensate diverges 
linearly with the cutoff. This is to be contrasted with the corresponding quadratic 
divergence at vanishing field; the linear behavior observed here is a direct consequence 
of the effective dimensional reduction introduced by the magnetic field.
\begin{equation}
\braket{\bar{q}q}_{m\ne 0}\rightarrow\text{(finite terms)}+m\Lambda.
\end{equation} 
Since the divergence is the same for all bare masses, it can be regularized by 
subtracting the chiral condensate of a heavy quark,
\begin{equation}
\braket{\bar{q}q}_R=\braket{\bar{q}q}_m-\frac{m}{m_{\text{heavy}}}\braket{\bar{q}q}_\text{heavy}.
\end{equation}
This procedure leaves a residual term of order $m/m_\text{heavy}$ in addition to the 
finite part of the quark condensate. This residual term can be neglected when the mass 
of the regulator quark gets sufficiently heavy.

\begin{figure}[t!]
\includegraphics[width=8.9cm]{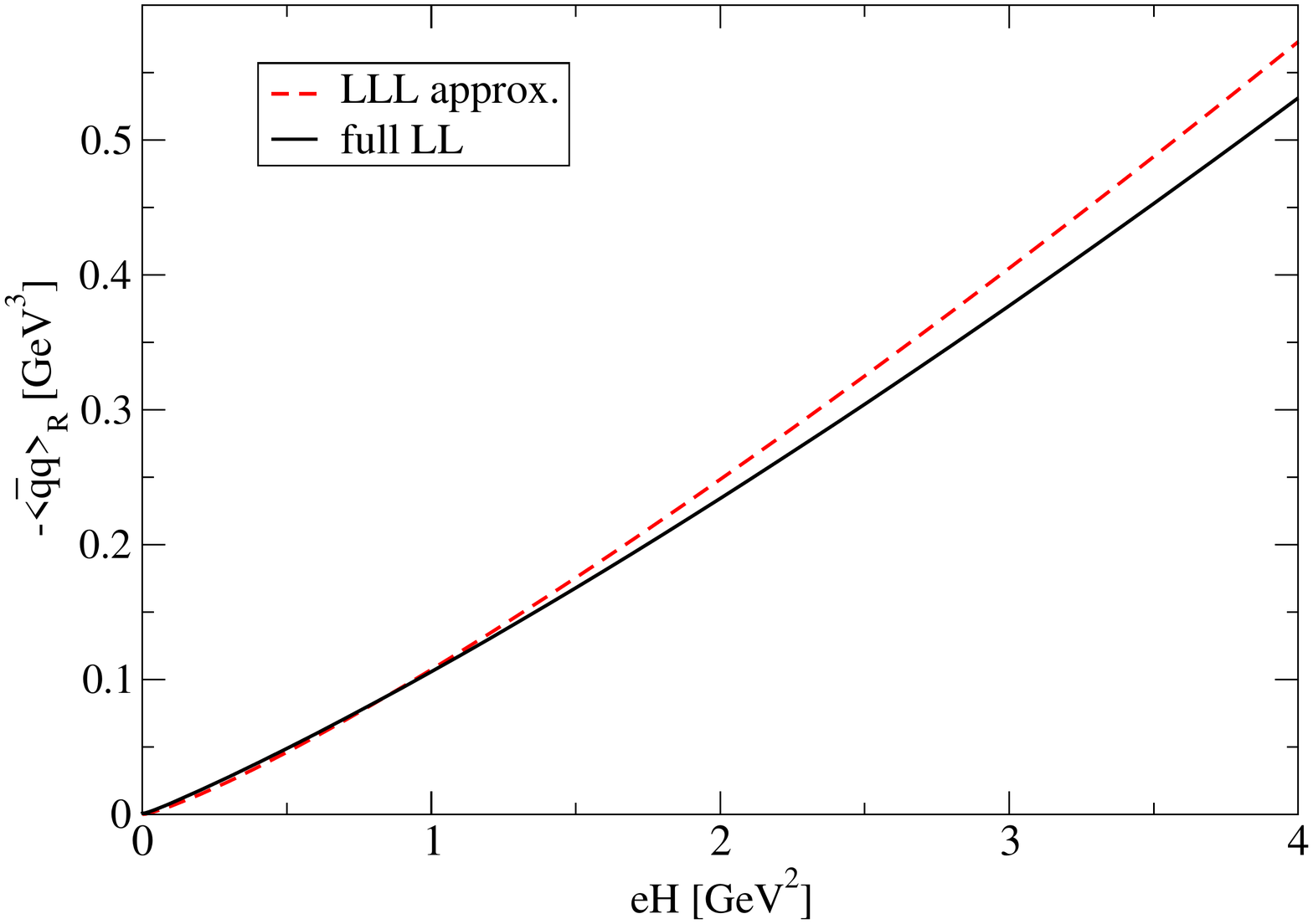}\hfill
\includegraphics[width=8.9cm]{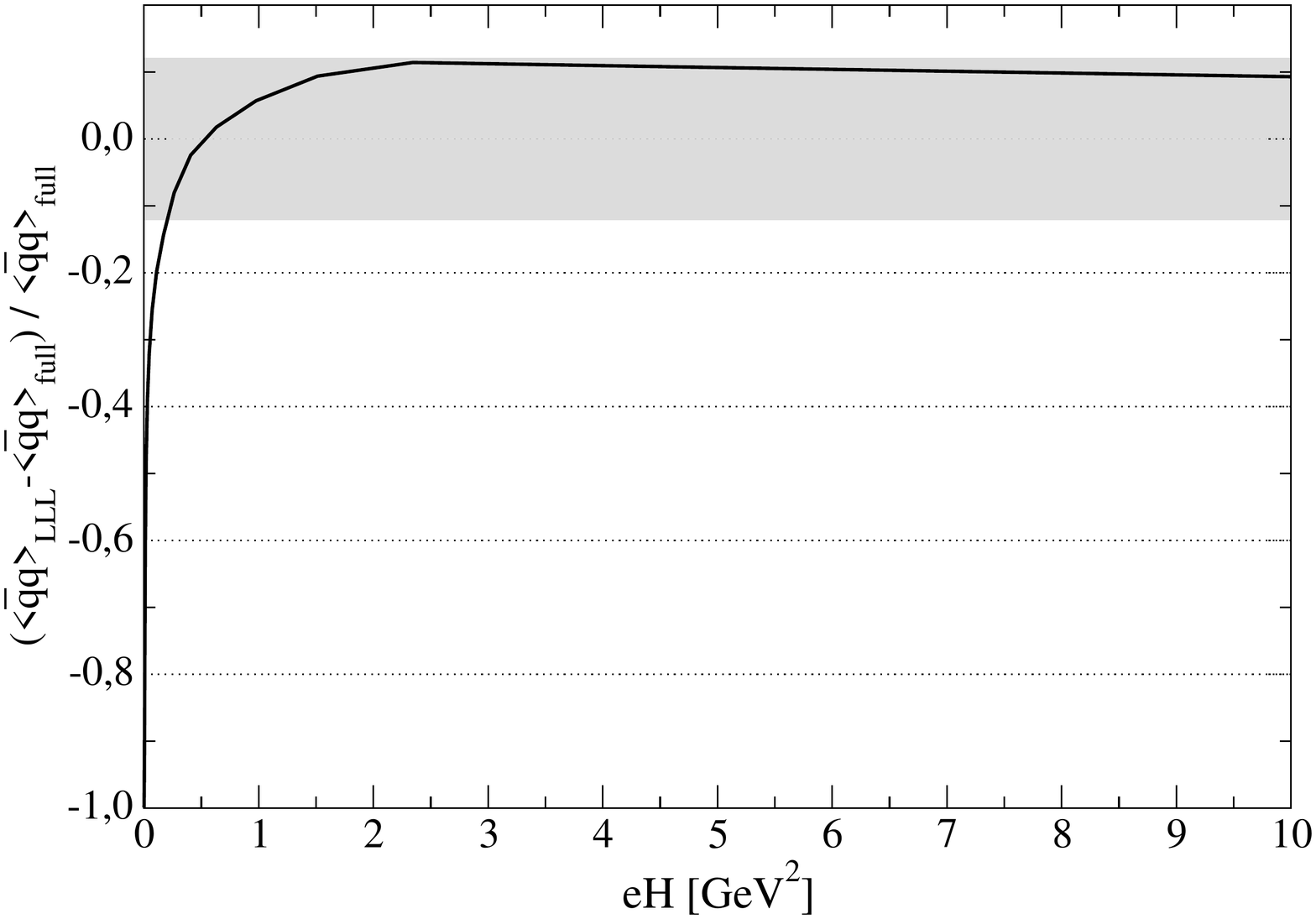}
\caption{Quark condensate for our full calculation compared with the lowest 
Landau level (LLL) approximation. On the left we display the condensate, on the right
the relative difference between the condensates. The shaded region indicates a twelve percent band
around zero.}
\label{fig:chiralCond}
\end{figure}

The regularized quark condensate for large magnetic fields is shown in the left diagram of 
\Fig{fig:chiralCond}. Clearly, the condensate grows for an increasing magnetic field. 
This behavior is in agreement with general expectations. In particular, chiral perturbation 
theory ($\chi$PT) predicts a quadratic rise for small fields $eH \ll m_\pi^2$, which then turns 
into a linear behavior for intermediate fields $m_\pi^2 < eH < \Lambda_\text{QCD}^2$ \cite{Cohen:2007bt}. 
Lattice simulations in general agree with this finding, see e.g. 
\cite{Buividovich:2008wf,D'Elia:2011zu,Bali:2012zg,D'Elia:2012tr,Ilgenfritz:2012fw}. The linear
growth of the condensate with magnetic field is also observed for fields much larger than
the $\chi$PT convergence radius $eH > \Lambda_\text{QCD}^2$ 
\cite{Gusynin:1995nb,Buividovich:2008wf,Braguta:2010ej,Bali:2012zg,Watson:2013ghq}. 
For asymptotically large fields, one furthermore expects a power law $\sim (eH)^{3/2}$
on dimensional grounds \cite{Shushpanov:1997sf}. 
In our calculation we cannot address the region of very small fields due to the approximations 
made in Eq.~(\ref{approx}). Consequently, our result for the quark condensate does not
approach the zero field limit $\langle \bar{q}q\rangle_R = 0.028 \,\mbox{GeV}^3$ but instead
goes to zero when $eH \rightarrow 0$. From the magnitude of the difference, one can infer 
that the approximation is probably good as long as 
$eH \ge 0.5 \,\text{GeV}^2 \approx \Lambda_\text{QCD}^2$, 
with $\Lambda_\text{QCD}$ evaluated in the MOM-scheme.
This ties in with the fact that our approximation of the quark DSE follows from an 
expansion in $k_\perp^2/2\left|eH \right|$, where the largest contributions to the 
quark self energy stems from momenta $k_\perp < \Lambda_{QCD}$. As a consequence we 
cannot see the quadratic rise of the condensate for small fields predicted by chiral 
perturbation theory.
However we do find the linear growth at intermediary fields which is supplemented by 
a term proportional to $(eB)^{3/2}$ for large fields in agreement with the expectations
discussed above. We come back to this discussion in section \ref{res:unquenched}, 
where we present corresponding unquenched results. 

In \Fig{fig:chiralCond} we also compare our full calculation with the lowest Landau level
(LLL) approximation. A particularly useful quantity is the relative difference between the two 
calculations shown in the right diagram of \Fig{fig:chiralCond}. As expected, for small 
fields the Landau levels are close to each other and the LLL approximation becomes 
unreliable. For fields larger than about $eH > 0.2 \,\text{GeV}^2$, the LLL becomes reliable
on the twelve percent level (indicated in grey in the diagram). This deviation persists 
to very large fields and decreases only very slowly: only for asymptotically large fields, 
the relative difference goes to zero and the LLL becomes exact.

Next, we discuss the connection of our result with the spin polarization structure of
QCD. It was shown in \cite{Ioffe:1983ju}, that external fields can give a handle on observables 
that could not be obtained otherwise. The presence of a magnetic field induces a nonzero 
expectation value for the tensor polarization operator $\sigma^{\mu\nu}$ as described e.g.
in Refs.~\cite{Bali:2012jv}. In the case of a field along the z-axis, 
$\braket{\sigma^{12}} \equiv \braket{\bar{q}\sigma^{12}q}$ 
will correspond to the average spin alignment along this quantization axis. Here, we find 
that $\braket{\sigma^{12}}$ obtains a non-zero value even in the absence of spin dependent 
tensor structures in the quark propagator. In such a case, the polarization of the QCD vacuum 
will be caused by the special role of the lowest Landau level. The expectation value of the 
operator can be pictorially represented as
\begin{figure}[!htb]
\includegraphics[scale=.40]{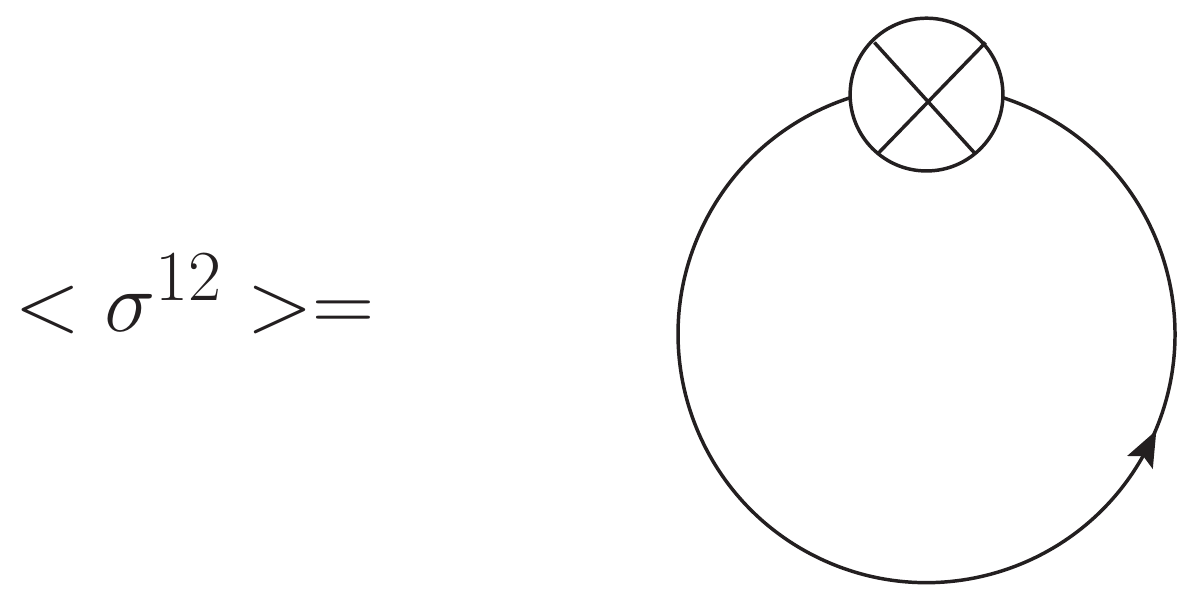}
\end{figure}\\
where the cross represents an insertion of $\sigma^{12}$. This quantity is explicitly written as
\begin{eqnarray}\nonumber
\braket{\sigma^{12}}&=&Z_2\,N_c\lim_{x\rightarrow 0}\,\,\mathclap{\displaystyle\int}\mathclap{\textstyle\sum}\text{ }\frac{d^4q}{(2\pi)^4}
\text{ Tr}\left[\sigma^{12}E_q(x)\frac{1}{\mb i\gamma \cdot q_{\|}\,A_{\|}(q)+\mb i\gamma \cdot q_{\perp}\,A_{\perp}(q)+B(q)}\bar{E}_q(0)\right]\\&=&
Z_2\,N_c\lim_{x\rightarrow 0}\,\,\mathclap{\displaystyle\int}\mathclap{\textstyle\sum}\text{ }\frac{d^4q}{(2\pi)^4}\sum\limits_{\underset{l_q > 0}{\sigma=\pm}} E_{q,\sigma}(x)E^\dagger_{q,\sigma}(0)
\text{ Tr}\left[\Delta(\sigma)\sigma^{12}\frac{B(q)}{B^2(q)+A_\parallel^2(q)q_\parallel^2+A_\perp^2(q)q_\perp^2}\right],
\end{eqnarray}
where $\bar{E}_q=\gamma^0E^\dagger_{q}\gamma^0$. All Landau levels, except for the 
lowest, are degenerate with respect to the two spin directions $\uparrow\downarrow$, 
which means that for a non-explicit spin-dependent propagator, the contributions to 
$\braket{\sigma^{12}}$ from higher Landau levels cancel on average, as can be seen from 
the form of the expectation value
\begin{equation}
\braket{\sigma^{12}}=Z_2 N_c \frac{eH}{2\pi^2}\sum\limits_{l_q=0}^\infty\int\limits_0^\infty dq_\parallel q_\parallel\sum\limits_{\underset{l_q > 0}{\sigma=\pm}} 
\left.\left(\frac{\sigma B(q)}{B^2(q)+A_\parallel^2(q)q_\parallel^2+A_\perp^2(q)q_\perp^2}\right)\right|_{l_q}
=Z_2 N_c \frac{eH}{2\pi^2}\int\limits_0^\infty dq_\parallel q_\parallel\frac{\Delta{\text{sgn}(eH)} B(q)}{B^2(q)+A_\parallel^2(q)q_\parallel^2}.
\end{equation}

This quantity behaves in analogy to the chiral condensate in terms of regularization, 
simply because the inserted operator $\sigma^{12}$ is dimensionless. Therefore the 
regularized quantity can be defined as
\begin{equation}
\braket{\sigma^{12}}_R=\braket{\sigma^{12}}_m
-\frac{m}{m_{\text{heavy}}}\braket{\sigma^{12}}_\text{heavy},
\end{equation}
where $m_\text{heavy}$ is a heavy mass as before.

\begin{figure}[t!]
\includegraphics[width=8.9cm]{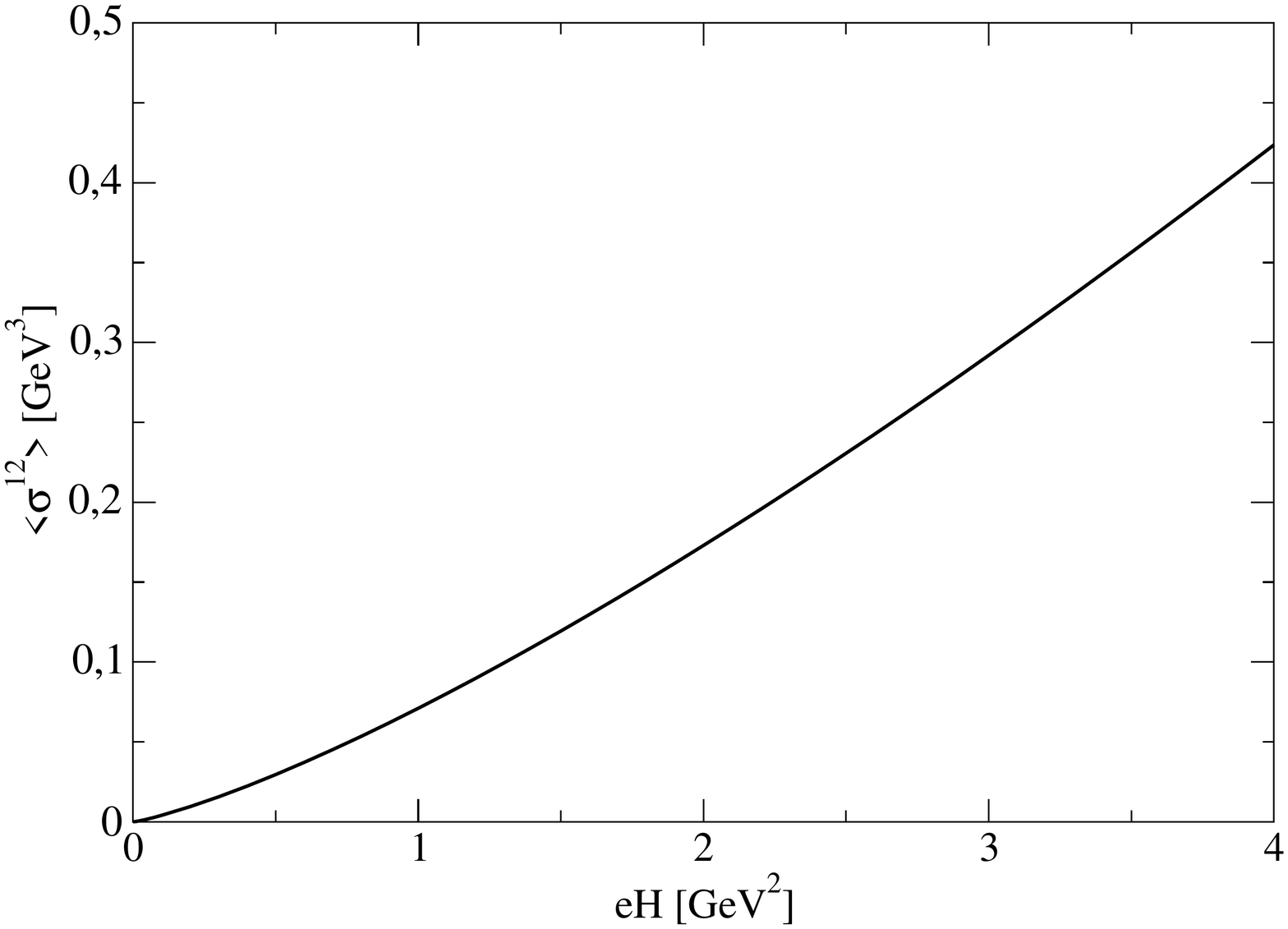}\hfill
\includegraphics[width=8.9cm]{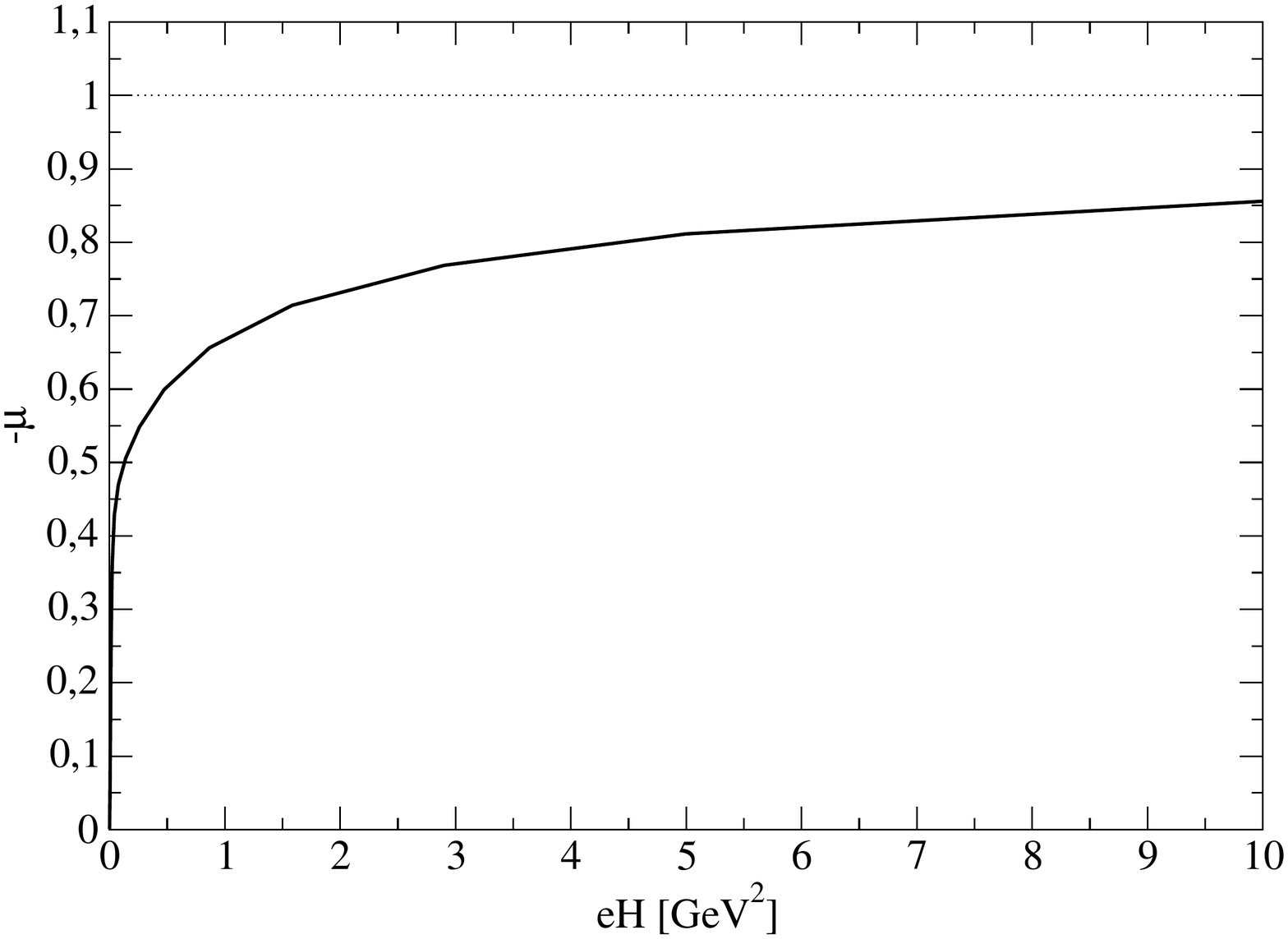}
\caption{Left hand side: Regularized expectation value of the spin polarization 
tensor $\braket{\sigma^{12}}$.
Right hand side: Regularized magnetic polarization $\mu$ of the QCD vacuum.}\label{fig:RegXi-Simple}
\end{figure}

Our results for $\braket{\sigma^{12}}$ as a function of the external field is shown in the
left diagram of \Fig{fig:RegXi-Simple}. Similar to the quark condensate, the magnetic
moment $\braket{\sigma^{12}}$ follows a power law with a linear term and a term 
$\sim (eH)^{3/2}$. The polarization $\mu$ of the QCD vacuum
\begin{equation}
\mu=\frac{\braket{\sigma^{12}}}{\braket{\bar{q}q}},
\end{equation}
tends to one in the large field limit, indicating the similarity of the coefficient in front
of the term $\sim (eH)^{3/2}$. Since $\braket{\sigma^{12}}$ extracts the contribution 
of the lowest Landau level to the chiral condensate, this limit is driven by the lowest
Landau level. In line with the results discussed above we find that this saturation only 
sets in at very large, if not asymptotic fields.

For completeness, note that the spin tensor expectation value can be expanded into operators
\begin{equation}
\braket{\sigma^{12}}=\chi\braket{\bar{q}q}eH+O(eH^2),
\end{equation}
where terms $\propto O(eH^0)$ need to vanish, since the QCD vacuum in the zero field 
case is isotropic and therefore unpolarized. For small fields $eH$, the magnetic 
susceptibility $\chi$ is given by 
\begin{equation}
\chi\approx\frac{\braket{\sigma^{12}}}{\braket{\bar{q}q}}\frac{1}{eH}=\frac{\mu}{eH}
\label{eq:mu-smalleH}.
\end{equation}
Since our approximation tends to break down in the small field limit, the magnetic susceptibility is 
not well accessible in our scheme and we will refrain from attempting to give an extrapolated 
result.

\section{Unquenched DSEs: Formalism}\label{sec:unquenched}

In this section, we establish the techniques necessary to formulate unquenched QCD in a magnetic 
field combining the Ritus method with the Dyson Schwinger approach. Although gluons do not 
couple directly to the external Abelian field, they are affected by its presence via the quark 
loop in the gluon DSE. Due to their coupling to charged quarks, the gluons inherit the anisotropy
introduced by the magnetic field. Indeed, a magnetic field will modify $\Pi^{\mu\nu}$ in a 
non-trivial way. We will use an orthogonal basis for the gluon polarization tensor 
\cite{Batalin:1971au,Shabad:1975ik,Dittrich:2000zu} that is well suited to accommodate for this effect.

\subsection{Gluon Polarisation Tensor}

There are four linear independent vectors which can be constructed from $k^\mu$, $F^{\mu\nu}$ and 
the dual field strength tensor ${}^*F^{\mu\nu}$
\begin{equation}
k^\mu,\qquad F^{\mu\nu}k_\nu, \qquad F^{\mu\nu}F_{\nu\alpha}k^\alpha,\qquad ^*F^{\mu\nu}k_\nu.
\label{eq:vectors}
\end{equation}
Similarly one can find four independent (pseudo-)scalar structures
\begin{equation}
\frac{1}{4}F^{\mu\nu}F_{\mu\nu},\qquad\frac{1}{4}{}^*F^{\mu\nu}F_{\mu\nu},\qquad k^2, 
\qquad (k_\nu F^{\nu\mu})^2\label{eq:scalarComp},
\end{equation}
which all contain an even number of the anti-symmetric tensors $F^{\mu\nu}$ and ${}^*F^{\mu\nu}$. 
The symmetric polarization tensor $\Pi^{\mu\nu}$ contains by construction ten independent 
components. Furthermore, Furry's theorem \cite{Furry:1937zz} tells us that all components of 
$\Pi^{\mu\nu}$ with an odd number of $F_{\mu\nu}$ vanish. The dressings of these components
depend only on the even structures \Eq{eq:scalarComp} and are therefore also even. Thus, it
follows that only even combinations of the vectors \Eq{eq:vectors} are allowed, which reduces 
the number of possible tensors to six. Finally, these have to satisfy the Ward-identity 
$\Pi^{\mu\nu}k_\nu=0$ and we are left with four possible linear independent basis 
tensors \cite{Dittrich:2000zu}.

Finding those is essentially an eigenvalue problem. $\Pi^{\mu\nu}$ has four orthogonal 
eigenvectors $b_{i}^\mu$ with corresponding eigenvalues
\begin{equation}
\kappa_{i}=\kappa_{i}\left( \frac{1}{4}F^{\mu\nu}F_{\mu\nu},\frac{1}{4}{}^*F^{\mu\nu}F_{\mu\nu}, k^2, (k_\nu F^{\nu\mu})^2\right).\label{EVkappa}
\end{equation}
Having solved the eigenvalue problem, the polarization tensor can be written in its eigenbasis
\begin{eqnarray}
\Pi^{\mu\nu}(k,k') &=&(2\pi)^4\delta^{(4)}(k'-k)\Pi^{\mu\nu}(k),\\
\Pi^{\mu\nu}(k) & = & \sum\limits_{i=0}^3 \kappa_{i}\frac{b_{i}^\mu b_{i}^\nu}{(b_{i})^2}.
\end{eqnarray}
The first eigenvector is $b_0^\mu=k^\mu$ with eigenvalue 0, since $\Pi^{\mu\nu}k_\nu=0$. The other eigenvectors 
are (see \cite{Batalin:1971au})
\begin{eqnarray}
b_1^\mu &=&(F^{\mu\nu}F_{\nu\rho}k^\rho)k^2-k^\mu(k_\nu F^{\nu\alpha}F_{\alpha\beta}k^\beta),\\
b_2^\mu &=&{}^*F^{\mu\nu}k_\nu,\\
b_3^\mu &=&{}F^{\mu\nu}k_\nu,
\end{eqnarray}
from which it is found that the projectors along those eigenvectors look like
\begin{equation}
\begin{split}
\frac{b_2^\mu b_2^\nu}{(b_2)^2}=\frac{\tilde{k}_\parallel^\mu \tilde{k}_\parallel^\nu}{\tilde{k}_\parallel^2}=\left(\delta_\parallel^{\mu\nu}-\frac{k_\parallel^\mu k_\parallel^\nu}{k_\parallel^2} \right)\equiv P_\parallel^{\mu\nu},\\
\frac{b_3^\mu b_3^\nu}{(b_3)^2}=\frac{\tilde{k}_\perp^\mu \tilde{k}_\perp^\nu}{\tilde{k}_\perp^2}=\left(\delta_\perp^{\mu\nu}-\frac{k_\perp^\mu k_\perp^\nu}{k_\perp^2} \right)\equiv P_\perp^{\mu\nu},\label{bla1}
\end{split}
\end{equation}
where we have defined the orthogonal momenta (being orthogonal to its corresponding partner, 
e.g. $\tilde{k}_\parallel\perp k_\parallel$ and similar for $\tilde{k}_\perp$)
\begin{eqnarray}
\tilde{k}_\parallel^{\alpha}=\epsilon_\parallel^{\alpha\beta}k^{\beta}_\parallel & \,\,\,\,\,\,\,\alpha,\beta=0,3,\\
\tilde{k}_\perp^{\alpha}=\epsilon_\perp^{\alpha\beta}k^{\beta}_\perp & \,\,\,\,\,\,\,\alpha,\beta=1,2,
\end{eqnarray}
with $\epsilon_{\perp}^{12}=-\epsilon^{12}_\perp=1$, $\epsilon_{\perp}^{11}=\epsilon^{22}_\perp=0$ and 
correspondingly $\epsilon_{\parallel}^{30}=-\epsilon^{03}_\parallel=1$, $\epsilon_{\parallel}^{33}=\epsilon^{00}_\parallel=0$. 
Obviously, $k_i^\mu \tilde{k}_j^\mu=0$  ($i, j=\perp,\parallel$) and further $k_i^2=\tilde{k}_i^2$. 
For a constant magnetic field, the last tensor structure is easily found
\begin{equation}
\frac{b_1^\mu b_1^\nu}{(b_1)^2}=\frac{(k_\perp^2 k_\parallel^\mu-k_\parallel^2k_\perp^{\mu})(k_\perp^2 k_\parallel^\nu-k_\parallel^2k_\perp^{\nu})}{k_\parallel^2 k_\perp^2 k^2}\equiv P_0^{\mu\nu}.\label{eq:P0}
\end{equation}
Note that because of the completeness of the basis, we can write 
\begin{equation}
\delta^{\mu\nu}=P_0^{\mu\nu}+P_\parallel^{\mu\nu}+P_\perp^{\mu\nu}+P_L^{\mu\nu},
\end{equation}
where $P_L^{\mu\nu}=k^\mu k^\nu / k^2$. The structures $P_\parallel^{\mu\nu}$, $P_\perp^{\mu\nu}$ 
and $P_0^{\mu\nu}$ constitue a complete orthonormal basis for the transverse subspace $P^{\mu\nu}$. 
The transverse subspace is defined by
\begin{equation}
P^{\mu\nu}=\delta^{\mu\nu}-\frac{k^\mu k^\nu}{k^2}=P_0^{\mu\nu}+P_\parallel^{\mu\nu}+P_\perp^{\mu\nu},
\end{equation}
and therefore, $P_0^{\mu\nu}$ has an alternative expression to \Eq{eq:P0}
\begin{equation}
P_0^{\mu\nu}=\delta^{\mu\nu}-\frac{k^\mu k^\nu}{k^2} - \frac{\tilde{k}_\parallel^\mu \tilde{k}_\parallel^\nu}{\tilde{k}_\parallel^2} - \frac{\tilde{k}_\perp^\mu \tilde{k}_\perp^\nu}{\tilde{k}_\perp^2}=\frac{k_\parallel^\mu k_\parallel^\nu}{k_\parallel^2}+\frac{k_\perp^\mu k_\perp^\nu}{k_\perp^2}-\frac{k^\mu k^\nu}{k^2},
\end{equation}
which is easier to use in certain calculations. The basis properties of the projectors 
found here can be seen from
\begin{eqnarray}
P_i^{\mu\alpha}P_j^{\alpha\nu}=\delta_{ij}P^{\mu\nu},\hspace*{5mm}
P_i^{\mu\mu}=1.\label{bla2}
\end{eqnarray}
With Eqs. (\ref{bla1}) - (\ref{bla2}), the most general form of the gluon propagator in 
the presence of an external magnetic field along the z-axis is given by
\begin{equation}
D^{\mu\nu}(k,k')=(2\pi)^4\delta^{(4)}(k'-k)\left( \frac{Z_0}{k^2}P^{\mu\nu}_0 (k) + \frac{Z_\parallel}{k^2}P_\parallel^{\mu\nu}(k) + \frac{Z_\perp}{k^2}P_\perp^{\mu\nu}(k) \right).
\end{equation}
The inverse propagator follows as
\begin{equation}
D^{-1\mu\nu}(k,k')=(2\pi)^4\delta^{(4)}(k'-k)k^2\left( Z_0^{-1}P^{\mu\nu}_0 (k) + Z_\parallel^{-1}P_\parallel^{\mu\nu}(k) + Z_\perp^{-1} P_\perp^{\mu\nu}(k) \right),
\end{equation}
with the gluon dressing functions $Z_i$. In terms of the eigenvalues $\kappa_i$ from 
\Eq{EVkappa}, these are formally given by
\begin{equation} 
Z_i\equiv\frac{1}{1-\kappa_i/k^2}\qquad i \in\{0,\parallel,\perp\}\,.
\end{equation}

This form of the gluon propagator illustrates an important effect, which was introduced as 
"vacuum birefringence" in \cite{Hattori:2012je}, denoting the non-degeneracy of the physical gluon 
modes. Stated otherwise, the refractive indices of different gluon polarizations deviate from 
each other.

%
%
\subsection{The Gluon Dyson--Schwinger Equation}

In order to solve the DSE for the gluon propagator in an external magnetic field, we resort
to an approximation introduced in Ref.~\cite{Fischer:2012vc} in the context of finite temperature and
chemical potential. There, the right hand side of the gluon DSE has been split into a
part containing the gluon self interaction and the coupling to a ghost anti-ghost pair 
('Yang--Mills part') and the quark-loop. The Yang--Mills part, together with the bare term,
has been approximated by quenched lattice results for the propagator, whereas the
quark loop has been treated dynamically together with the quark DSE. Within the framework
of Ref.~\cite{Fischer:2003rp}, this approximation can be compared with the fully 
back-coupled result and is found to be accurate on the five percent level. For the exploratory
calculation presented in this work, this is certainly acceptable. 

\begin{figure}[t]
\includegraphics[width=12cm]{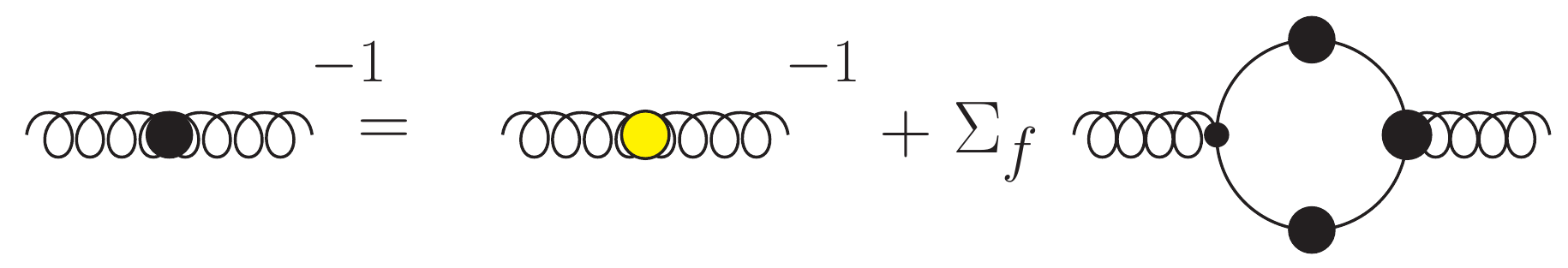}
\caption{Pictographic representation of the approximation of Ref.~\cite{Fischer:2012vc} for the 
unquenched gluon propagator.}\label{fig:jan}
\end{figure}

The resulting Dyson--Schwinger equation for the gluon propagator is displayed in 
\Fig{fig:jan} and given by  
\begin{equation}
D^{-1}_{\mu\nu}(k)=D^{-1}_{(0)\mu\nu}(k)+\Pi^{\text{g}}_{\mu\nu}(k)
+\Pi^{\text{q}}_{\mu\nu}(k)\approx D^{-1\text{ eff}}_{\mu\nu}(k) 
+\Pi^\text{q}_{\mu\nu}(k)\label{eq:effProp}\,.
\end{equation}  
Here, the quenched contributions are denoted by $D^{-1\text{ eff}}_{\mu\nu}(k)$ and the
yellow dot in \Fig{fig:jan}. Within this approximation, the effective propagator in 
\Eq{eq:effProp} is taken to be isotropic wrt. its polarization which is well justified 
when realizing that there is no direct appearance of charged particles in this sector. 

In order to include the effect of the magnetic field onto the gluon sector, the Ritus 
method will be employed for the quark loop in the gauge boson self energy. The quark
part of the gluon polarization reads
\begin{equation}
\Pi^\text{q}_{\mu\nu}(x,y)=-\frac{g^2}{2}\text{ }\Tr{\gamma_\mu S(x,y)\Gamma_\nu (y) S(y,x)}\,.
\end{equation}
Any part of the gluon self energy is diagonal in Fourier space
\begin{equation}
\Pi_{\mu\nu}^q(k,k')=\int\dint{x}\dint{y}e^{-\mb i(kx-k'y)}\Pi^q_{\mu\nu}(x,y)
=(2\pi)^4\delta^{(4)}(k-k')\Pi^{\mu\nu}(k)\,,
\end{equation}
and explicitly given in terms of the Ritus representation for the quark propagator $S(x,y)$ 
(cf. \Eq{eq:quarkRitus}),
\begin{equation}
\Pi^{\mu\nu}(k,k')=-\frac{g^2}{2}\,\,\text{ }\mathclap{\displaystyle\int}\mathclap{\textstyle\sum}\text{   }\, 
\frac{\dint{q}}{(2\pi)^4}\frac{\dint{q'}}{(2\pi)^4} 
\text{tr}\left(\left[\int\dint{x}\bar{E}_{q'}(x)\gamma^\mu E_q(x)e^{-\mb ikx}\right]S(q) 
\left[ \int\dint{y}\bar{E}_q(y)\Gamma^\nu(y)E_{q'}(y)e^{\mb ik'y}\right]S(q')\right)\,.
\end{equation}
Here, $S(q)$ denotes the quark propagator in Ritus space. Again, the simplifications leading to 
\Eq{eq:vertex-Simple} are employed, rendering also the gluon polarization tensor potentially
unreliable at small magnetic fields,
\begin{eqnarray}\label{eq:pt}
\Pi^{\mu\nu}(k,k')&=&-(2\pi)^4\delta^{(3)}(k-k')\frac{g^2}{2}\sum\limits_{l,l'}\int\frac{ d^2q_\parallel}{(2\pi)^4}
\int\limits_{-\infty}^{\infty} dq_2\text{ }e^{-\frac{k_\perp^2+k_\perp '^2}{4|eH|}}
e^{\mb i(k_1'-k_1)q_2/eH} e^{-\mb i(k_1'-k_1)k_2'/2eH}\\&&\times
\sum\limits_{\sigma_1,\sigma_2,\sigma_3,\sigma_4}\delta_{n_1(l',\sigma_1)n_2(l,\sigma_2)}
\delta_{n_3(l',\sigma_3)n_4(l,\sigma_4)}\Tr{\Delta_1\gamma^\mu\Delta_2 S(q)
\Delta_3 \gamma^\nu
\Delta_4 S(q')}\Gamma(\hat{q_{\|}})\,.  \nonumber
\end{eqnarray}
with $ \hat{q_{\|}} = q_{\|}^2 + (q'_{\|})^2 + k_{\perp}^2 $ and $\Delta_i = \Delta(\sigma_i)$.
For the quark-gluon vertex, we used the same truncation as in the quenched case, 
$\Gamma_\nu(q,q') = \gamma_\nu \Gamma(\hat{q_{\|}})$. However, 
the argument of the model dressing function $\Gamma$ was adapted such that the vertex
is symmetric under the exchange of the two quarks and the equation 
remains multiplicative renormalizable, see Ref.~\cite{Fischer:2003rp} for details.
The expression \Eq{eq:pt} is apparently diagonal in $k_\parallel$ and $k_2$. By using
\begin{equation}
\int\limits_{-\infty}^{\infty} dq_2\text{ }e^{\mb i(k_1'-k_1)q_2/eH}=2\pi\delta(k_1'-k_1)eH,
\end{equation}
the anticipated diagonality can be made more obvious and we obtain
\begin{eqnarray}
\Pi^{\mu\nu}(k,k')&=&(2\pi)^4\delta^{(4)}(k'-k)\Pi^{\mu\nu}(k)\\
\Pi^{\mu\nu}(k) &=& 2\pi \frac{g^2}{2}eH\sum\limits_{l,l'}\int\frac{ d^2q_\parallel}{(2\pi)^4}
\Bigg\{ e^{-k_\perp^2/2|eH|}\Gamma(\hat{q_{\|}})\sum\limits_{\{\sigma_i \}}
\delta_{n_1n_2}\delta_{n_3n_4}\Tr{\Delta_1\gamma^\mu\Delta_2 S(q)\Delta_3 \gamma^\nu
\Delta_4 S(q')}\Bigg\}.\label{eq:gluon:self1}
\end{eqnarray}
The relationship between $q$ and $q'$ is given by
\begin{equation}
q'_0=q_0-k_0\,, \hspace*{5mm} q'_3=q_3-k_3\,, \hspace*{5mm} 
q'_\perp=\sqrt{2|eH|l'}\,,\hspace*{5mm} q_\perp=\sqrt{2|eH|l}.
\end{equation}
We define
\begin{equation}
\mathcal{D}(q,q')=[B^2(q)+A_\parallel^2(q)q_\parallel^2+A_\perp^2(q)q_\perp^2][B^2(q')+A_\parallel^2(q')q'^2_\parallel+A_\perp^2(q')q'^2_\perp]\label{eq:denom}.
\end{equation}
The trace in \Eq{eq:gluon:self1} can be performed easily, yielding
\begin{equation}
\Tr{\Delta_1 \gamma^\mu\Delta_2 S(q)\Delta_3\gamma^\nu
\Delta_4 S(q')}=\frac{T_1^{\mu\nu}+T_2^{\mu\nu}+T_3^{\mu\nu}}{\mathcal{D}(q,q')},
\end{equation}
where
\begin{eqnarray}
T_1^{\mu\nu}=&2B(q)B(q')\left(\delta_\parallel^{\mu\nu}\delta_{\sigma_1,\sigma_2}
\delta_{\sigma_3,\sigma_4}\delta_{\sigma_1,\sigma_3}+\delta_\perp^{\mu\nu}\delta_{\sigma_1,-\sigma_2}
\delta_{\sigma_3,-\sigma_4}\delta_{\sigma_1,-\sigma_3} \right)
\\T_2^{\mu\nu}=&2A_\perp(q)A_\perp(q')\left( q_\perp q_\perp' \delta_\parallel^{\mu\nu} \delta_{\sigma_1,\sigma_2}\nonumber
\delta_{\sigma_3,\sigma_4} \delta_{\sigma_1,-\sigma_3} + [q_\perp q'_\perp \delta_\perp^{\mu\nu} -2  q_\perp^\mu q'^\nu_\perp ] \delta_{\sigma_1,-\sigma_2}
\delta_{\sigma_3,-\sigma_4} \delta_{\sigma_1,\sigma_3} \right)
\\T_3^{\mu\nu}= &2A_\parallel(q)A_\parallel(q')\left( [q_\parallel\cdot q'_\parallel\delta_\parallel^{\mu\nu} - 2q_\parallel^\nu q'^{\mu}_\parallel ]\delta_{\sigma_1,\sigma_2}
\delta_{\sigma_3,\sigma_4}\delta_{\sigma_1,\sigma_3} + q_\parallel\cdot  q'_\parallel\delta_\perp^{\mu\nu}\delta_{\sigma_1,-\sigma_2}
\delta_{\sigma_3,-\sigma_4}\delta_{\sigma_1,-\sigma_3}  \right)\nonumber.
\end{eqnarray}
Inserting these three expressions in \Eq{eq:gluon:self1}, we find similar properties as 
for the quark self-energy above: when combining the Kronecker deltas, the Landau level 
transitions appear
\begin{eqnarray}
\delta_{n_1(l',\sigma_1)n_2(l,\sigma_2)}\delta_{n_3(l',\sigma_3)n_4(l,\sigma_4)}\delta_{\sigma_1,\sigma_2}
\delta_{\sigma_3,\sigma_4}\delta_{\sigma_1,\pm\sigma_3}&\propto \delta_{l,l'},\nonumber
\\
\delta_{n_1(l',\sigma_1)n_2(l,\sigma_2)}\delta_{n_3(l',\sigma_3)n_4(l,\sigma_4)}\delta_{\sigma_1,-\sigma_2}
\delta_{\sigma_3,-\sigma_4}\delta_{\sigma_1,\mp\sigma_3}&\propto \delta_{l+\sigma_1\text{sgn}(eH),l'}.
\end{eqnarray}
Thus, either the gluon splits into a quark-antiquark pair on the same Landau level, or it
induces a transition from one Landau level to the next. Other cases are not compatible 
with the spin-one-boson nature of the gluon.

Putting everything together, the gluon DSE reads
\begin{eqnarray}
k^2\left( Z_0^{-1}(k)P_0^{\mu\nu} + Z_\parallel^{-1}(k)P_\parallel^{\mu\nu}+Z_\perp^{-1}(k)P_\perp^{\mu\nu} \right)=&&k^2 Z^{-1}(k)P^{\mu\nu}-\pi g^2\,eH\, e^{-k_\perp^2/2|eH|}
\nonumber\\
&&\times \sum\limits_{l,l'}\int\frac{ d^2q_\parallel}{(2\pi)^4}\Gamma(\hat{q}_\parallel^2)\sum\limits_{\{\sigma_i\}}\delta_{n_1n_2}\delta_{n_3n_4}
\frac{T_1^{\mu\nu}+T_2^{\mu\nu}+T_3^{\mu\nu}}{\mathcal{D}(q,q')}.\label{DSE-GLUE}
\end{eqnarray}

The equation can be decomposed into its contributions from the polarization subspaces 
denoted by $P^{\mu\nu}_\perp$, $P^{\mu\nu}_\parallel$ and $P^{\mu\nu}_0$. In the following,  
$Z(k)$ stands for the dressing function of the quenched isotropic gluon propagator. 
The resulting equations for the dressing functions for the full gluon propagator read 
in a compact notation (here we have one quark flavor, $N_f=1$, with charge $q_f = e$ 
for brevity, although later on we solve for $N_f=1+1$ up- and down-quarks with charges
$q_f=+2/3 \,e$ and $q_f=-1/3 \,e$ respectively).
\begin{eqnarray}
Z_\parallel^{-1}(k)&=& Z^{-1}(k)-\beta\sum\limits_{l}
\int\frac{ d^2q_\parallel}{(2\pi)^4}\,\frac{\chi(l)}{2}\,\frac{M_\parallel(q,q')}{\mathcal{D}(q,q')}
\Big|_{l'=l} \, \Gamma(\hat{q}_\parallel^2),\label{x1}\\
Z_\perp^{-1}(k)&=& Z^{-1}(k)-\beta\sum_{l}\int\frac{ d^2q_\parallel}{(2\pi)^4}  
\sum_{l'=l\pm 1, l'\ge 0}\frac{N_\perp(q,q')}{\mathcal{D}(q,q')}\,\Gamma(\hat{q}_\parallel^2),\\
Z_0^{-1}(k)&=&Z^{-1}(k)-\beta\sum\limits_{l}\int\frac{ d^2q_\parallel}{(2\pi)^4}
\Bigg\{\,\frac{\chi(l)}{2}\,\frac{M_0(q,q')}{\mathcal{D}(q,q')}\Big|_{l'=l}
+\sum_{l'=l\pm 1, l'\ge 0}\frac{N_0(q,q')}{\mathcal{D}(q,q')}\Bigg\}\,\Gamma(\hat{q}_\parallel^2)\,.\label{x3}
\end{eqnarray}
Here, $\beta=\beta(k,eH)\equiv2\pi g^2 q_f H e^{-k_\perp^2/2|eH|}$, 
$q_\parallel'=q_\parallel-k_\parallel$ and $\mathcal{D}(q,q')$ as in \Eq{eq:denom}. The factor 
$\frac{\chi(l)}{2}$ again accounts for the spin degeneracy of the Landau levels, which is equal to one for 
the lowest level but equal to two otherwise. We have defined
\begin{eqnarray}
M_\parallel(q,q')&=& A_\perp(q)A_\perp(q')q_\perp {q'}_\perp+A_\parallel(q)A_\parallel(q')\left( q_\parallel\cdot{q'}_\parallel-2 q^2_\parallel\sin^2(\phi) \right),\\
N_\perp(q,q')&=&A_\perp(q)A_\perp(q')q_\perp{q'}_\perp\left(1-2\frac{k_2^2}{k_\perp^2}\right)+A_\parallel(q)A_\parallel(q')q_\parallel\cdot {q'}_\parallel,\\
M_0(q,q')&=&A_\perp(q)A_\perp(q')q_\perp{q'}_\perp\frac{k_\perp^2}{k^2}
+A_\parallel(q)A_\parallel(q')\left( q_\parallel\cdot{q'}_\parallel\frac{k_\perp^2}{k^2}-2\frac{q_\parallel\cdot k_\parallel {q'}_\parallel\cdot k_\parallel}{k^2
}\frac{k_\perp^2}{k_\parallel^2} \right),\\
N_0(q,q')&=&A_\perp(q)A_\perp(q')q_\perp{q'}_\perp\left( \frac{k_\parallel^2}{k^2} - 2\frac{k_1^2}{k^2}\frac{k_\parallel^2}{k_\perp^2}\right)
+A_\parallel(q)A_\parallel(q')q_\parallel\cdot {q'}_\parallel\frac{k_\parallel^2}{k^2}.
\end{eqnarray}
Naively, also terms proportional to $B(q)B(q')$ may appear. However, it is clear from the
$H=0$ case that these terms disappear after renormalization \cite{Fischer:2003rp}, such 
that we dropped them in the first place.
Note that $Z_\parallel(k)$ only gets contributions from similar Landau levels $l'=l$, 
whereas $Z_\perp(k)$ only gets contributions from the neighboring ones, where $l'=l\pm 1$. 
The third dressing function $Z_0$ receives contributions from both cases.

The gluon polarization tensor decomposition affects the structure of the quark self energy, too.
With the abbreviations 

$\int_q \equiv \int\frac{ d^2q_\parallel}{(2\pi)^4}\int\limits_{-\infty}^\infty  dq_2 dk_1$
and $D_q(q) \equiv B^2(q)+A_\parallel^2(q)q_\parallel^2+A_\perp^2(q)q_\perp^2$, the
quark DSE then reads
\begin{eqnarray}
B(p) &=& m + g^2C_F \int_q \frac{B(q)}{D_q(q)} e^{-k_\perp^2/2|eH|}\Gamma(k^2)
\left( \frac{Z_\parallel(k)}{k^2} +\frac{k_\perp^2}{k^2}\frac{Z_0(k)}{k^2}   \right)\label{y1}\\
&&+\frac{2}{\chi(l)}g^2C_F\sum_{l_q=l\pm 1, l_q\ge 0}
\int_q\frac{B(q)}{D_q(q)} e^{-k_\perp^2/2|eH|}\Gamma(k^2)
\left( \frac{Z_\perp(k)}{k^2} +\frac{k_\parallel^2}{k^2}\frac{Z_0(k)}{k^2}   \right)\,,\nonumber\\
A_\parallel(p) &=& 1 - g^2C_F\int_q \frac{A_\parallel(q)}{D_q(q)}
\frac{e^{-k_\perp^2/2|eH|}}{p_\parallel^2}\Gamma(k^2) \left( \frac{Z_\parallel(k)}{k^2} \,K_1(p,q)\, 
+ \frac{Z_0(k)}{k^2}\,K_2(p,q)\, 
\right)\nonumber\\
&&+\frac{2}{\chi(l)}g^2C_F\sum_{l_q=l\pm 1, l_q\ge 0} \int_q \frac{A_\parallel(q)}{D_q(q)} 
\frac{e^{-k_\perp^2/2|eH|}}{p_\parallel^2}\Gamma(k^2)
\left( \frac{Z_\perp(k)}{k^2}p_\parallel\cdot q_\parallel
+\frac{Z_0(k)}{k^2}p_\parallel\cdot q_\parallel\frac{k_\parallel^2}{k^2}   \right),
\end{eqnarray}
with kernels
\begin{eqnarray}
K_1(p,q) &=& 2\frac{(p_\parallel q_\parallel \sin(\phi))^2}{k_\parallel^2}-p_\parallel\cdot q_\parallel\\
K_2(p,q) &=& 2\frac{k_\perp^2}{k_\parallel^2}\frac{p_\parallel\cdot k_\parallel q_\parallel\cdot k_\parallel}{k^2}-p_\parallel\cdot q_\parallel\frac{k_\perp^2}{k^2}
\end{eqnarray}
Furthermore,
\begin{eqnarray}
A_\perp(p) &=& 1 + g^2C_F\int_q \frac{A_\perp(q)}{D_q(q)}\frac{e^{-k_\perp^2/2|eH|}}{p_\perp^2}\Gamma(k^2)\left( \frac{Z_\parallel(k)}{k^2}p_\perp q_\perp+\frac{Z_0(k)}{k^2}p_\perp q_\perp\frac{k_\perp^2}{k^2}  \right)\\
&+&\frac{2}{\chi(l)}g^2C_F\!\!\!\sum_{l_q=l\pm 1, l_q\ge 0} \,\int_q\frac{A_\perp(q)}{D_q(q)}
 \frac{e^{-k_\perp^2/2|eH|}}{p_\perp^2}\Gamma(k^2)\left( \frac{Z_\perp(k)}{k^2}p_\perp q_\perp \left( 1-2\frac{k_2^2}{k_\perp^2}\right) + \frac{Z_0(k)}{k^2}p_\perp q_\perp \left(\frac{k_\parallel^2}{k^2}-2\frac{k_\parallel^2 k_1^2}{k_\perp^2 k^2}  \right) \right)\,.\nonumber \label{y3}
\end{eqnarray}
Eqs. (\ref{x1})-(\ref{x3}) and (\ref{y1})-(\ref{y3}) are coupled and need to be solved simultaneously. The dressing functions $A_\parallel$, $A_\perp$ and $B$ are functions of the scalar variables $p_\parallel^2$ and $p_\perp^2$, whereas the gluon dressing functions depend on $k_\parallel^2$, $k_1$ 
and $k_2$.

For the large fields studied here, the lowest Landau level approximation is trustworthy
on the ten percent level, cf. the discussion in section \ref{res:quenched}. In order to
limit the huge numerical effort necessary to solve the coupled gluon and quark DSE 
self-consistently we restrict ourselves to the following scheme: we back-couple 
only the lowest Landau level of the quark onto the lowest Landau level of the gluon 
propagator and treat all other Landau levels of the gluon in quenched approximation. 
In this way we consistently unquench only the lowest Landau level of the gluon propagator.
For the dressing functions in Eqs.~(\ref{x1}-\ref{x3}) this means that $Z_\perp$ stays 
quenched completely (since it receives only contributions from neighboring Landau levels), 
whereas in $Z_\parallel$ the lowest Landau level becomes modified. The same contribution
for $Z_0$ needs a separate discussion: In order to solve three equations for the gluon 
dressing functions numerically, they need to be properly regularized. To this end, we use 
the results of \cite{Leung:2005xz}, where the fermion-loop with bare propagators are 
discussed. It is found that the $M_0$-term in Eq.~(\ref{x3}) is cancelled by the 
regularization procedure. We adopt this prescription also here and explicitly set 
$M_0=0$. Within our approximation scheme, this then entails that also the lowest Landau 
level of $Z_0$ is unaffected by unquenching and $Z_\parallel$ is the only dressing function
that is modified. The remaining equation for $Z_\parallel$ is finite due to dimensional 
reduction and needs no further regularisation.

\section{Results for full QCD}\label{res:unquenched}

Here we present results for the unquenched system of two up/down quarks back-coupled
to the Yang-Mills sector in the above described approximation. To this end we need to
take into account the different charges of the quarks. The magnetic background field
then breaks the isospin symmetry of the system by coupling differently to the
charges $+2/3 \,e$ of the up-quark and $-1/3 \,e$ of the down quark. We take this fully
into account by solving for two quark DSEs for the up- and down-quark. Correspondingly,
in Eq.~(\ref{x1}-\ref{x3}) we take into account one quark-loop for the up- and
one for the down-quark with respective charges.

\begin{figure}[t!]
\subfloat[$Z_\parallel(k1,k_2)$ \,\,($eH=$ 0 GeV$^2$)]
{\includegraphics[width=5.9cm]{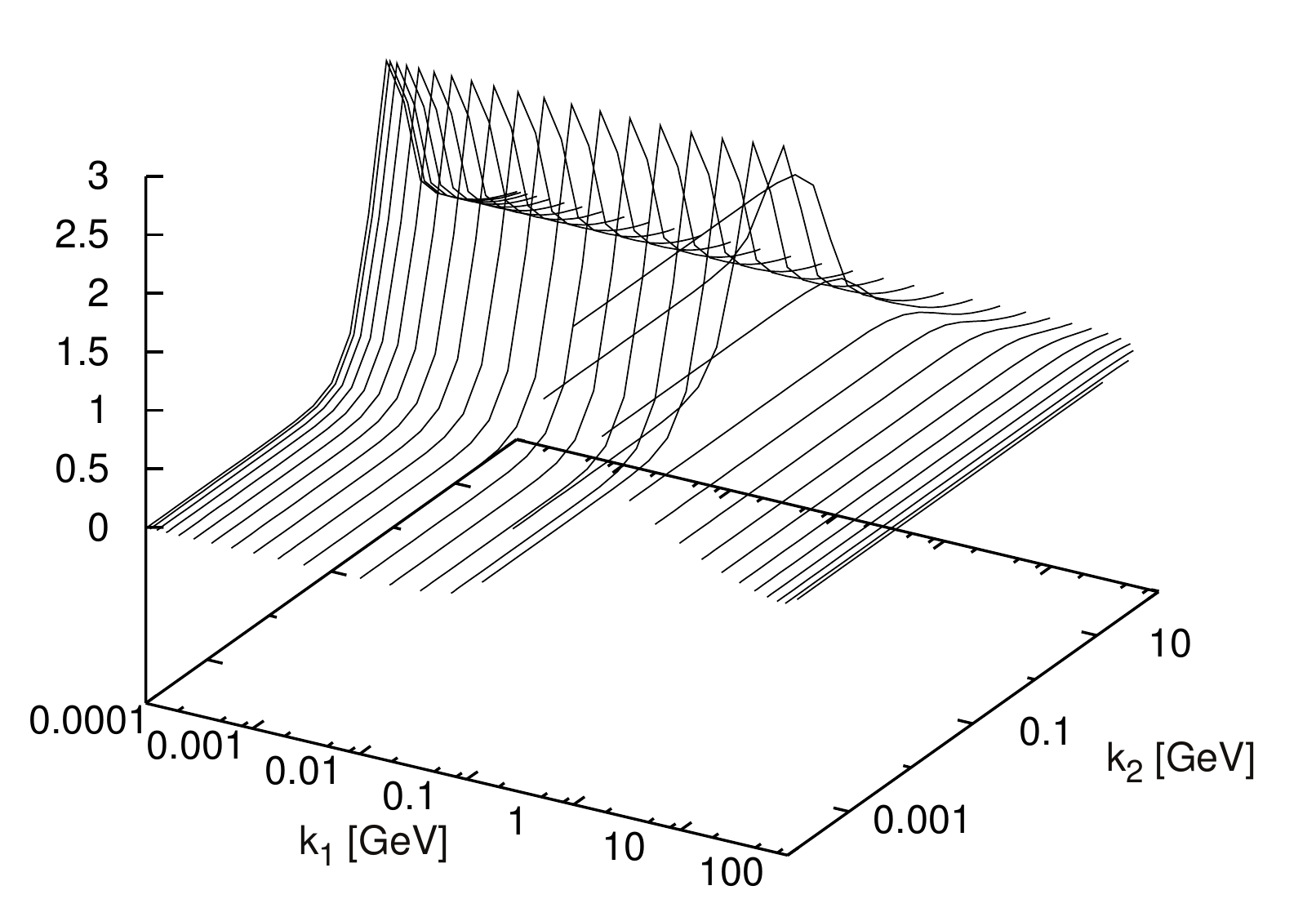}}
\subfloat[$Z_\parallel(k1,k_\parallel)$ \,\,($eH=$ 0 GeV$^2$)]
{\includegraphics[width=5.9cm]{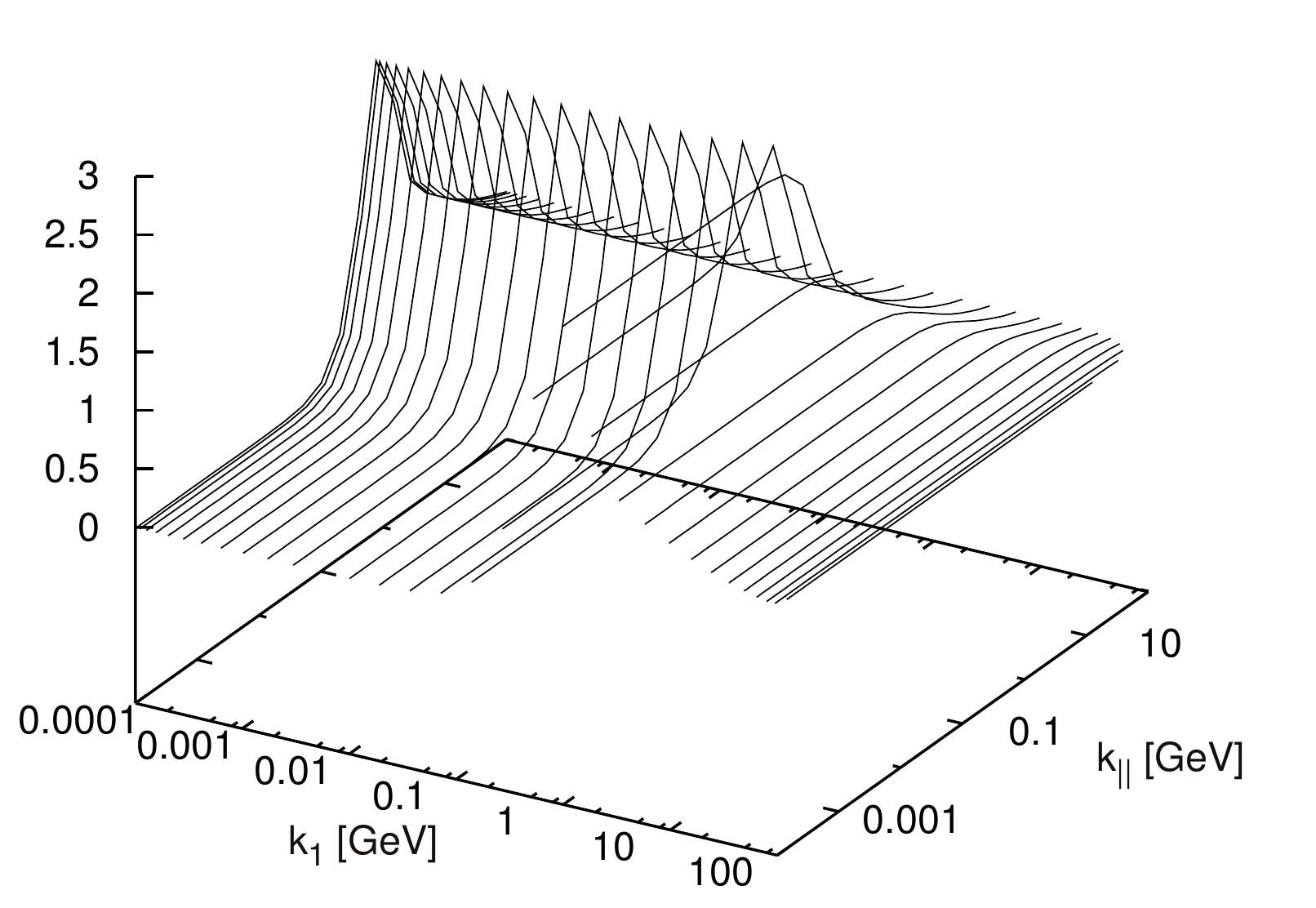}}
\subfloat[$Z_\parallel(k2,k_\parallel)$ \,\,($eH=$ 0 GeV$^2$)]
{\includegraphics[width=5.9cm]{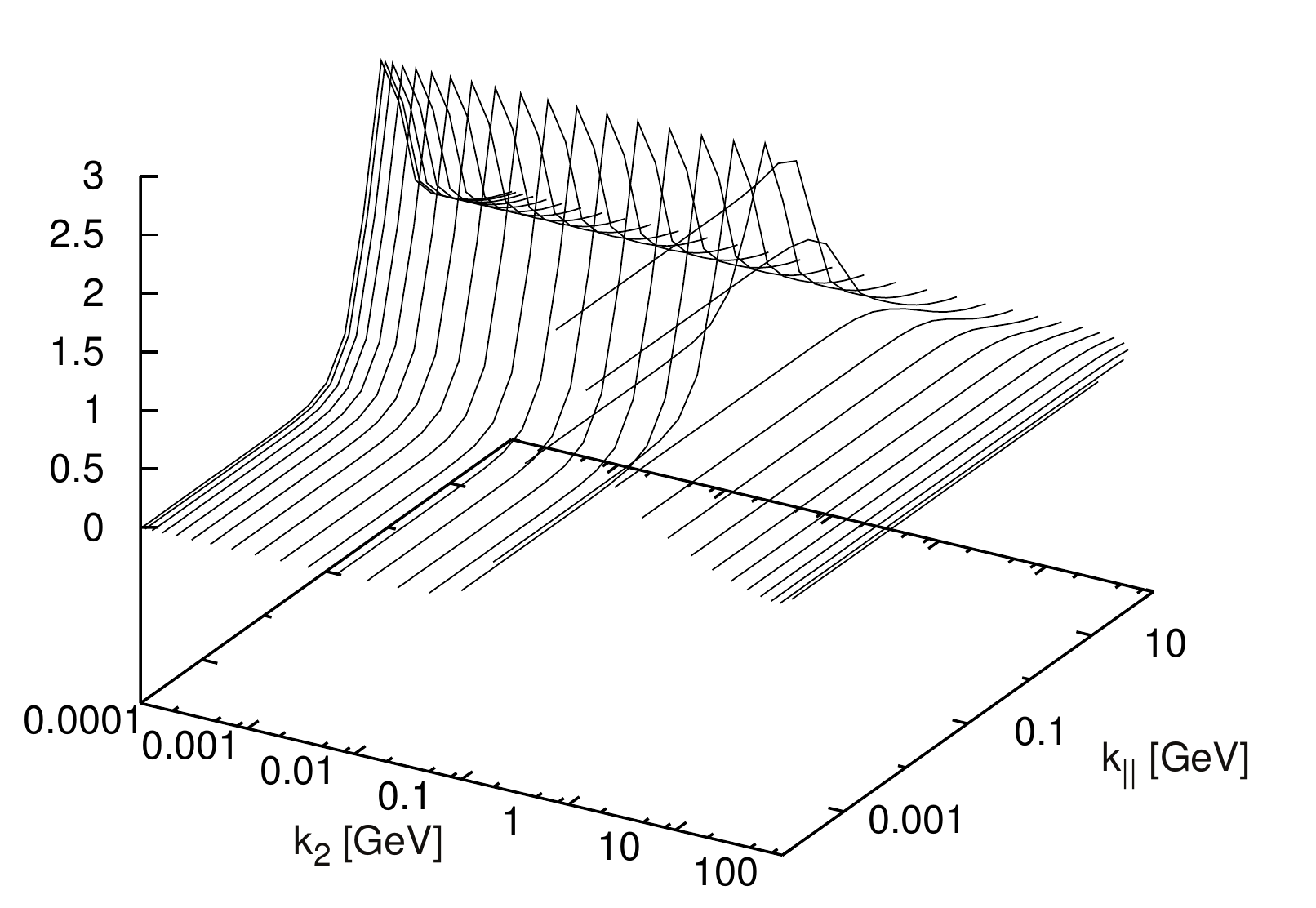}}\\
\subfloat[$Z_\parallel(k1,k2)$ {[}GeV{]} \,\,($eH=$ 0.5 GeV$^2$)]
{\includegraphics[width=5.9cm]{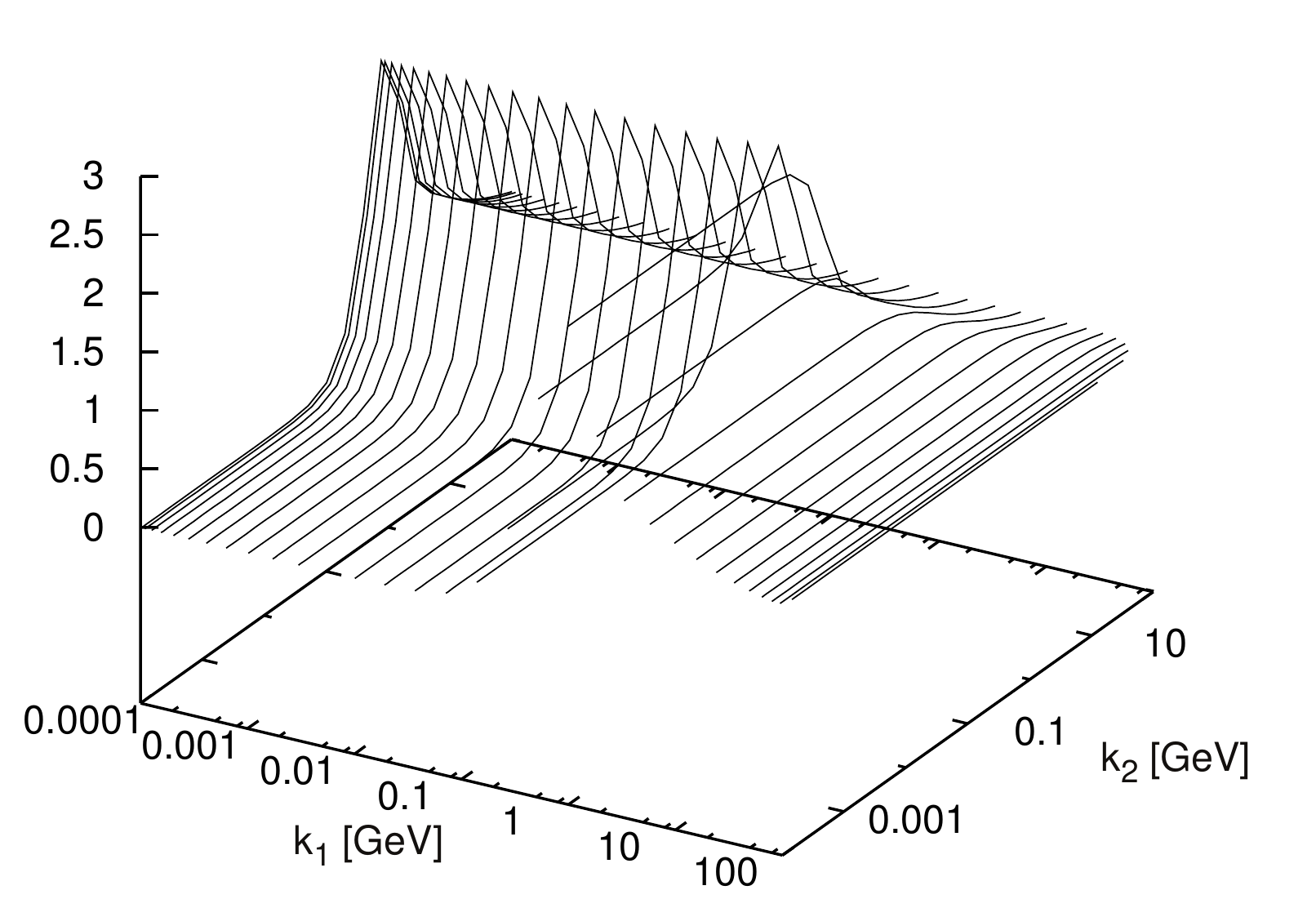}}
\subfloat[$Z_\parallel(k1,k_\parallel)$ \,\,($eH=$ 0.5 GeV$^2$)]
{\includegraphics[width=5.9cm]{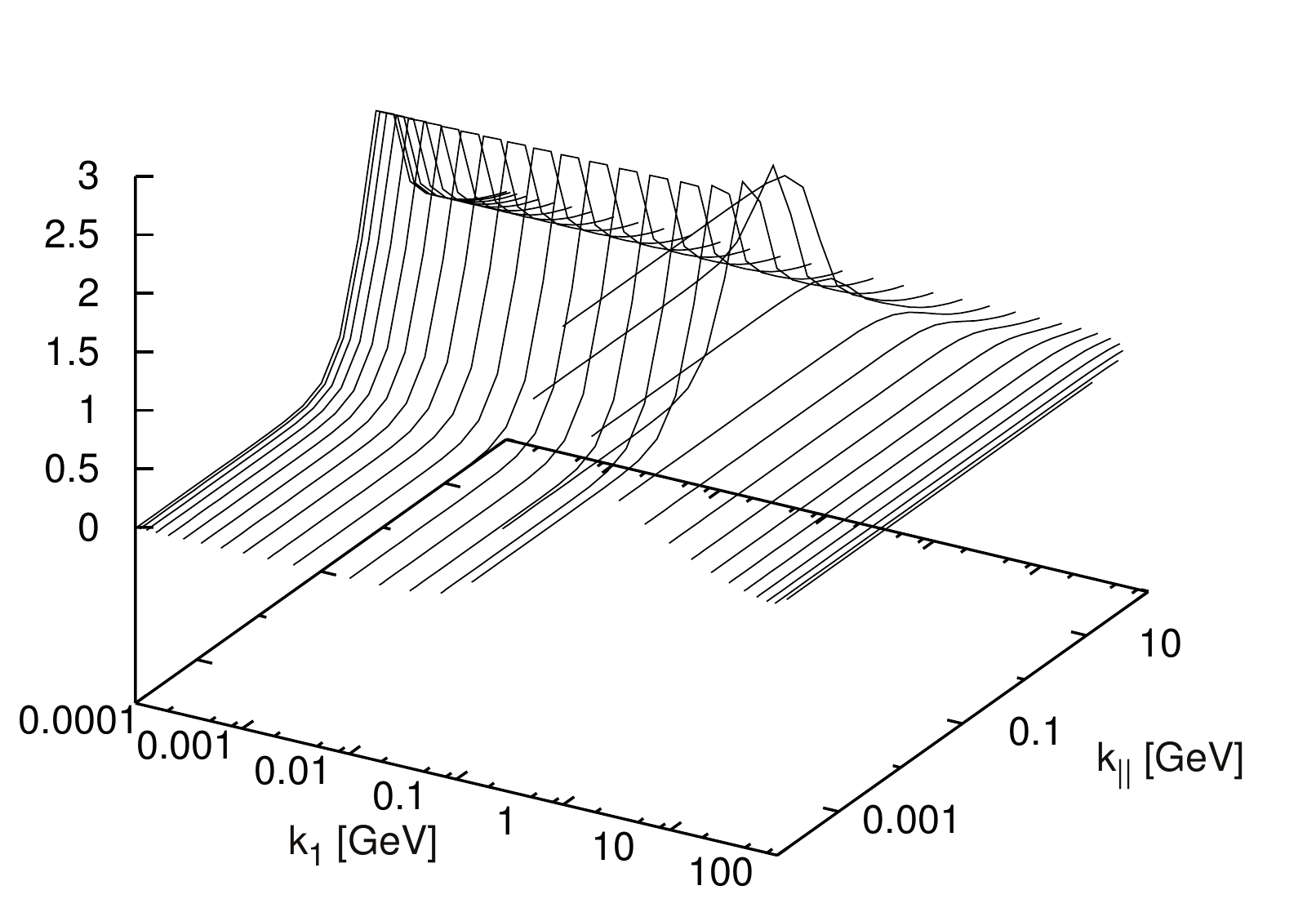}}
\subfloat[$Z_\parallel(k2,k_\parallel)$ \,\,($eH=$ 0.5 GeV$^2$)]
{\includegraphics[width=5.9cm]{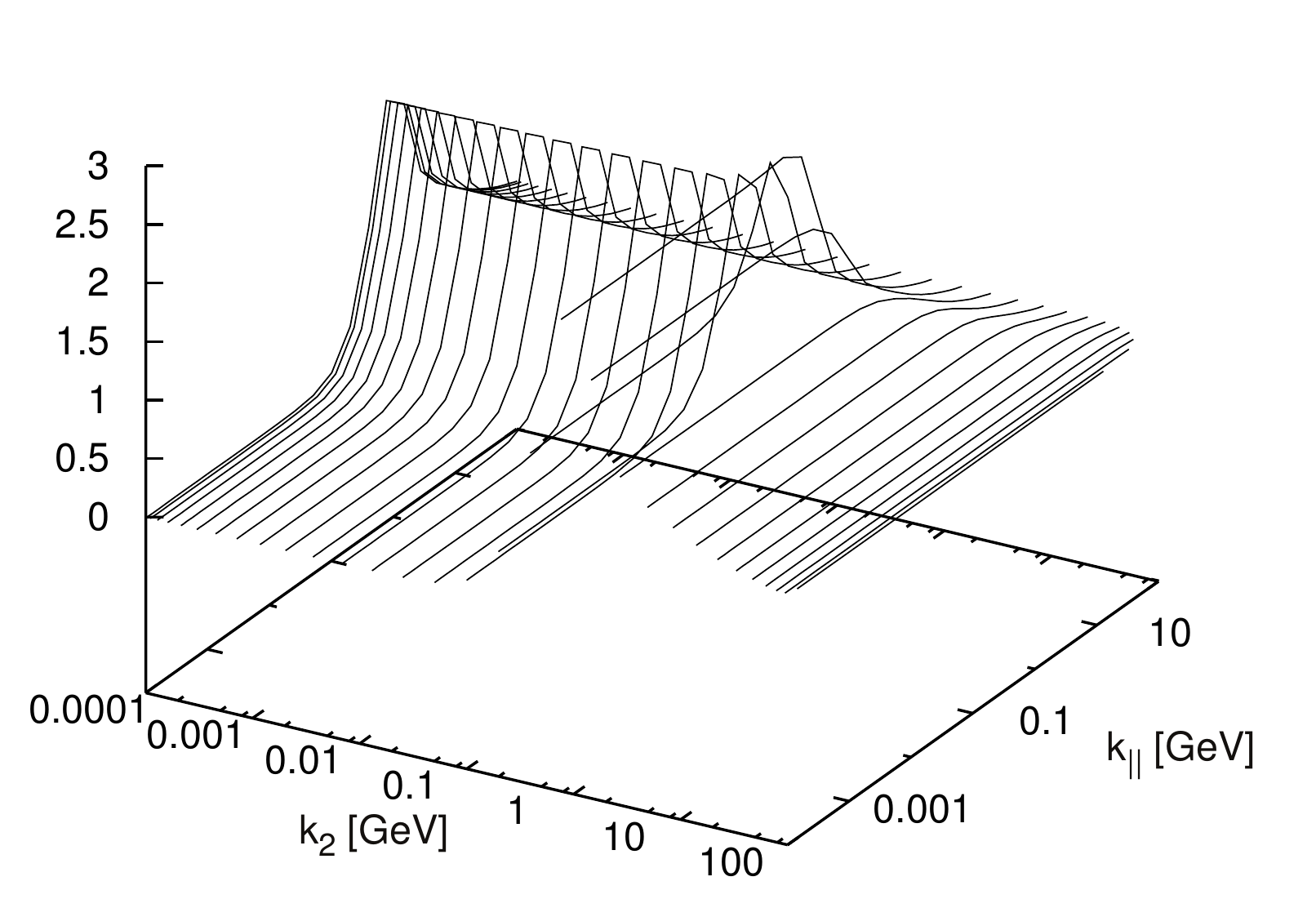}}\\
\subfloat[$Z_\parallel(k1,k2)$ {[}GeV{]} \,\,($eH=$ 1 GeV$^2$)]
{\includegraphics[width=5.9cm]{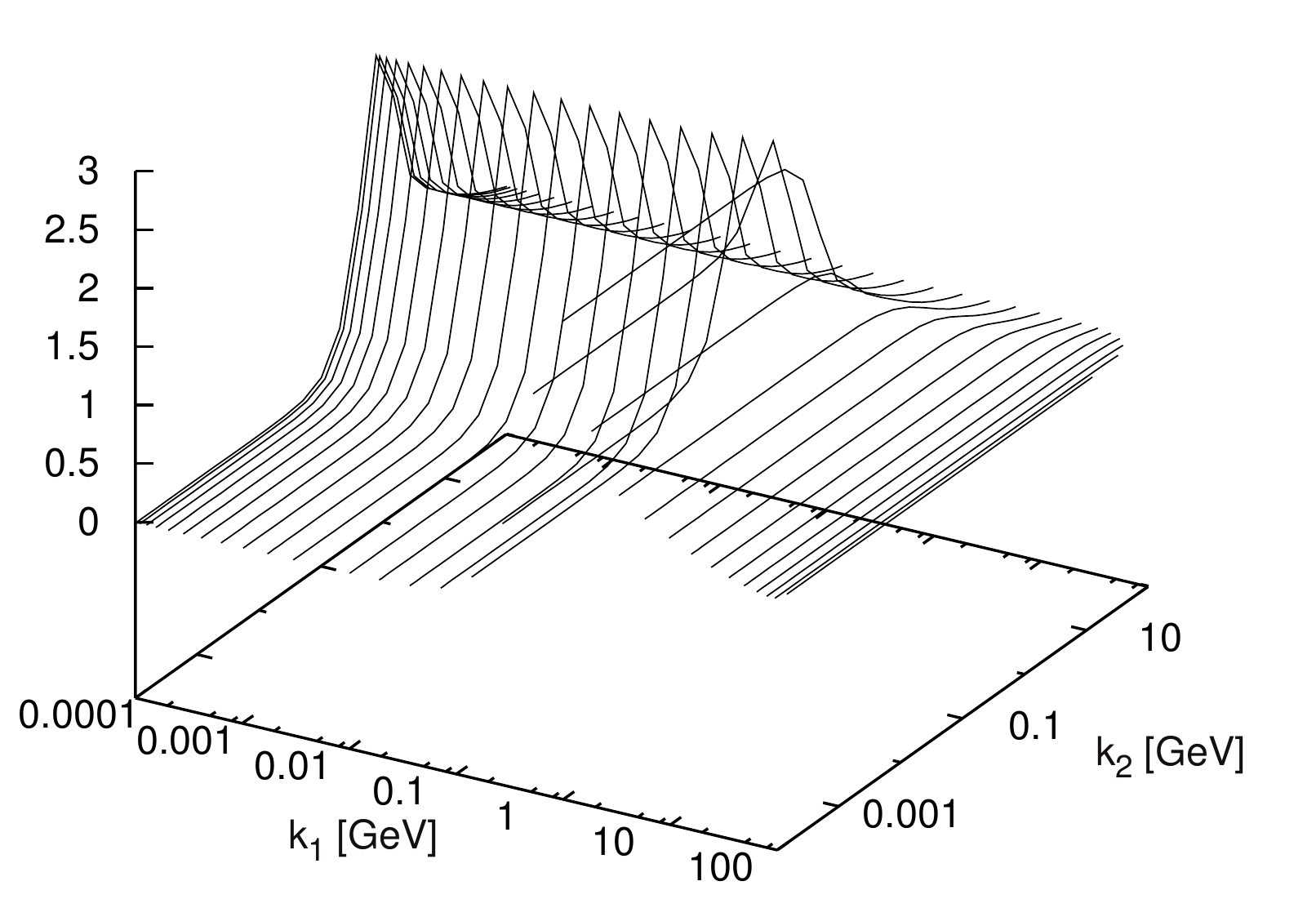}}
\subfloat[$Z_\parallel(k1,k_\parallel)$ \,\,($eH=$ 1 GeV$^2$)]
{\includegraphics[width=5.9cm]{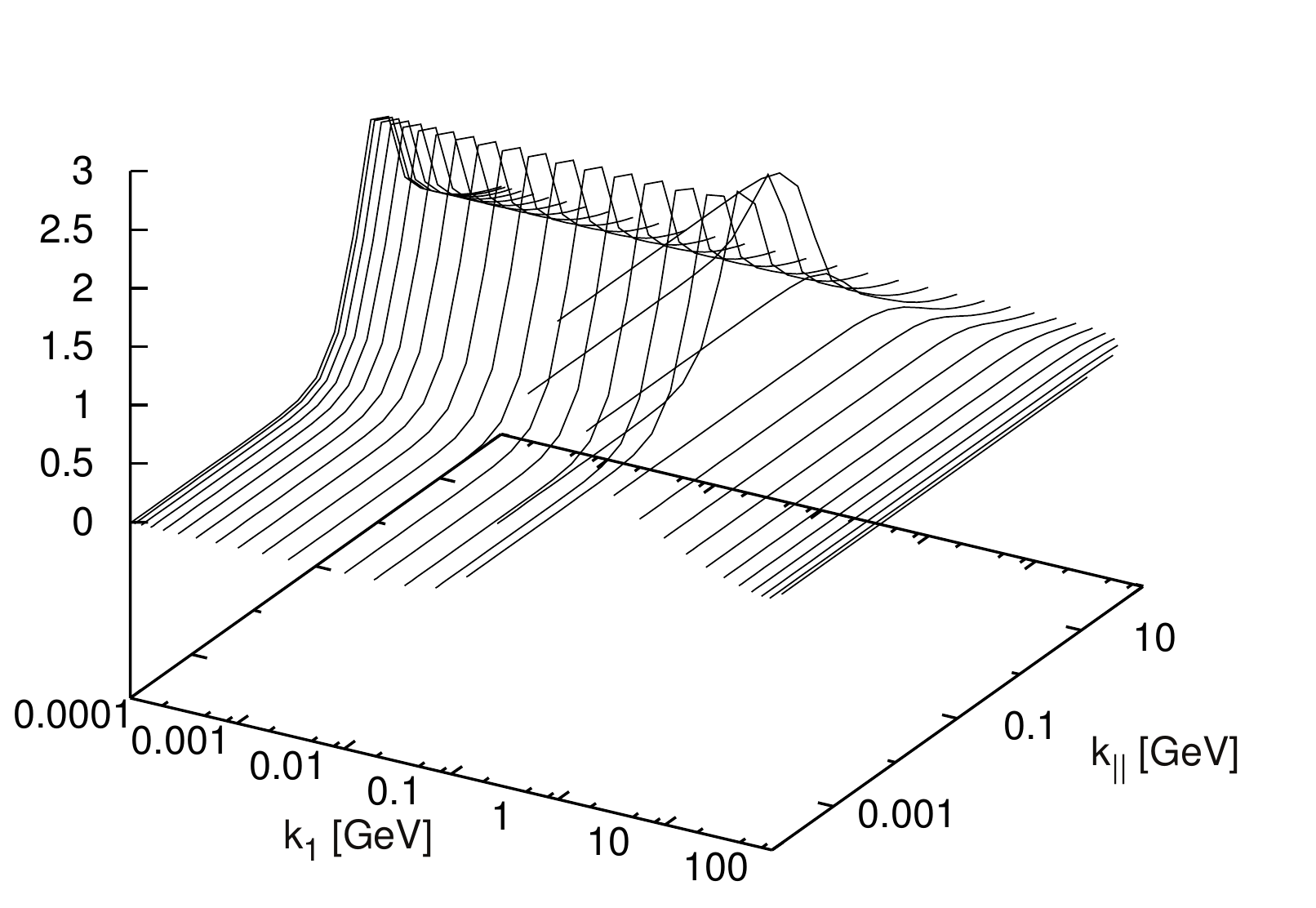}}
\subfloat[$Z_\parallel(k2,k_\parallel)$ \,\,($eH=$ 1 GeV$^2$)]
{\includegraphics[width=5.9cm]{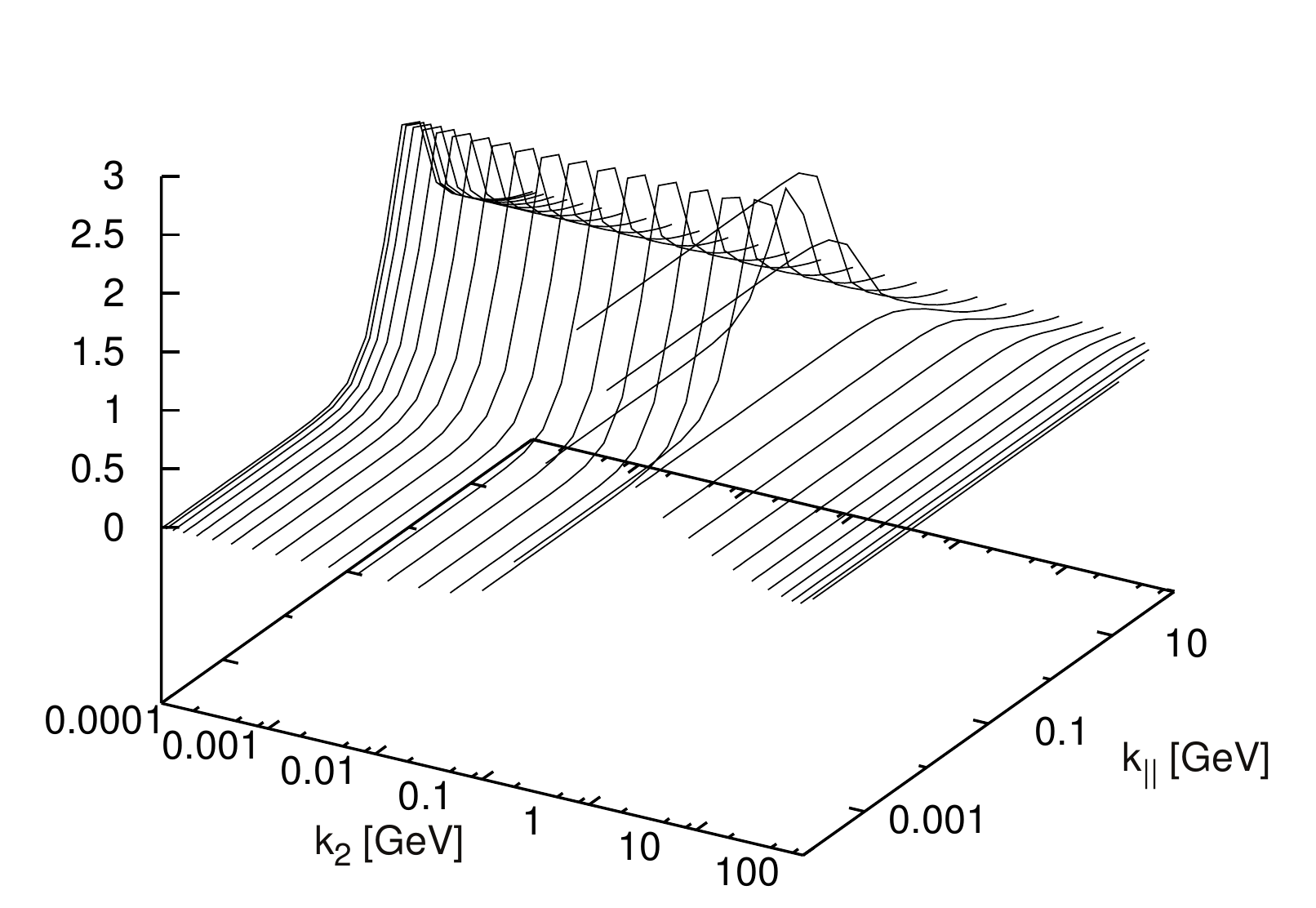}}\\
\subfloat[$Z_\parallel(k1,k2)$ {[}GeV{]} \,\,($eH=$ 4 GeV$^2$)]
{\includegraphics[width=5.9cm]{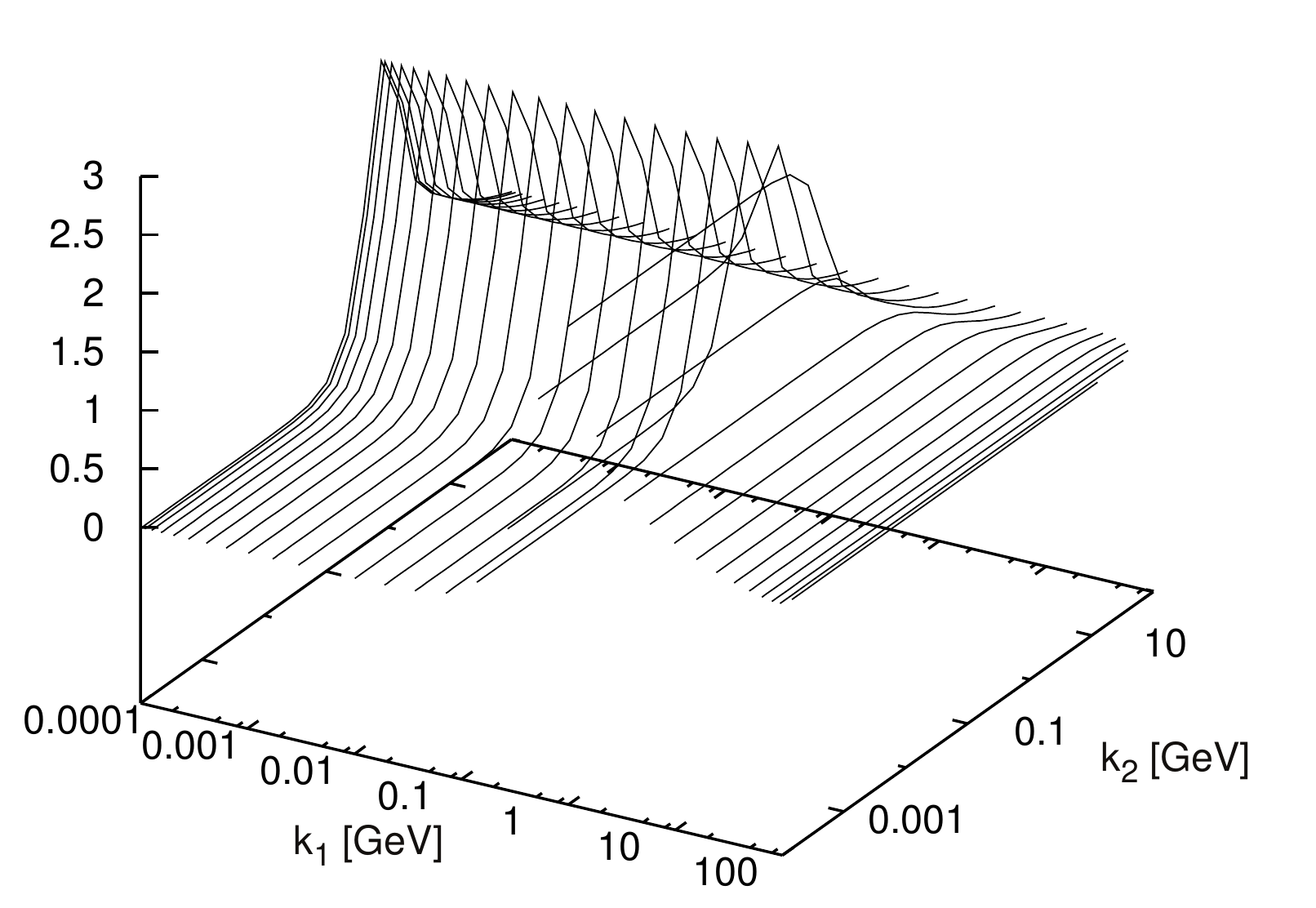}}
\subfloat[$Z_\parallel(k1,k_\parallel)$ \,\,($eH=$ 4 GeV$^2$)]
{\includegraphics[width=5.9cm]{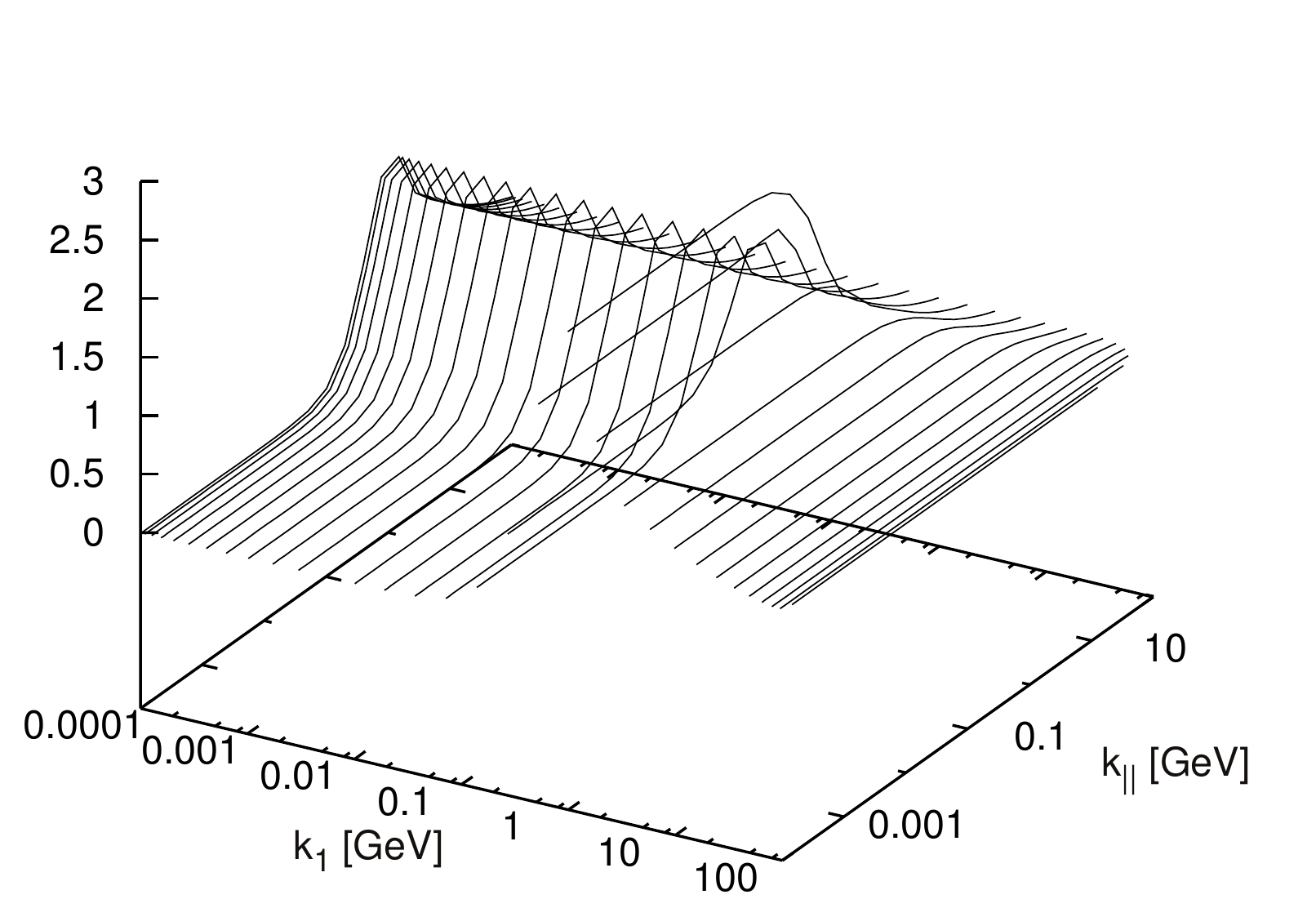}}
\subfloat[$Z_\parallel(k2,k_\parallel)$ \,\,($eH=$ 4 GeV$^2$)]
{\includegraphics[width=5.9cm]{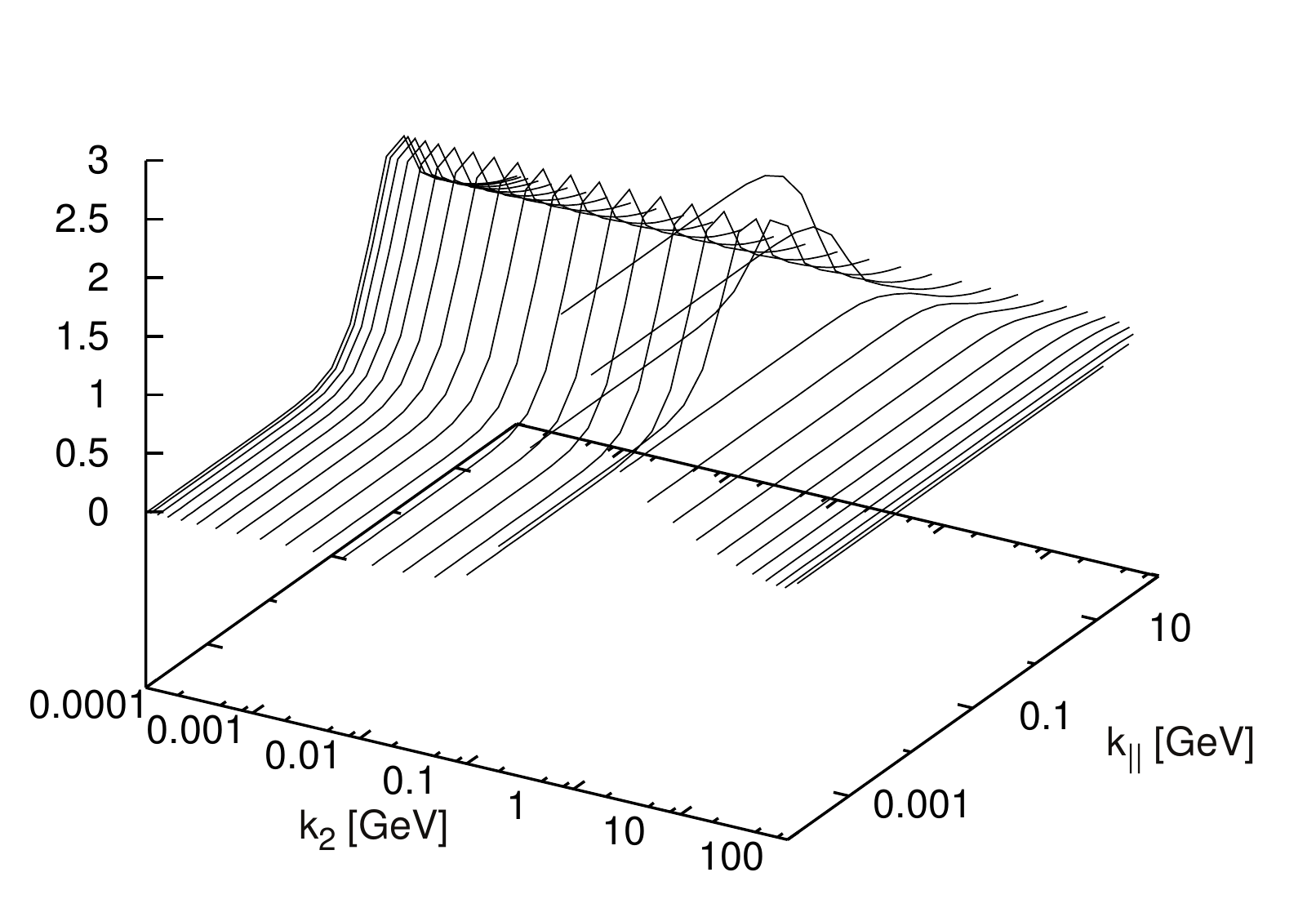}}\\
\caption{Gluon dressing function $Z_\parallel(k_1,k_2,k_\parallel)$ for 
$eH=$ 0 $\text{GeV}^2$ (quenched, first line),
$eH=$ 0.5 $\text{GeV}^2$ (second line), $eH=$ 1 $\text{GeV}^2$ 
(third line) and $eH=$ 4 $\text{GeV}^2$ (fourth line) and
different momentum slices, where the third momentum is set to zero
respectively.}
\label{fig:unq_gluon}
\end{figure}

Let us firstly discuss the effects of the magnetic field in the Yang-Mills sector. 
The only non-trivial (i.e. unquenched) longitudinal part $Z_\parallel(k_1,k_2,k_\parallel)$ 
of the gluon dressing is displayed in Fig.~(\ref{fig:unq_gluon}) for different momentum 
slices along $k_1 \equiv k_\perp^1$, $k_2 \equiv k_\perp^2$ and $k_\parallel$.
Overall, we find that the changes of the gluon propagator due to the magnetic field 
are very much dependent on the kinematics. For $Z_\parallel(k_1,k_2)$ almost nothing 
happens, whereas unquenching effects are largest for the low- and mid-momentum behavior
in $Z_\parallel(k_1,k_\parallel)$ and $Z_\parallel(k_2,k_\parallel)$, where the 
typical 'bump' in the gluon dressing function gets reduced by the presence of the
quarks. In general, this reduction is typical for unquenched systems and has been 
observed for the case of zero magnetic field in lattice as well as Dyson-Schwinger
studies (see e.g. \cite{Fischer:2003rp,Bowman:2004jm,Fischer:2005en,Aguilar:2013hoa}).
For stronger magnetic fields with growing effects due to dynamical chiral symmetry 
breaking, this reduction gets ever stronger, notably in the $k_1$ and $k_2$ directions
of the plots. In contrast, the $k_\parallel$-directions as well as the high ultraviolet 
behavior of the gluon dressing functions are hardly affected by the magnetic field. 

\begin{figure}[t!]
\includegraphics[width=8.5cm]{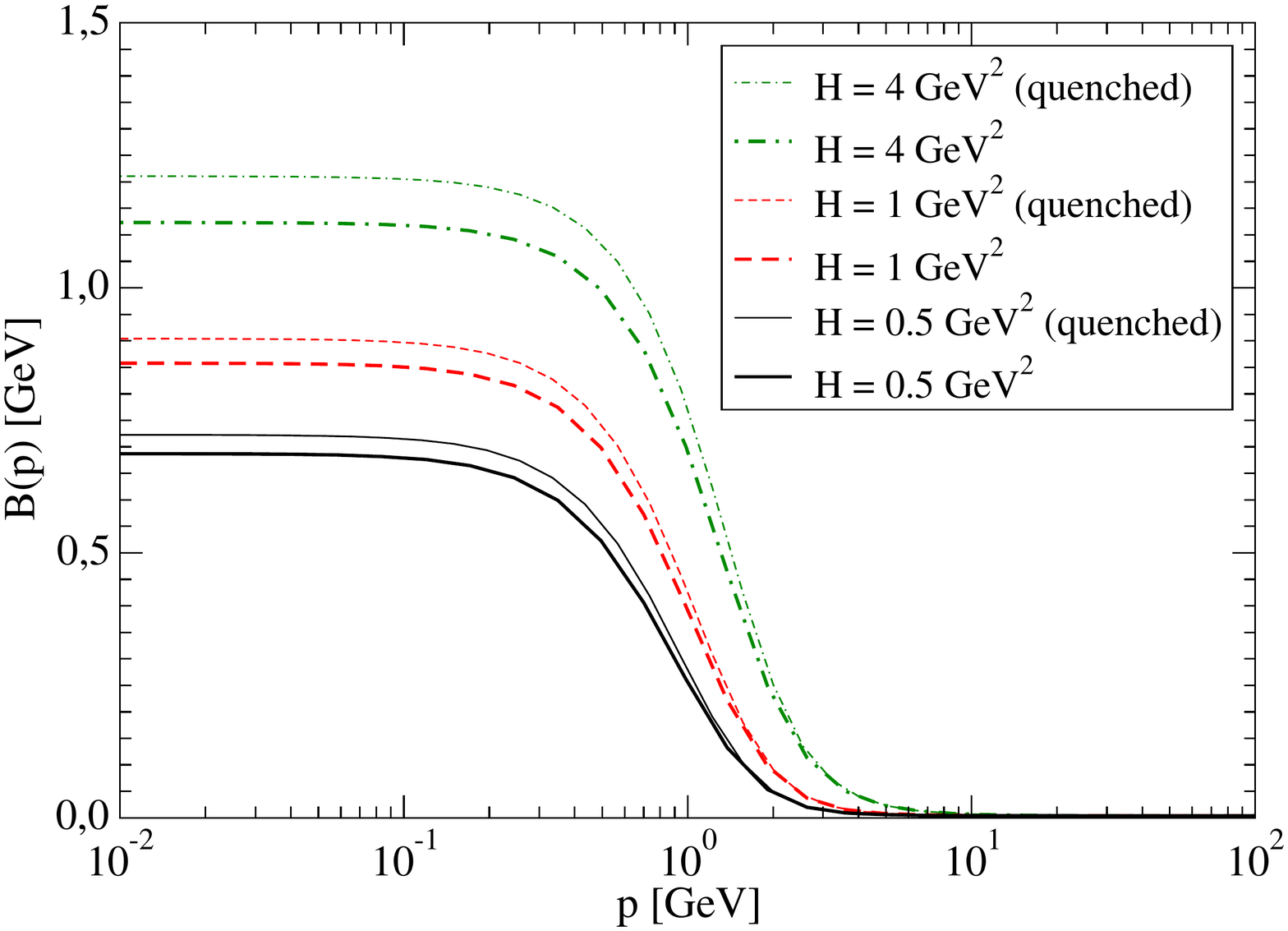}\hfill
\includegraphics[width=8.5cm]{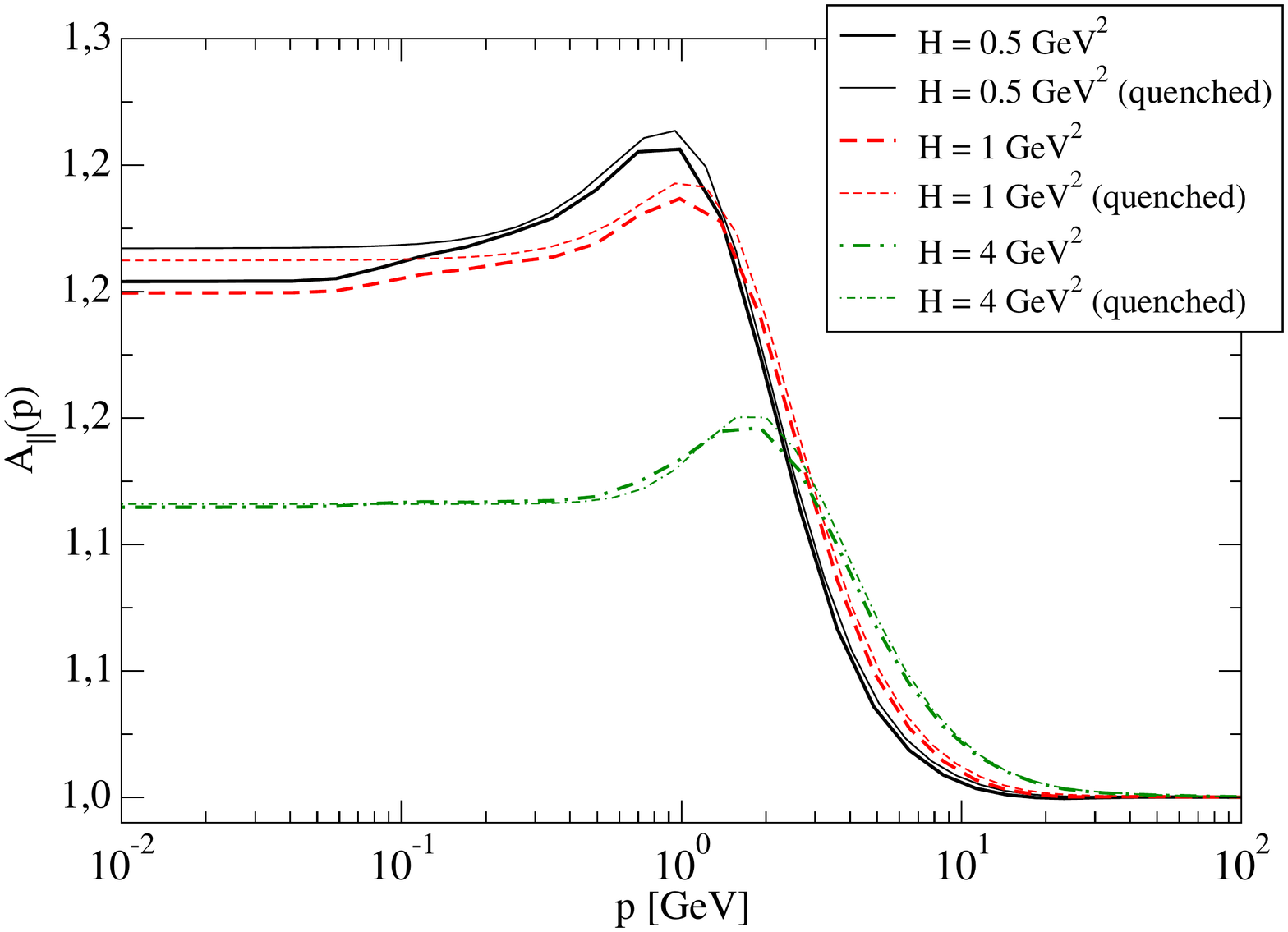}
\caption{Unquenched dressing functions $B$ and $A_\parallel$ of the up quark 
propagator (lowest Landau level) for different magnetic fields as a function 
of $p\equiv p_\parallel$ at a bare quark mass of m = 3.7 MeV at $\mu =$ 100 GeV. 
The dressing function $A_\perp$ is not defined on the lowest Landau
level.}
\label{fig:unqDress}
\includegraphics[width=8.5cm]{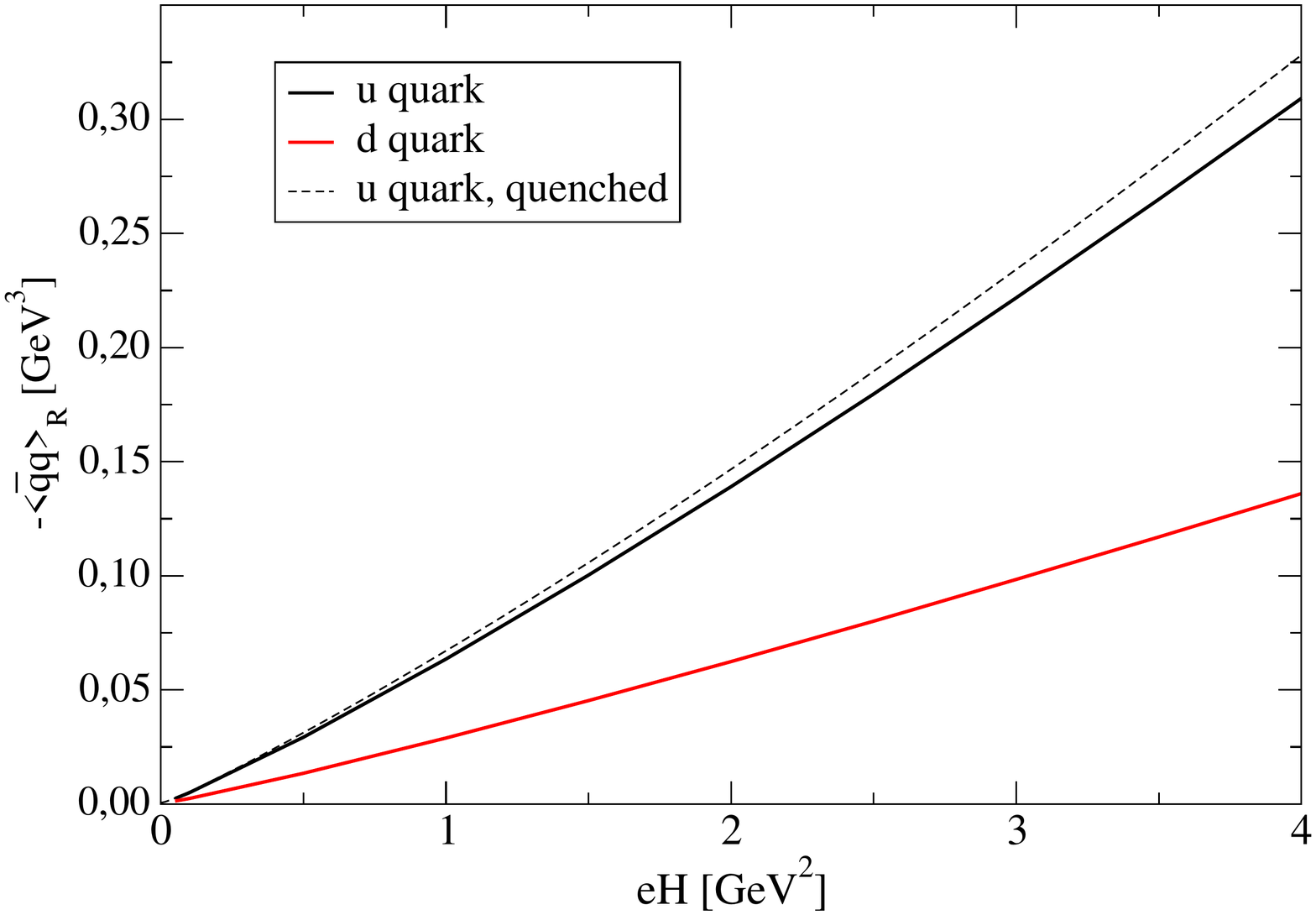}
\caption{Unquenched quark condensate for the up- and down-quark.}
\label{fig:unqchiralCond}
\end{figure}%

The corresponding dressing functions for the quark propagator are shown in 
Fig.~\ref{fig:unqDress}. We display the dressing functions $B$ and $A_\parallel$
for an up quark with charge $q_f = +2/3 e$ on the lowest Landau level as a 
function of momentum $p_\parallel$ and compare with the corresponding quenched
result. Note that $A_\perp$ is not shown, since it is only defined for higher
Landau levels. Also in the unquenched case we observe that the scalar dressing 
function $B$ grows with larger magnetic field. However, this growth is less 
pronounced as in the quenched case. Clearly, the reduction of the gluon dressing
function due to the quark-loop leads to reduced interaction strength in the
quark DSE as compared to the quenched case and this reduces the amount of chiral
symmetry breaking. For $A_\parallel$, displayed on the right hand side of 
Fig.~\ref{fig:unqDress}, we find only small changes. Similar to the quenched 
case we find an increase in $A_\parallel(0)$ as a function of magnetic field for 
smaller fields (not shown in the plot) with a maximum at $|eH| = 0.5$ GeV$^2$. 
As can be seen in Fig.~\ref{fig:unqDress}, for larger fields, $A_\parallel(0)$ 
decreases again, but the rate is considerably smaller than for the quenched case. 
For extremely large magnetic field we find that $A_\parallel(0) \approx 1$. 
This suggests that the different behavior found in section \ref{res:quenched} 
is particular to the quenched approximation.

Next we discuss the behavior of the quark condensate as a function of the external
field as displayed in Fig.~\ref{fig:unqchiralCond}. Here, we observe the breaking of isospin
symmetry due to the different charges of the up and down quarks, resulting in a different
slope of the condensate as a function of $eH$. Similar to the 
quenched case we find a power law behavior of the condensate proportional to a linear 
term and a term $\sim (eH)^{3/2}$ compatible with the expected behavior for the
condensate for fields $eH > m_\pi^2$ and at asymptotically large values of the field.
The unquenching effects in the condensate are small but qualitatively significant. 
On the one hand, the amount of condensate generated is decreased in accord with our
results for the scalar dressing function discussed above; the back-reaction of
the quarks onto the gluon leads to a {\it reduced amount of magnetic catalysis}
as compared to the quenched case. This finding agrees with the results of 
Ref.~\cite{Miransky:2002rp}. On the other hand, the range of magnetic fields which 
are dominated by the linear behavior of the condensate is of the same order. 
Whereas for the quenched case, the $(eH)^{3/2}$
term in the condensate becomes comparable in size with the linear one for
fields around $eH \sim 12 \,\mbox{GeV}^2$, this happens in the unquenched case 
around $eH \sim 14 \,\mbox{GeV}^2$. The corresponding fits to the up-quark condensate 
of the form
\begin{equation}
\langle \bar{q}q\rangle \sim a_1 |eH| + a_2 |eH|^{3/2},
\end{equation}
are given by $a_1=0.052 \,\text{GeV}, a_2=0.015$ in the quenched case and
$a_1=0.0503 \,\text{GeV}, a_2=0.0136$ for the unquenched case.

\begin{figure}[t!]
\includegraphics[width=8.5cm]{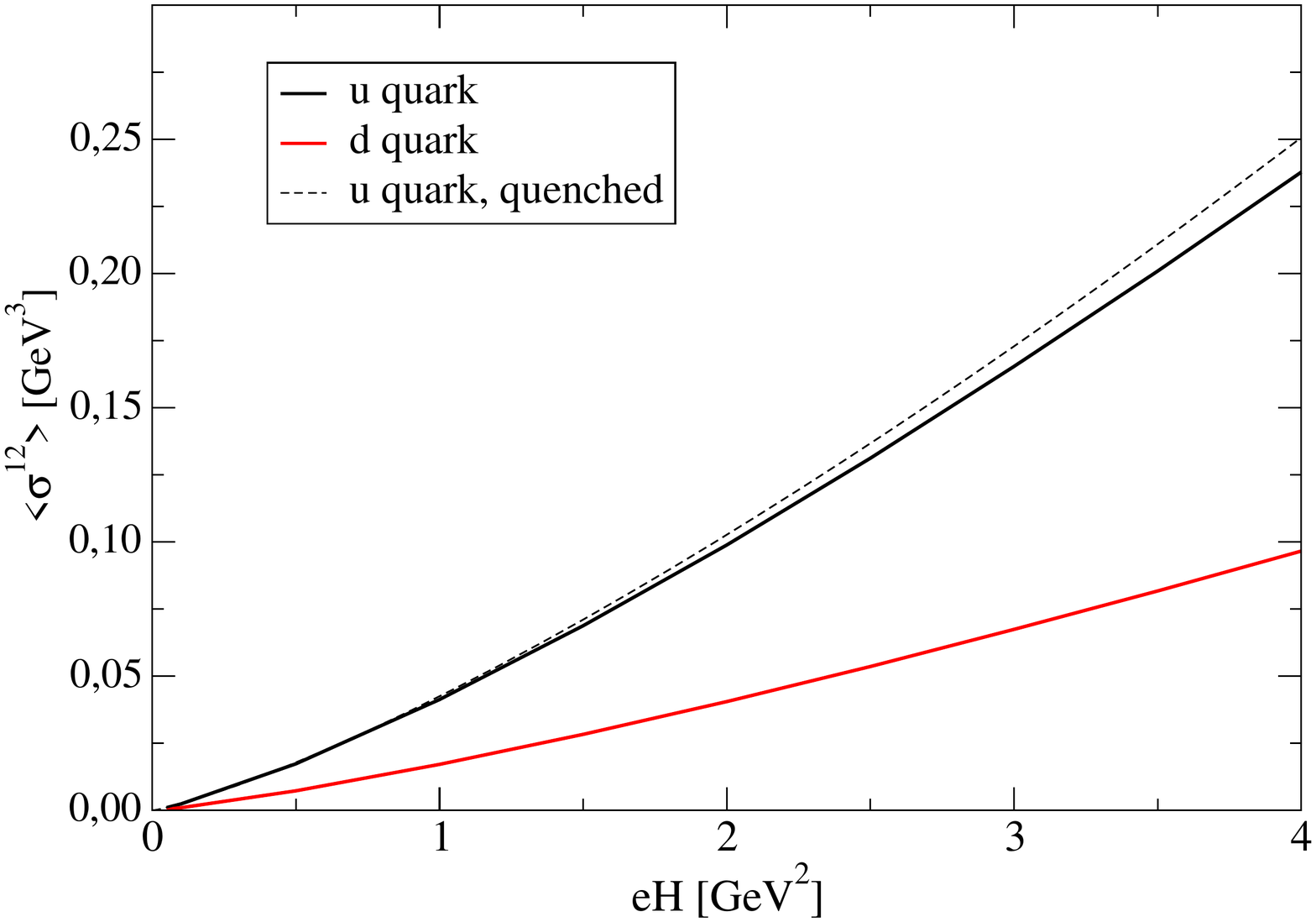}\hfill
\includegraphics[width=8.5cm]{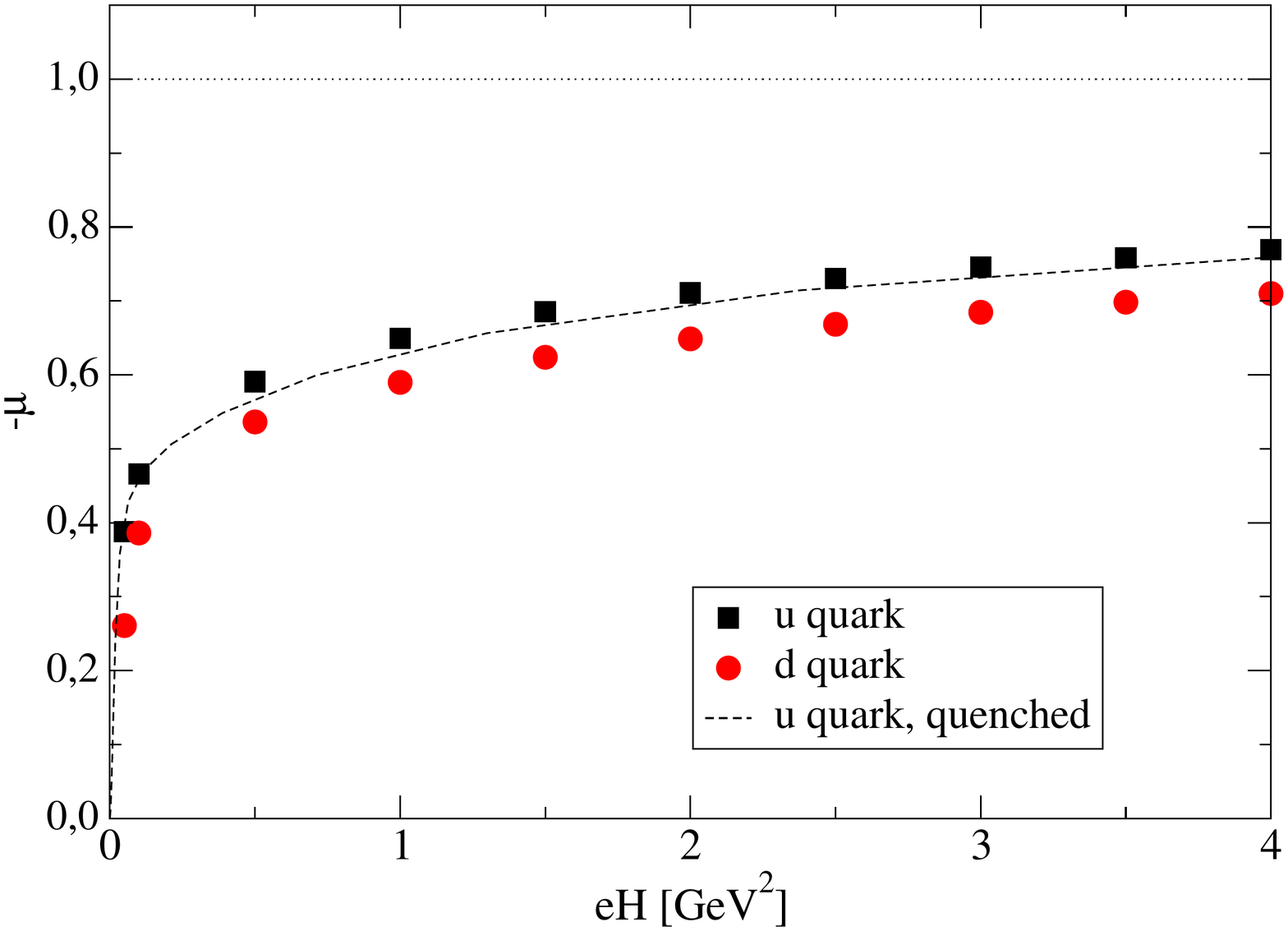}
\caption{Left hand side: regularized expectation value of the spin polarization 
tensor $\braket{\sigma^{12}}$.
Right hand side: regularized magnetic polarization $\mu$ of the QCD vacuum.}
\label{fig:unqpol}
\end{figure}

Similar effects as for the condensate can be observed for the expectation value of 
the spin polarization shown in the left diagram of Fig.~\ref{fig:unqpol}. The 
unquenching effects are quantatively similar as for the quark condensate. This
can also be seen in the magnetic polarization of the vacuum, shown in the
right diagram of Fig.~\ref{fig:unqpol}. Since the unquenching effects in the
condensate and spin polarisation are almost similar, the ratio of the two
is not drastically affected. Especially for large fields, the quenched results are
very close to the unquenched one, whereas for small fields we observe unquenching
corrections of the order of ten percent. Similar to the quenched case, the
polarizability rises only slowly with magnetic field and approaches its asymptotic
limit $\mu \rightarrow 1$ only for extremely large fields.

\section{Summary and conclusions} \label{summary}

In this work we studied the influence of (strong) magnetic fields onto the quark
and gluon propagators of Landau gauge QCD and the associated quark condensate and
spin polarization. Our most important observation is the {\it decrease 
in magnetic catalysis} induced by the back-reaction of the quarks onto the Yang-Mills
sector. We find a considerable reduction of the gluon dressing function $Z_\parallel$
in the mid-momentum region due to the magnetic field induced changes in the
quark-loop of the gluon DSE. Compared to the quenched case, this reduces the
interaction strength in the quark DSE and leads to a smaller amount of chiral
symmetry breaking, reducing the corresponding order parameters, i.e. the scalar
quark dressing function and the quark condensate. Unquenching effects in the 
gluon sector therefore contribute to {\it magnetic inhibition} in addition to the 
{\it magnetic catalysis} effects in the quark sector. This finding agrees with 
the interpretation of {\it inverse magnetic catalysis} due to magnetic effects on
the gluonic background given in the context of recent lattice studies 
\cite{Bali:2012zg,Bali:2013esa}.

For the quenched and unquenched quark condensate, we find a linear dependence on 
the magnetic field for $eH \ge \Lambda_{\text{QCD}}^2$, which gradually develops 
additional components $\sim (eH)^{3/2}$ for larger fields. This additional component 
becomes dominant only for extremely large magnetic fields indicating the asymptotic 
nature of this component.

Our framework takes into account also effects from higher Landau levels and therefore
enables us to assess the validity of the lowest Landau level (LLL) approximation. In 
general we observe sizable contributions from the higher Landau levels such that the 
LLL approximation, although valid on the ten percent level, becomes exact only at
asymptotically large fields. 

Finally, we like to emphasize that unquenching effects due to the hadronic 
back-reaction onto the system are not yet included in our truncation scheme.
These effects would show up in the details of the quark-gluon vertex \cite{Fischer:2007ze}
which need to be resolved diagrammatically for that purpose. In the model
study of Ref.~\cite{Fukushima:2012kc}, effects from neutral mesons are found 
to reduce the amount of quark condensate generated and therefore contribute 
qualitatively similar to the {\it magnetic inhibition} of the system as the
effects in the gluon sector discussed in this work. It remains to be seen in a 
more general study, how the unquenching effects in the gluon and meson sectors 
compare on a quantitative basis. Very recent results indeed suggest, that meson 
effects alone are not sufficient to explain inverse magnetic catalysis at finite 
temperature \cite{Kamikado:2013pya}.

\section*{Acknowledgement}
CF thanks the Yukawa Institute for Theoretical Physics, Kyoto University, 
where this work was completed during the YITP-T-13-05 workshop
on 'New Frontiers in QCD'. This work was supported by the Helmholtz 
International Center for FAIR within the LOEWE program of the State of 
Hesse and the Studienstiftung des deutschen Volkes.


\begin{appendix}
\section{Gluon propagator and quark-gluon vertex}\label{gluonvertex}
In this study we employ a truncation scheme for the quark-gluon vertex based on results found 
in \cite{Fischer:2010fx} with some minor modifications. There, for the quenched gluon propagator,
a fit to lattice data has been employed. It is given by 
\begin{equation}
Z(k^2)=\frac{q^2\Lambda^2}{(q^2+\Lambda^2)^2}\left[ \left( \frac{c}{q^2+a\Lambda^2}\right)^b + \frac{q^2}{\Lambda^2}\left( \frac{\beta_0\alpha(\mu)\log{q^2/\Lambda^2+1}}{4\pi}\right)^\gamma  \right]\label{eq:gluonQuenched}
\end{equation}
with the parameters
\begin{eqnarray}
a=0.60 & b = 1.36 & \Lambda =  1.4\text{ GeV}\\
c = 11.5\text{ GeV}^2 & \beta_0 = 11N_c/3 & \gamma = - 13/22
\end{eqnarray}
where $\alpha(\mu)=0.3$. Since the quenched gluon propagator does not get modified by the presence of 
an external magnetic field, this form is exact within the limits of the systematic error of the lattice 
data. In our calculations of the unquenched gluon propagator, this form acts as a seed which is supplemented
by the quark-loop, see the main text for details.

For the quark-gluon vertex we use the approximation $\Gamma^\nu\rightarrow\gamma^\nu\Gamma(k^2)$ with
\begin{equation}
\Gamma(k^2)=\frac{d_1}{d_2+q^2}+\frac{q^2}{\Lambda^2+q^2}
\left( \frac{\beta_0\alpha(\mu)\log{q^2/\Lambda^2+1}}{4\pi} \right)^{2\delta}\,,
\end{equation}
where $k$ is the gluon momentum. The parameters used are
\begin{eqnarray}
d_1=7.9\text{ GeV}^2 & d_2=0.5\text{ GeV}^2 \\  \delta = -18/88 & \Lambda =  1.4\text{ GeV}
\end{eqnarray}
The form of the ansatz is similar than in \cite{Fischer:2010fx}. However, there this ansatz has been employed
together with the first term of the Ball-Chiu form of the vertex. Here, we use it together with a bare vertex,
which results in a change of the strength parameter $d_1$, which is $d_1=7.9 \text{ GeV}^2$ instead of 
$d_1=4.6 \text{ GeV}^2$ as in the reference. The other parameter $d_2$ represents a scale, which is adjusted 
to the scale inherent in the lattice data for the gluon propagator and remains unchanged as compared with 
\cite{Fischer:2010fx}.
Note that the vertex above is given in terms of the gluon momentum only, which in the Ritus case is still a 
"physical" momentum (in contrast to the Ritus eigenvalues). This makes it particularly simple and renders our
study feasible. In future work, a more refined vertex construction may involve the Ward-identity in the
presence of magnetic fields \cite{Ferrer:1998vw}. 
\end{appendix}

\end{document}